\documentclass[journal,romanappendices]{IEEEtran-v17}
\pdfoutput=1

\usepackage{xspace}
\usepackage{url}
\usepackage[cmex10]{amsmath}
\usepackage{bbm}
\usepackage{graphicx}

\usepackage{tikz}

\usepackage{fancyref} % for \fref
\usepackage{amsmath}
\usepackage{amssymb}
\usepackage{dsfont}
\usepackage{footnote}
\usepackage[nosort]{cite}

\usepackage{color}
\definecolor{links}{rgb}{0.7,0,0}   % red
\definecolor{urls}{rgb}{0,0,0.8}    % blue
\definecolor{cites}{rgb}{0,0,0.8}   % blue

\usepackage[colorlinks,hyperindex,linkcolor=links,citecolor=cites,urlcolor=urls]{hyperref}

\usepackage{bm}
\usepackage{upgreek}
\usepackage{amssymb}
\usepackage{amsfonts}
\usepackage{mathrsfs}
\usepackage{xspace}

\newcommand{\safemath}[2]{\newcommand{#1}{\ensuremath{#2}\xspace}}
\safemath{\opE}{\mathbb{E}}
\newcommand{\Ex}[2]{\ensuremath{\opE_{#1}\lefto[#2\right]}} 	% expectation
\safemath{\prob}{\mathbb{P}}
\safemath{\bigO}{\mathcal{O}}
\safemath{\littleo}{\mathit{o}}

\newtheorem{thm}{Theorem}
\newtheorem{lemma}[thm]{Lemma}

\newtheorem{rem}{Remark}
\newcommand{\indfun}[1]{\mathbbmss{1}\mathopen{}\left\{#1\right\}}

\safemath{\matA}{\mathsf{A}}
\safemath{\matB}{\mathsf{B}}
\safemath{\matC}{\mathsf{C}}
\safemath{\matD}{\mathsf{D}}
\safemath{\matE}{\mathsf{E}}
\safemath{\matF}{\mathsf{F}}
\safemath{\matG}{\mathsf{G}}
\safemath{\matH}{\mathsf{H}}
\safemath{\matI}{\mathsf{I}}
\safemath{\matJ}{\mathsf{J}}
\safemath{\matK}{\mathsf{K}}
\safemath{\matL}{\mathsf{L}}
\safemath{\matM}{\mathsf{M}}
\safemath{\matN}{\mathsf{N}}
\safemath{\matO}{\mathsf{O}}
\safemath{\matP}{\mathsf{P}}
\safemath{\matQ}{\mathsf{Q}}
\safemath{\matR}{\mathsf{R}}
\safemath{\matS}{\mathsf{S}}
\safemath{\matT}{\mathsf{T}}
\safemath{\matU}{\mathsf{U}}
\safemath{\matV}{\mathsf{V}}
\safemath{\matW}{\mathsf{W}}
\safemath{\matX}{\mathsf{X}}
\safemath{\matY}{\mathsf{Y}}
\safemath{\matZ}{\mathsf{Z}}

\safemath{\randveca}{\bm{A}}
\safemath{\randvecb}{\bm{B}}
\safemath{\randvecc}{\bm{C}}
\safemath{\randvecd}{\bm{D}}
\safemath{\randvece}{\bm{E}}
\safemath{\randvecf}{\bm{F}}
\safemath{\randvecg}{\bm{G}}
\safemath{\randvech}{\bm{H}}
\safemath{\randveci}{\bm{I}}
\safemath{\randvecj}{\bm{J}}
\safemath{\randveck}{\bm{K}}
\safemath{\randvecl}{\bm{L}}
\safemath{\randvecm}{\bm{M}}
\safemath{\randvecn}{\bm{N}}
\safemath{\randveco}{\bm{O}}
\safemath{\randvecp}{\bm{P}}
\safemath{\randvecq}{\bm{Q}}
\safemath{\randvecr}{\bm{R}}
\safemath{\randvecs}{\bm{S}}
\safemath{\randvect}{\bm{T}}
\safemath{\randvecu}{\bm{U}}
\safemath{\randvecv}{\bm{V}}
\safemath{\randvecw}{\bm{W}}
\safemath{\randvecx}{\bm{X}}
\safemath{\randvecy}{\bm{Y}}
\safemath{\randvecz}{\bm{Z}}

\safemath{\randmatA}{\mathbb{A}}
\safemath{\randmatB}{\mathbb{B}}
\safemath{\randmatC}{\mathbb{C}}
\safemath{\randmatD}{\mathbb{D}}
\safemath{\randmatE}{\mathbb{E}}
\safemath{\randmatF}{\mathbb{F}}
\safemath{\randmatG}{\mathbb{G}}
\safemath{\randmatH}{\mathbb{H}}
\safemath{\randmatI}{\mathbb{I}}
\safemath{\randmatJ}{\mathbb{J}}
\safemath{\randmatK}{\mathbb{K}}
\safemath{\randmatL}{\mathbb{L}}
\safemath{\randmatM}{\mathbb{M}}
\safemath{\randmatN}{\mathbb{N}}
\safemath{\randmatO}{\mathbb{O}}
\safemath{\randmatP}{\mathbb{P}}
\safemath{\randmatQ}{\mathbb{Q}}
\safemath{\randmatR}{\mathbb{R}}
\safemath{\randmatS}{\mathbb{S}}
\safemath{\randmatT}{\mathbb{T}}
\safemath{\randmatU}{\mathbb{U}}
\safemath{\randmatV}{\mathbb{V}}
\safemath{\randmatW}{\mathbb{W}}
\safemath{\randmatX}{\mathbb{X}}
\safemath{\randmatY}{\mathbb{Y}}
\safemath{\randmatZ}{\mathbb{Z}}

\safemath{\pdff}{f}
\safemath{\pdfp}{p}
\safemath{\pdfq}{q}

\safemath{\cdfF}{F}
\safemath{\cdfP}{P}
\safemath{\cdfQ}{Q}

\safemath{\veca}{\bm{a}}
\safemath{\vecb}{\bm{b}}
\safemath{\vecc}{\bm{c}}
\safemath{\vecd}{\bm{d}}
\safemath{\vece}{\bm{e}}
\safemath{\vecf}{\bm{f}}
\safemath{\vecg}{\bm{g}}
\safemath{\vech}{\bm{h}}
\safemath{\veci}{\bm{i}}
\safemath{\vecj}{\bm{j}}
\safemath{\veck}{\bm{k}}
\safemath{\vecl}{\bm{l}}
\safemath{\vecm}{\bm{m}}
\safemath{\vecn}{\bm{n}}
\safemath{\veco}{\bm{o}}
\safemath{\vecp}{\bm{p}}
\safemath{\vecq}{\bm{q}}
\safemath{\vecr}{\bm{r}}
\safemath{\vecs}{\bm{s}}
\safemath{\vect}{\bm{t}}
\safemath{\vecu}{\bm{u}}
\safemath{\vecv}{\bm{v}}
\safemath{\vecw}{\bm{w}}
\safemath{\vecx}{\bm{x}}
\safemath{\vecy}{\bm{y}}
\safemath{\vecz}{\bm{z}}

\safemath{\matSigma}{\bm{\Sigma}}

\safemath{\setA}{\mathcal{A}}
\safemath{\setB}{\mathcal{B}}
\safemath{\setC}{\mathcal{C}}
\safemath{\setD}{\mathcal{D}}
\safemath{\setE}{\mathcal{E}}
\safemath{\setF}{\mathcal{F}}
\safemath{\setG}{\mathcal{G}}
\safemath{\setH}{\mathcal{H}}
\safemath{\setI}{\mathcal{I}}
\safemath{\setJ}{\mathcal{J}}
\safemath{\setK}{\mathcal{K}}
\safemath{\setL}{\mathcal{L}}
\safemath{\setM}{\mathcal{M}}
\safemath{\setN}{\mathcal{N}}
\safemath{\setO}{\mathcal{O}}
\safemath{\setP}{\mathcal{P}}
\safemath{\setQ}{\mathcal{Q}}
\safemath{\setR}{\mathcal{R}}
\safemath{\setS}{\mathcal{S}}
\safemath{\setT}{\mathcal{T}}
\safemath{\setU}{\mathcal{U}}
\safemath{\setV}{\mathcal{V}}
\safemath{\setW}{\mathcal{W}}
\safemath{\setX}{\mathcal{X}}
\safemath{\setY}{\mathcal{Y}}
\safemath{\setZ}{\mathcal{Z}}
\safemath{\emptySet}{\varnothing}

\safemath{\veczero}{\mathbf{0}} %vector font of 0,

\safemath{\diag}{\mathrm{diag}}
\safemath{\jpg}{\mathcal{CN}}			% jointly proper Gaussian
\safemath{\complexset}{\mathbb{C}}
\safemath{\realset}{\mathbb{R}}

\usepackage{bm}
\usepackage{upgreek}
\usepackage{amssymb}
\usepackage{amsfonts}
\usepackage{mathrsfs}
\usepackage{xspace}

\safemath{\noise}{Z}
\safemath{\vecnoise}{\randvecz}

\safemath{\optF}{F^*}

\safemath{\avg}{\mathrm{lt}}
\safemath{\st}{\mathrm{st}}
\safemath{\lt}{\mathrm{lt}}
\safemath{\awgn}{\mathrm{awgn}}
\safemath{\qs}{\mathrm{qs}}
\safemath{\Pout}{P_{\mathrm{out}}}
\safemath{\thres}{\mathrm{th}}
\safemath{\avgg}{\bar{g}_{\error}}
\safemath{\avggn}{\bar{g}_{\bl}}
\safemath{\func}{q}

\safemath{\Finvs}{F_{\mathrm{inv}}}

\safemath{\binent}{H_{\mathrm{b}}}
\safemath{\gth}{g_{\mathrm{th}}}

\safemath{\Cawgn}{C}
\safemath{\Ceawgn}{ C_{\error}}
\safemath{\Vawgn}{V}
\safemath{\Veawgn}{V_{\error}}

\safemath{\powercontst}{\hat{\normxre}}

\safemath{\Ceqsst}{C^{\qs}_{\error}}
\safemath{\ceqslt}{C^{\qs,\lt}_{\error}}
\safemath{\Cerror}{C_\error}
\safemath{\Rquasi}{R_{\qs,\lt}}
\safemath{\Rawgn}{R_{\mathrm{awgn},\lt}}
\safemath{\Rawgnst}{R_{\mathrm{awgn}}}
\safemath{\Rquasils}{R_{\qs,\mathrm{ls}}}
\safemath{\Rquasist}{R_{\qs}}

\safemath{\Rawgnnalt}{R_{\mathrm{awgn},\lt}^{\mathcal{N}}}
\safemath{\Rawgnnast}{R_{\mathrm{awgn},\st}^{\mathcal{N}}}
\safemath{\Rqsltna}{R_{\qs,\lt}^{\mathcal{N}}}
\safemath{\Ceqs}{C_{\error}^{\qs}}
\safemath{\Veqs}{V_{\error}^{\qs}}
\safemath{\Ceqsls}{C_{\error}^{\qs}}

\safemath{\powalloc}{v}

\safemath{\polyset}{\setA}
\safemath{\extpoint}{\veca^*}
\safemath{\extpointi}{a^*}

\safemath{\normx}{\Pi}
\safemath{\normy}{V}
\safemath{\lognormy}{U}

\safemath{\normxre}{\pi}
\safemath{\tangentp}{\omega_0}
\safemath{\normyre}{u}
\safemath{\optP}{P_{\normx}^*}

\safemath{\NonnegReal}{\mathbb{R}_{+}}

\safemath{\indist}{\cdfP} %input distribution
\safemath{\outdist}{\cdfQ} %output distribution
\safemath{\inpdf}{\pdfp} %input pdf
\safemath{\outpdf}{\pdfq} %output pdf
\safemath{\testdist}{\cdfP} %the distribution of test P_{Z|W}

\safemath{\inset}{\setF} %the input power set

\safemath{\encoder}{f} % the encoder of the code
\safemath{\decoder}{g} % the decoder of the code
\safemath{\msg}{J} % the message
\safemath{\csir}{\mathrm{r}}
\safemath{\csit}{\mathrm{t}}
\safemath{\csi}{\mathrm{csi}}
\safemath{\csirt}{\mathrm{rt}}
\safemath{\Rcsirt}{R_{\csirt}} %rate with CSIT
\safemath{\Rcsir}{R_{\csir}}
\safemath{\Rcsit}{R_{\csit}}

\safemath{\Vnocsit}{V_{\error}^{\mathrm{no}}}
\safemath{\Vcsit}{V_{\error}^{\mathrm{rt}}}

\safemath{\cadist}{F_C} %capacity distribution

\safemath{\deF}{d_0} %the derivative of F_C(\xi) %\left.\frac{d\cadist(\argpn)}{d\argpn}\right|_{\argpn=C_\error}

\safemath{\bl}{n} %block length
\safemath{\error}{\epsilon} %prob of error
\safemath{\NumCode}{M}
\safemath{\RXant}{r}
\safemath{\rxant}{r}
\safemath{\snr}{\rho}

\safemath{\BoundFU}{k_{\delta}}

\safemath{\argpn}{\gamma} % the argument in the probability p(R)

\safemath{\angletest}{Z} %the angle test used in achievability part

\safemath{\errorach}{P_\mathrm{e}} %the error in intuition part

\newcommand{\given}{\,\vert\,}				% conditioning
\safemath{\define}{\triangleq}			% definition
\safemath{\fnorm}{\mathrm{F}}

\safemath{\altbl}{\tilde{\bl}}
\safemath{\constL}{k_{\mathrm{L}}}
\safemath{\constU}{k_{\mathrm{U}}}
\safemath{\funcL}{\tilde{q}}
\safemath{\funcU}{q}
\safemath{\ConstThm}{k_0}
\safemath{\randrevec}{\randvecy}
\safemath{\revec}{\vecy}
\safemath{\trcwd}{\vecx}
\safemath{\randtrcwd}{\randvecx}
\safemath{\randnoisevec}{\randvecw}
\safemath{\transmitcwd}{\vecx_1} %the transmitted codeword in achievability
\safemath{\pickcwdnoch}{\vecx_0} %the chosen codeword x=[P,P,...,P]
\safemath{\pickcwd}{\vecx_0} %the chosen codeword x=[P,P,...,P]

\safemath{\inseqrand}{\randvecx}
\safemath{\inseq}{\vecx}

\safemath{\outseq}{\matY}
\safemath{\outseqrand}{\randmatY}

\safemath{\altT}{\widetilde{T}}
\safemath{\altU}{\widetilde{U}}
\safemath{\altmean}{\tilde{\mu}}
\safemath{\altvar}{\tilde{\sigma}}
\safemath{\altf}{\funcL}
\safemath{\altg}{\tilde{g}}
\safemath{\altgamma}{\tilde{\gamma}}
\safemath{\altdelta}{\tilde{\delta}}
\safemath{\altk}{\tilde{k}}
\safemath{\constant}{\tilde{k}}

\safemath{\pdfG}{dP_G}

\begin{document}
\IEEEoverridecommandlockouts

\title{Optimum Power Control at Finite Blocklength}

\author{\IEEEauthorblockN{Wei Yang,~\IEEEmembership{Student Member,~IEEE}, Giuseppe Caire,~\IEEEmembership{Fellow,~IEEE},\\ Giuseppe Durisi,~\IEEEmembership{Senior Member,~IEEE}, and Yury Polyanskiy,~\IEEEmembership{Senior Member,~IEEE}}
\thanks{This work was supported in part by the Swedish Research Council (VR) under grant no. 2012-4571.
The material of this paper was presented in part at the IEEE International Symposium on Information Theory (ISIT), Honolulu, HI, USA, July 2014.}
\thanks{W. Yang and G. Durisi are with the Department of Signals and Systems, Chalmers University of Technology, 41296,  Gothenburg, Sweden (e-mail: \{ywei, durisi\}@chalmers.se).}
\thanks{G. Caire is with the Department of Electrical Engineering, University of Southern California, Los Angeles, CA, 90089 USA (e-mail: caire@usc.edu).}
\thanks{Y. Polyanskiy is with the Department of Electrical Engineering and Computer Science, MIT, Cambridge, MA, 02139 USA (e-mail: yp@mit.edu).}

\thanks{Copyright (c) 2014 IEEE. Personal use of this material is permitted.  However, permission to use this material for any other purposes must be obtained from the IEEE by sending a request to pubs-permissions@ieee.org.}
}

\maketitle

\begin{abstract}
This paper investigates the maximal channel coding rate achievable at a given blocklength $\bl$ and error probability $\error$, when the codewords are subject to a long-term (i.e., averaged-over-all-codeword) power constraint.
The second-order term in the large-\bl expansion of the maximal channel coding rate is characterized both for additive white Gaussian noise (AWGN) channels and for quasi-static fading channels with perfect channel state information available at both the transmitter and the  receiver.
It is shown that in both cases the second-order term is proportional to $\sqrt{\bl^{-1}\ln\bl}$.
For the quasi-static fading case, this second-order term is achieved by \emph{truncated channel inversion}, namely, by concatenating
a dispersion-optimal code for an AWGN channel subject to a short-term power constraint, with a power controller that inverts the channel whenever the fading gain is above a certain threshold.
Easy-to-evaluate approximations of the maximal channel coding rate are developed for both the AWGN and the quasi-static fading case.

\end{abstract}

\section{Introduction}

\label{sec:intro}

% Recent focus on finite-blocklength results in information theory has uncovered a number of interesting implications, which are important both for theorists and system designers.
% %
% Examples include: feedback is useless for increasing the capacity of memoryless channels, but is immensely useful non-asymptotically~\cite{polyanskiy11-08};
% separate source channel coding achieves the first-order optimality, but is rather suboptimal in the dispersion term~\cite{kostina13-05};
% large coherence time of the noncoherent fading channel improves capacity but decreases the non-asymptotic channel coding rate~\cite{yang12-09};
% independent isotropic Gaussian-like signals are capacity-optimal in multi-antenna systems, but are not dispersion-optimal~\cite{collins14-07}, etc.
Recent works in finite-blocklength information theory have shed additional light on a number of cases where asymptotic results yield inaccurate engineering insights on the design of communication systems
once a constraint on the codeword length is imposed.
For example, feedback does not increase the capacity of memoryless discrete-time channels, but is exceedingly useful at finite blocklength~\cite{polyanskiy11-08c};
separate source-channel coding is first-order but not second-order optimal~\cite{kostina13-05}; the capacity of block-memoryless fading channels in the noncoherent setting increases monotonically with the coherence time, whereas for the nonasymptotic coding rate there exists a rate maximizing coherence time after which the coding rate starts decreasing~\cite{yang12-09,durisi14-12}; independent isotropic Gaussian-like signals achieve the capacity of multi-antenna channels under the assumption of perfect channel state information (CSI) at the receiver, but are not dispersion optimal~\cite{collins14-07}.
While some of the insights listed above were known already from earlier works on error exponents (see, e.g.,~\cite{burnashev76-04,csiszar80-05}), analyses under the assumption of finite blocklength and nonvanishing error probability may be more relevant for the design of modern communication systems.

In this paper, we analyze a scenario, namely communication over a quasi-static fading channel subject to a \textit{long-term} power constraint, for which the asymptotically optimal transmission strategy turns out to perform well also at finite blocklength.
Specifically, we consider a quasi-static single-antenna fading channel with input-output relation given by
\begin{equation}
\randvecy = H \vecx + \vecnoise.
\label{eq:channel-io-qs}
\end{equation}
Here, $\vecx \in \complexset^{\bl}$ is the transmitted codeword; $H$ denotes the complex fading coefficient, which is random but remains constant for all~$\bl$ channel uses; and $\vecnoise\sim \jpg(\mathbf{0}, \matI_{\bl})$ is the additive white Gaussian noise vector.
We study the maximal channel coding rate $\Rquasi^*(\bl,\error)$ achievable at a given blocklength~$\bl$ and average error probability~$\error$ over the channel~\eqref{eq:channel-io-qs}.
We assume that both the transmitter and the receiver have perfect CSI, i.e., perfect knowledge of the fading gain $H$.
To exploit the benefit of transmit CSI (CSIT), we consider the scenario where the codewords are subject to a long-term power constraint, i.e., the average power of the transmitted codewords, averaged over all messages and all channel realizations, is limited.
This is in contrast to the conventional \emph{short-term} power constraint, where the power of each transmitted codeword is limited.
From a practical perspective, a long-term power constraint is useful in situations where the power limitation comes from energy efficiency considerations.
For example, it captures the relatively long battery life of mobile terminals (at least compared to the duration of a codeword) in the  uplink of cellular communication systems~\cite{berry02-05}.
The notion of long-term power constraint is widely used in the wireless communication literature (see, e.g.,~\cite{yates97-09,goldsmith97-11a,hanly98-11}) as it opens up the possibility to perform a dynamic allocation of power and rate based on the current channel state (also known as link adaptation~\cite{lozano10-09a}).

\begin{figure}[t]
\centering
\includegraphics[scale=0.57]{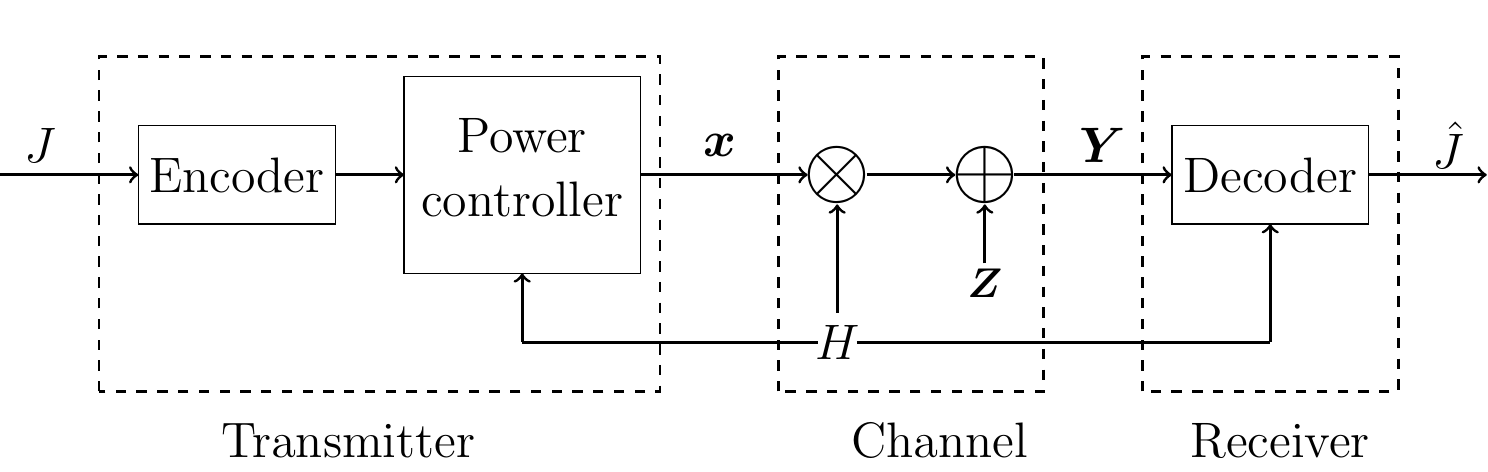}
\caption{\label{fig:arch-sept} A single-codebook, variable-power scheme achieves~\eqref{eq:epsilon-c-lt} (see~\cite{caire99-05}).}
\end{figure}

For the scenario described above, the asymptotic limit $\lim_{\bl\to\infty} \Rquasi^*(\bl,\error)$, which gives the so called $\epsilon$-capacity (also known as outage capacity), was characterized in~\cite{caire99-05}.
Specifically, it follows from~\cite[Props.~1 and~4]{caire99-05} that for quasi-static single-antenna fading channels subject to the long-term power constraint\footnote{This holds under regularity conditions on the probability distribution of $G$. A sufficient condition is that $\Cawgn(\snr/ \avgg)$, or equivalently, $\Finvs(\error)$ defined in~\eqref{eq:def-finvs}, is continuous in $\error$~\cite{verdu94-07a}. A more general condition is provided in Theorem~\ref{thm:quasi-static-second-order}.} $\snr$,
\begin{IEEEeqnarray}{rCl}
\Rquasi^*(\bl,\error) = C\mathopen{}\left(\frac{\snr}{\avgg}\right) +\littleo(1), \quad \bl\to\infty
\label{eq:epsilon-c-lt}
\end{IEEEeqnarray}
where%\footnote{All logarithms, $\ln$, and exponents, $\exp$, in this paper are taken with respect to an arbitrary fixed base, unless otherwise specified. \todo{WEI: you cannot take $\exp$ with respect to an arbitrary basis! I do not like that sometimes the $\log$ has no basis specified, sometimes is the natural logarithm. Better using the natural logarithm throughout $\ln(\cdot)$}}
\begin{IEEEeqnarray}{rCL}
\Cawgn(\snr)  &\define & \ln(1+\snr)\label{eq:def-capacity-awgn}
\end{IEEEeqnarray}
denotes the channel capacity of a complex-valued additive white Gaussian noise (AWGN) channel under the short-term power constraint $\snr$, and
\begin{IEEEeqnarray}{rCl}
\avgg \define \Ex{}{\frac{1}{G} \indfun{ G > \Finvs(\error)}} + \frac{ \prob[G \leq \Finvs(\error)] -\error}{\Finvs(\error)} \IEEEeqnarraynumspace
\label{eq:avgg-def}
\end{IEEEeqnarray}
with $G\define|H|^2$ denoting the fading gain, $\indfun{\cdot}$ standing for the indicator function, and $\Finvs: [0,1] \to \NonnegReal$ being defined as
\begin{IEEEeqnarray}{rCl}
\Finvs (t) \define \sup \{g:  \prob[G < g] \leq t\}.
\label{eq:def-finvs}
\end{IEEEeqnarray}
As shown in~\cite{caire99-05} and illustrated in Fig.~\ref{fig:arch-sept}, the $\error$-capacity~$\Cawgn(\snr/\avgg)$ can be achieved by concatenating a fixed Gaussian codebook with a power controller that works as follows: it performs channel inversion when the fading gain~$G$ is above~$\Finvs(\error)$; it turns off transmission when the fading gain is below~$\Finvs(\error)$.
This single-codebook, variable-power scheme, which is sometimes referred to as \emph{truncated channel inversion}~\cite{goldsmith97-11a},\cite[Sec.~4.2.4]{goldsmith05}, is attractive from an implementation point of view, as it eliminates the need of adapting the codebook to the channel state (for example by multiplexing several codebooks)~\cite{caire99-05}.

In this paper, we show that this single-codebook, variable-power scheme is also second-order optimal.
Specifically, we prove i) that
\begin{IEEEeqnarray}{rCl}
 \Rquasi^\ast (\bl,\error)  = \Cawgn\mathopen{}\left(\frac{\snr}{\avgg}\right) - \sqrt{\Vawgn \mathopen{}\left(\frac{\snr}{\avgg} \right)} \sqrt{\frac{\ln\bl}{\bl}}+ \bigO\mathopen{}\left(\frac{1}{\sqrt{\bl}}\right)\IEEEeqnarraynumspace
\label{eq:intro-R-qs-lt}
\end{IEEEeqnarray}
where
 \begin{IEEEeqnarray}{rCL}
\Vawgn(\snr) &\define & \frac{\snr(\snr+2)}{(\snr+1)^2}\label{eq:def-dispersion-awgn}
\end{IEEEeqnarray}
denotes the \emph{dispersion}~\cite[Def.~1]{polyanskiy10-05} of a complex-valued AWGN channel subject to the short-term power constraint $\snr$, and~ii) that truncated channel inversion achieves~\eqref{eq:intro-R-qs-lt}.
% The presence of the AWGN dispersion in~\eqref{eq:intro-R-qs-lt} suggests that its proof involves channel inversion.
% %
% Indeed,~\eqref{eq:intro-R-qs-lt} is achieved by employing the best (capacity-dispersion optimal) code for the AWGN channel together with the simple truncated channel inversion power controller with an appropriate threshold.

%
%
%

A single-codebook, variable-power scheme turns out to be second-order optimal also for the (simpler) scenario of AWGN channel subject to a long-term power constraint.
Indeed, for this scenario we show that
%\begin{IEEEeqnarray}{rCl}
% \Rawgn^* (\bl,\error)%}  \notag\\ \quad
%  &=& \Cawgn\mathopen{}\left(\frac{\snr}{1-\error}\right)   - \sqrt{\Vawgn\mathopen{}\left(\frac{\snr}{1-\error}\right)} \sqrt{\frac{\ln\bl}{\bl}} + \bigO\mathopen{}\left(\frac{1}{\sqrt{\bl}}\right) \IEEEeqnarraynumspace \label{eq:intro-R-awgn-lt}
%\end{IEEEeqnarray}
\begin{equation}
 \Rawgn^* (\bl,\error)%}  \notag\\ \quad
  = \Cawgn\mathopen{}\left(\frac{\snr}{1-\error}\right)   - \sqrt{\Vawgn\mathopen{}\left(\frac{\snr}{1-\error}\right)} \!\sqrt{\frac{\ln\bl}{\bl}} + \bigO\mathopen{}\left(\frac{1}{\sqrt{\bl}}\right)
  \label{eq:intro-R-awgn-lt}
\end{equation}
and that~\eqref{eq:intro-R-awgn-lt} is achieved by concatenating a  dispersion-optimal codebook designed for an AWGN channel subject to a short-term power constraint, with a power controller that sets the power of the transmitted codeword to zero with probability $\error -\bigO(1/\sqrt{\bl\ln\bl})$ and to $\snr/(1-\error) + \bigO(1/\sqrt{\bl\ln\bl})$ otherwise.
%
%It is useful to observe the following connections between the quasi-static fading case and the AWGN case:
% \begin{itemize}
%\item The AWGN channel can be seen as a quasi-static fading channel with fading gain $G=1$ with probability one. Hence, the result in~\eqref{eq:intro-R-awgn-lt} can be obtained from~\eqref{eq:intro-R-qs-lt} by using that $\avgg =1-\error$ (see~\eqref{eq:avgg-def});
%    \item The truncated channel inversion power controller transforms the quasi-static fading channel into an AWGN channel.
%\end{itemize}
%
The asymptotic expansion in~\eqref{eq:intro-R-awgn-lt} refines a result reported in~\cite[Sec.~4.3.3]{polyanskiy10}.

\paragraph*{Proof Techniques}
The asymptotic expressions in~\eqref{eq:intro-R-qs-lt} and~\eqref{eq:intro-R-awgn-lt} are obtained by deriving achievability and converse bounds that match up to second order.
The achievability bounds rely on the truncated channel inversion scheme described above.
The converse bounds are based on the meta-converse theorem~\cite[Th.~26]{polyanskiy10-05} with auxiliary channel chosen so that it depends on the transmitted codewords only through their power.
In deriving the converse bounds, we also exploit that the solution of the following minimization problem
\begin{IEEEeqnarray}{rCl}
\inf_{\normx\sim P_{\normx}} \,\,\Ex{}{Q\mathopen{}\left(\sqrt{\bl}\frac{\Cawgn(\normx) - \gamma }{\sqrt{\Vawgn(\normx)}}\right)}
\label{eq:opt-problem-orig-intro}
\end{IEEEeqnarray}
is a two-mass-point distribution (with one mass point located at the origin), provided that~$\gamma$ is chosen appropriately and $\bl$ is sufficiently large.
In~\eqref{eq:opt-problem-orig-intro},~$Q(\cdot)$ stands for the Gaussian $Q$-function,~$\gamma$ is a positive number, and the infimum is over all probability distributions $P_\normx$ on~$\NonnegReal$ satisfying $\Ex{P_{\normx}}{\normx}\leq \snr$.
The minimization in~\eqref{eq:opt-problem-orig-intro} arises when optimizing the $\epsilon$-quantile of the information density over all power allocations.
%
%
%The proof of the converse  part of~\eqref{eq:intro-R-qs-lt} builds upon that of~\eqref{eq:intro-R-awgn-lt}.
%
%In particular, the achievability part of~\eqref{eq:intro-R-qs-lt} is based on the truncated channel inversion scheme (with an appropriate threshold), which transforms the quasi-static channel into an AWGN channel.
%
%This implies that truncated channel inversion achieves the first two terms in the large-$\bl$ expansion of $\Rquasi^*(\bl,\error)$.

The remainder of this paper is organized as follows.
In Section~\ref{sec:awgn-case}, we focus on the AWGN setup and prove the asymptotic expansion~\eqref{eq:intro-R-awgn-lt}.
We then move to the quasi-static fading case in Section~\ref{sec:qs-case} and establish~\eqref{eq:intro-R-qs-lt} building upon~\eqref{eq:intro-R-awgn-lt}.
In both Section~\ref{sec:awgn-case} and Section~\ref{sec:qs-case}, we also develop easy-to-evaluate approximations for $\Rawgn^* (\bl,\error)$ and $\Rquasi^\ast (\bl,\error)$, respectively, and compare them against nonasymptotic converse and achievability bounds.
Finally, we summarize our main findings in Section~\ref{sec:conclusion}.

\paragraph*{Notation}
Upper case letters such as $X$ denote scalar random variables and their realizations are written in lower case, e.g., $x$.
We use boldface upper case letters to denote random vectors, e.g., $\randvecx$, and boldface lower case letters for their realizations, e.g., $\vecx$.
Upper case letters of a special font are used to denote deterministic matrices, e.g., $\matX$.
%
%
%The superscripts~$\tp{}$ and $\herm{}$ stand for transposition and Hermitian transposition, respectively.
%
%
For two functions~$f(x)$ and~$g(x)$, the
notation~$f(x) = \bigO(g(x))$, $x\to \infty$, means that
$\lim \sup_{x\to\infty}\bigl|f(x)/g(x)\bigr|<\infty$, and
$f(x) = \littleo(g(x))$, $x\to \infty$, means that $\lim_{x\to\infty}\bigl|f(x)/g(x)\bigr|=0$.
We use $\matI_{a}$ to denote the identity matrix of size $a\times a $.
The distribution of a circularly symmetric complex
Gaussian random vector with covariance matrix~$\matA$ is denoted by $\jpg(\mathbf{0}, \matA)$.
The symbol $\NonnegReal$ stands for the nonnegative real line and $\ln(\cdot)$ denotes the natural logarithm.
The indicator function is denoted by $\indfun{\cdot}$, and $|\cdot|^{+} \define \max\{\,\cdot\,, 0\}$.
Given two probability distributions $\indist$ and $\outdist$ on a common measurable space $\setW$, we define a randomized test between~$\indist$ and~$\outdist$ as a random transformation $\testdist_{Z\given W}: \setW\to\{0,1\}$ where $0$ indicates that the test chooses~$\outdist$. We shall need the following performance metric for the test between~$\indist$ and~$\outdist$:
\begin{IEEEeqnarray}{rCl}
\label{eq:def-beta}
\beta_\alpha(\indist,\outdist) \define \min\int \testdist_{Z\given W}(1\given w)  \outdist(d w)
\end{IEEEeqnarray}
where the minimum is over all probability distributions $\testdist_{Z\given W}$
satisfying
\begin{IEEEeqnarray}{rCl}
 \int \testdist_{Z\given W} (1\given w) \indist(dw)\geq \alpha.
\end{IEEEeqnarray}

\section{The AWGN Channel}
\label{sec:awgn-case}

In this section, we consider the AWGN channel
\begin{equation}
\randvecy = \vecx + \vecnoise.
\label{eq:channel-io-awgn}
\end{equation}
An $(\bl, \NumCode,\error)_{\avg}$ code for the AWGN channel~\eqref{eq:channel-io-awgn} consists of:
\begin{enumerate}
\item an encoder $\encoder$: $ \{1,\ldots,\NumCode\} \to \complexset^\bl$ that maps the message $\msg \in \{1,\ldots,\NumCode\}$ to a codeword $\vecx \in \{\vecc_1,\ldots,\vecc_{\NumCode}\}$ satisfying the power constraint
\begin{IEEEeqnarray}{rCl}
\frac{1}{\NumCode}\sum\limits_{j=1}^{\NumCode} \| \vecc_j \|^2 \leq \bl\snr.
 \label{eq:power-constr-awgn}
\end{IEEEeqnarray}
\item A decoder $\decoder$: $\complexset^{\bl} \to\{1,\ldots,\NumCode\}$ satisfying the average error probability constraint
\begin{equation}
\prob[\decoder(\randvecy) \neq \msg ] \leq \error.
\label{eq:avg-error-awgn}
\end{equation}
Here, $\msg$ is assumed to be equiprobable on $\{1,\ldots,\NumCode\}$, and $\randvecy$ denotes the channel output induced by the transmitted codeword according to~\eqref{eq:channel-io-awgn}.
\end{enumerate}
We shall refer to~\eqref{eq:power-constr-awgn} as long-term power constraint~\cite{caire99-05}, as opposed to the more common and  more stringent short-term power constraint
\begin{equation}
\|\vecc_j\|^2 \leq \bl\snr, \quad j=1,\ldots,\NumCode.
\label{eq:awgn-st-pc}
\end{equation}

The maximal channel coding rate is defined as
\begin{IEEEeqnarray}{rCl}
\Rawgn^*(\bl,\error) \define \sup\mathopen{}\left\{\frac{\ln\NumCode}{\bl}: \,\, \exists\, (\bl,\NumCode,\error)_{\avg} \,\,\,\text{code}\right\}. \IEEEeqnarraynumspace
\end{IEEEeqnarray}
This quantity was characterized up to first order in~\cite[Th.~77]{polyanskiy10}, where it was shown that
\begin{IEEEeqnarray}{rCl}
\lim\limits_{\bl\to\infty}\Rawgn^*(\bl,\error) = \Cawgn\mathopen{}\left(\frac{\snr}{1-\error}\right), \quad 0<\error <1.
\label{eq:epsilon-capacity-awgn}
\end{IEEEeqnarray}
The asymptotic expression~\eqref{eq:epsilon-capacity-awgn} implies that the strong converse~\cite[p.~208]{cover06-a} does not hold for AWGN channels subject to a long-term power constraint.
Note that if we replace~\eqref{eq:power-constr-awgn} with~\eqref{eq:awgn-st-pc} or the average error probability constraint~\eqref{eq:avg-error-awgn} with the maximal error probability
 constraint
\begin{equation}
\max\limits_{1\leq j\leq \NumCode} \prob[\decoder(\randvecy) \neq \msg \given \msg =j ] \leq \error
\end{equation}
the strong converse applies and~\eqref{eq:epsilon-capacity-awgn} ceases to be valid.

Theorem~\ref{thm:awgn-second-order} below characterizes the first two terms in the asymptotic expansion of $\Rawgn^*(\bl,\error)$ for fixed $0<\error<1$ and $\bl$ large.
\begin{thm}
\label{thm:awgn-second-order}
For the AWGN channel~\eqref{eq:channel-io-awgn} subject to the long-term power constraint~$\snr$ and for $0<\error<1$, the maximal channel coding rate $\Rawgn^*(\bl,\error)$ is
\begin{equation}
 \Rawgn^* (\bl,\error)
 = \Cawgn\mathopen{}\left(\!\frac{\snr}{1-\error}\!\right)   - \sqrt{\Vawgn\mathopen{}\left(\!\frac{\snr}{1-\error}\!\right)} \sqrt{\frac{\ln\bl}{\bl}} +  \bigO\mathopen{}\left(\frac{1}{\sqrt{\bl}}\right)
  \label{eq:thm-awgn-R}
\end{equation}
%\begin{equation}
%
%
where the functions $\Cawgn(\cdot)$ and $\Vawgn(\cdot)$ are defined in~\eqref{eq:def-capacity-awgn} and~\eqref{eq:def-dispersion-awgn}, respectively.
\end{thm}
\begin{rem}
The $\bigO\mathopen{}\left({1}/{\sqrt{\bl}}\right)$ term in the expansion~\eqref{eq:thm-awgn-R} can be strengthened to $\littleo(1/\sqrt{\bl })$ by replacing the Berry-Esseen theorem in the proof of the converse part (see Section~\ref{sec:proof-awgn-sec-order-conv}) with a Cramer-Esseen-type central-limit theorem (see~\cite[Th.~VI.1]{petrov75}).
\end{rem}

\begin{IEEEproof}
See Sections~\ref{sec:proof-awgn-sec-order-conv} and \ref{sec:proof-awgn-sec-order-ach} below.
\end{IEEEproof}

Before proving~\eqref{eq:thm-awgn-R}, we motivate its validity through a heuristic argument, which also provides an outline of the proof.
For AWGN channels subject to the short-term power constraint $\normxre$, the maximal channel coding rate $\Rawgnst^*(\bl,\error)$ roughly satisfies~\cite[Sec.~IV]{polyanskiy10-05}
\begin{IEEEeqnarray}{rCl}
\error \approx Q\mathopen{}\left(\sqrt{\bl}\frac{\Cawgn(\normxre) - \Rawgnst^*(\bl,\error) }{\sqrt{\Vawgn(\normxre)}}\right).
\end{IEEEeqnarray}
In the long-term power constraint case, the codewords need not  be of equal power.
Fix an arbitrary code with rate $R$ that satisfies the long-term power constraint~\eqref{eq:power-constr-awgn}, and let $P_{\normx}$ be the probability distribution induced by the code on the normalized codeword power $\normx \define \|\randvecx\|^2/\bl$.
We shall refer to $P_{\normx}$ as \emph{power distribution}.
By~\eqref{eq:power-constr-awgn}, the nonnegative random variable $\Pi$ must satisfy
\begin{IEEEeqnarray}{rCl}
\Ex{P_{\normx}}{\Pi} \leq \snr.
\label{eq:power-constraint-normx}
\end{IEEEeqnarray}
Through a random coding argument, one can show that the following relation must hold for the best among all codes with rate $R$ and power distribution $P_{\normx}$:
\begin{IEEEeqnarray}{rCl}
\epsilon(P_{\normx}) \approx \Ex{P_{\normx}}{Q\mathopen{}\left(\sqrt{\bl}\frac{\Cawgn(\normx) - R }{\sqrt{\Vawgn(\normx)}}\right)}.
\label{eq:norm-app-heuristic}
\end{IEEEeqnarray}
Here, $\epsilon(P_{\normx})$ denotes  the minimum error probability achievable under the power distribution $P_{\normx}$.
This error probability can be further reduced by minimizing~\eqref{eq:norm-app-heuristic} over all power distributions $P_{\normx}$ that satisfy~\eqref{eq:power-constraint-normx}.
%
% Thus, the crucial problem now is to minimize the RHS of~\eqref{eq:norm-app-heuristic} over all probability distributions $P_{\normx}$ satisfying~\eqref{eq:power-constraint-normx}.
% This is the content of Lemma~\ref{lemma:solusion-awgn-prob}. In particular, we obtain that (for a given $R$)
% the optimal $P^*_\normx$ is a two-mass-point distribution:
It turns out that, for sufficiently large~$n$, the power distribution $P^*_{\normx}$ that minimizes the right-hand side (RHS) of~\eqref{eq:norm-app-heuristic} is the following two-mass-point distribution:
\begin{IEEEeqnarray}{rCl}
P^*_{\normx}(0) = 1-\frac{\snr}{\tangentp},\text{ and }  P^*_{\Pi}(\tangentp) = \frac{\snr}{\tangentp}
\label{eq:opt-p-normx}
\end{IEEEeqnarray}
with $\tangentp$ satisfying
\begin{IEEEeqnarray}{rCl}\label{eq:omega_approx}
\sqrt{\bl}\frac{\Cawgn(\tangentp) -R}{\sqrt{\Vawgn(\tangentp)}} \approx \sqrt{\ln\bl}.
\end{IEEEeqnarray}
Substituting~\eqref{eq:opt-p-normx} into~\eqref{eq:norm-app-heuristic}, setting $\error(P^{*}_{\normx}) =\error$, and then using~\eqref{eq:omega_approx}, we obtain
\begin{IEEEeqnarray}{rCL}
\error &\approx&  \frac{\snr}{\tangentp} Q\mathopen{}\left(\sqrt{\ln\bl}\right) + 1 - \frac{\snr}{\tangentp}\\
 &\approx & 1 - \frac{\snr}{\tangentp}\label{eq:first_expression_for_rquasi}
\end{IEEEeqnarray}
where the last approximation is accurate when $\bl$ is large.
Since~\eqref{eq:first_expression_for_rquasi} implies that $\tangentp \approx \snr/(1-\error)$, we see from~\eqref{eq:opt-p-normx} that the optimal strategy is to transmit at power $\snr/(1-\error)$ with probability approximately $1-\error$, and to transmit nothing otherwise.
Substituting~\eqref{eq:first_expression_for_rquasi} into~\eqref{eq:omega_approx} and solving for $R$, we obtain the desired result
 \begin{IEEEeqnarray}{rCl}
 \Rawgn^* (\bl,\error) \approx \Cawgn\mathopen{}\left(\frac{\snr}{1-\error}\right)  - \sqrt{\Vawgn\mathopen{}\left(\frac{\snr}{1-\error}\right)} \sqrt{\frac{\ln \bl}{\bl}}. \IEEEeqnarraynumspace
 \end{IEEEeqnarray}
 We next provide a rigorous justification for these heuristic steps.
% %
% A formal proof of Theorem~\ref{thm:awgn-second-order}, presented in Sections~\ref{sec:proof-awgn-sec-order-conv} and \ref{sec:proof-awgn-sec-order-ach} below, will provide
% rigorous justification for the above heuristic.}
%}

\subsection{Proof of the Converse Part}
\label{sec:proof-awgn-sec-order-conv}
Consider an arbitrary $(\bl,\NumCode, \error)_{\avg}$ code. Let~$P_{\randvecx}$ denote the probability distribution on the channel input~$\randvecx$ induced by the code.
To upper-bound $ \Rawgn^* (\bl,\error) $, we use the meta-converse theorem~\cite[Th.~26]{polyanskiy10-05} with the following auxiliary channel $Q_{\randvecy \given \randvecx}$:
\begin{IEEEeqnarray}{rCl}
Q_{\randvecy \given \randvecx=\vecx} = \jpg\mathopen{}\left(\mathbf{0}, \left(1+  \|\vecx\|^2/{\bl}\right)\matI_{\bl}\right).
\label{eq:def-Q-awgn}
\end{IEEEeqnarray}
The choice of letting the auxiliary channel in~\eqref{eq:def-Q-awgn} depend on the transmit codeword through its power, is inspired by a similar approach used in~\cite[Sec.~4.5]{polyanskiy10} to characterize the maximal channel coding rate for the case of parallel AWGN channels subject to a short-term power constraint, and in~\cite{yang14-07a} for the case of quasi-static multiple-antenna fading channels subject to a short-term power constraint.
With this choice, we have~\cite[Th.~26]{polyanskiy10-05}
\begin{IEEEeqnarray}{rCl}
%\inf_{P_{\randvecx}}
\beta_{1-\error}(P_{ \randvecx \randvecy}, P_{\randvecx} Q_{\randvecy\given \randvecx }) \leq 1-\error'
\label{eq:meta-converse-awgn}
\end{IEEEeqnarray}
where $\beta_{(\cdot)}(\cdot,\cdot)$ was defined in~\eqref{eq:def-beta} and $\error'$ is the average probability of error incurred by using the selected $(\bl,\NumCode, \error)_{\avg}$ code over the auxiliary channel $Q_{\randvecy \given \randvecx}$.

Next, we lower-bound the left-hand side (LHS) of \eqref{eq:meta-converse-awgn}.
Let~$ \normx = \|\randvecx\|^2/\bl$.
Under~$P_{ \randvecx \randvecy }$, the random variable $ \ln \frac{ d P_{\randvecx\randvecy}}{ d(P_{\randvecx} Q_{\randvecy\given \randvecx})}$ has the same distribution as (see~\cite[Eq.~(205)]{polyanskiy10-05})
\begin{IEEEeqnarray}{rCl}
S_\bl(\normx) &\define& \bl \Cawgn(\normx)  +  \sum\limits_{i=1}^{\bl}\left(1-\frac{\big|\sqrt{\normx}Z_i-1\big|^2}{1 + \normx  }\right) \IEEEeqnarraynumspace
\label{eq:info_density_awgn}
\end{IEEEeqnarray}
where $\{Z_i\}$, $i=1,\ldots,\bl$, are independent and identically distributed (i.i.d) $\jpg(0,1)$ random variables, which are also independent of $\normx$.
Using~\cite[Eq.~(102)]{polyanskiy10-05} and~\eqref{eq:info_density_awgn}, we obtain the following lower bound
\begin{IEEEeqnarray}{rCl}
 \beta_{1-\error}(P_{ \randvecx\randvecy}, P_{\randvecx} Q_{\randvecy\given \randvecx }) &\geq& e^{-\bl \gamma} \big|\prob[S_\bl(\normx) \leq \bl\gamma] -\error\big|^{+} \IEEEeqnarraynumspace
\label{eq:lower-bd-beta-PXY}
\end{IEEEeqnarray}
which holds for every $\gamma>0$.

As proven in Appendix~\ref{app:proof-awgn-converse-q}, the RHS of~\eqref{eq:meta-converse-awgn} can be upper-bounded  as follows:
\begin{equation}
1- \error' \leq \frac{1}{\NumCode}\bigg(1+ \sqrt{\frac{\bl}{2\pi}} \ln(1+\NumCode \snr)\bigg).
\label{eq:converse-q-awgn}
\end{equation}
Since, by Fano's inequality~\cite[Th.~2.10.1]{cover06-a}, $\ln\NumCode \leq  (\bl \Cawgn(\snr) + \binent(\error))/(1-\error)$, where $\binent(\cdot)$ denotes the binary entropy function, we conclude that
\begin{IEEEeqnarray}{rCl}
 \ln( 1-\error') \leq -\ln \NumCode + \bl \varrho_\bl
\label{eq:converse-Q-expansion}
\end{IEEEeqnarray}
where
\begin{IEEEeqnarray}{rCl}
\varrho_\bl  &\define& \frac{1}{\bl} \ln\mathopen{}\left(1+  \sqrt{\frac{\bl}{2\pi}}\ln\mathopen{}\left(1 + \snr\exp\mathopen{}\left(\frac{\bl C(\snr) + \binent(\error)}{1-\error}\right) \right)\right)\notag\\
\end{IEEEeqnarray}
does not depend on the chosen code.
Substituting~\eqref{eq:lower-bd-beta-PXY} and~\eqref{eq:converse-Q-expansion} into~\eqref{eq:meta-converse-awgn}, we obtain
\begin{IEEEeqnarray}{rCl}
\ln\NumCode \leq \bl \gamma - \ln\mathopen{}\big|\prob[S_\bl(\normx) \leq \bl\gamma] -\error\big|^{+}  + \bl \varrho_\bl.\IEEEeqnarraynumspace
\label{eq:ub-M-PX-awgn}
\end{IEEEeqnarray}
Note that the RHS of~\eqref{eq:ub-M-PX-awgn} depends on the chosen code only through the probability distribution~$P_{\normx}$ that the code induces on $\normx=\|\randvecx\|^2/\bl$.

Let $\Omega$ be the set of probability distributions $P_{\normx}$ on $\NonnegReal$ that satisfy~\eqref{eq:power-constraint-normx}.
%\begin{IEEEeqnarray}{rCl}
%\int \normxre P_{\normx}(d\normxre) \leq \snr.
%\label{eq:def-Omega-set}
%\end{IEEEeqnarray}
%
%
%
Maximizing the RHS of~\eqref{eq:ub-M-PX-awgn} over all $P_{\normx} \in \Omega$ and then dividing both terms by $\bl$, we obtain the following upper bound on $\Rawgn^*(\bl,\error )$:
\begin{equation}
 \Rawgn^*(\bl,\error )\leq  \gamma - \frac{1}{\bl} \ln\mathopen{}\left|\inf\limits_{P_{\normx}\in \Omega}\prob[S_\bl(\normx) \leq \bl\gamma] -\error\right|^{+} + \varrho_\bl.
\label{eq:ub-R-PX-awgn}
\end{equation}
This bound holds for every $\gamma>0$.

Next, we study the asymptotic behavior of the RHS of~\eqref{eq:ub-R-PX-awgn} in the limit $\bl\to\infty$.
To this end, we first lower-bound $\prob[S_\bl(\normx) \leq \bl\gamma]$.
Let
\begin{IEEEeqnarray}{rCl}
T_i(\normx) \define \frac{1 }{\sqrt{\Vawgn(\normx)}} \left(1- \frac{|\sqrt{\normx} Z_{i} -1|^2 }{1+ \normx}\right), \,\, i=1,\ldots,\bl. \IEEEeqnarraynumspace
\end{IEEEeqnarray}
The random variables $\{T_i\}$, $i=1,\ldots,\bl$, have zero mean and unit variance, and they are conditionally i.i.d. given $\normx$.
Furthermore, one can rewrite $\prob[S_\bl(\normx) \leq \bl\gamma] $ using the $\{T_i\}$ as follows:
\begin{IEEEeqnarray}{rCl}
\prob[S_\bl(\normx) \leq \bl \gamma] = \prob\mathopen{}\left[ \frac{1}{\sqrt{\bl}} \sum\limits_{i=1}^{\bl} T_i(\normx) \leq \sqrt{\bl}  \frac{\gamma - \Cawgn(\normx)}{\sqrt{\Vawgn(\normx)}}\right]. \IEEEeqnarraynumspace
\label{eq:cdf-sn-iid}
\end{IEEEeqnarray}
Using the Berry-Esseen Theorem (see, e.g.,~\cite[Th.~44]{polyanskiy10-05}), we next relate the cumulative distribution function of the random variable $\bl^{-1/2}\sum\nolimits_{i=1}^{\bl}T_i(\normx)$ on the RHS of~\eqref{eq:cdf-sn-iid} to that of a Gaussian random variable.
For a given $\normx=\normxre$, we obtain
\begin{IEEEeqnarray}{rCl}
\IEEEeqnarraymulticol{3}{l}{
\prob\mathopen{}\left[ \frac{1}{\sqrt{\bl}} \sum\limits_{i=1}^{\bl} T_i(\normxre) \leq \sqrt{\bl}  \frac{\gamma - \Cawgn(\normxre)}{\sqrt{\Vawgn(\normxre)}}\right] }\notag\\
\quad & \geq & \func_{\bl,\gamma}(\normxre) - \frac{6\Ex{}{\big|T_1(\normxre)\big|^3 }}{\sqrt{\bl}}\IEEEeqnarraynumspace
\label{eq:prob-Sn-first-ineq}
\end{IEEEeqnarray}
where
\begin{equation}
\func_{\bl,\argpn}(x) \define Q\mathopen{}\left( \sqrt{\bl} \frac{\Cawgn(x) - \argpn}{\sqrt{\Vawgn(x)}}\right).
\label{eq:def-f-func}
\end{equation}
It follows from~\cite[Eq.~(179)]{yang14-07a} that for all $\normxre >0$
\begin{IEEEeqnarray}{rCl}
\Ex{}{\big|T_1(\normxre)\big|^3 } \leq 3^{3/2}.
\label{eq:bound-B-wg}
\end{IEEEeqnarray}
Substituting~\eqref{eq:bound-B-wg} into~\eqref{eq:prob-Sn-first-ineq} and then averaging~\eqref{eq:prob-Sn-first-ineq} over $\normx$, we conclude that
\begin{IEEEeqnarray}{rCl}
\prob[S_\bl(\normx) \leq \bl \gamma] \geq \Ex{}{\func_{\bl,\gamma}(\normx)} -\frac{6 \cdot 3^{3/2}}{\sqrt{\bl}}.
\label{eq:prob-Sn-2nd-ineq}
\end{IEEEeqnarray}

To eliminate the dependency of the RHS of~\eqref{eq:prob-Sn-2nd-ineq} on $P_{\normx}$, we next minimize the first term on the RHS of~\eqref{eq:prob-Sn-2nd-ineq} over all $P_{\normx} $ in~$\Omega$, i.e., we solve the optimization problem
\begin{IEEEeqnarray}{rCl}
\inf_{P_{\normx} \in \Omega} \,\,\Ex{P_\normx}{\func_{\bl,\argpn}(\normx)}
\label{eq:opt-problem-orig}
\end{IEEEeqnarray}
which is identical to the one stated in~\eqref{eq:opt-problem-orig-intro}.
The solution of~\eqref{eq:opt-problem-orig} is given in the following lemma.

\begin{lemma}
\label{lemma:solusion-awgn-prob}
Let $\argpn >0$ and assume that $\bl \geq 2 \pi (e^{2\argpn}-1)\argpn^{-2}$.
Then,
\begin{enumerate}
\item there exists a unique $\tangentp=\tangentp(\bl,\gamma)$ in the interval $[e^\argpn-1,\infty)$ satisfying both
 \begin{IEEEeqnarray}{rCl}
\frac{\func_{\bl,\argpn}(\tangentp)-1}{\tangentp} = \func_{\bl,\argpn}'(\tangentp)
\label{eq:def-x0-mass-point}
\end{IEEEeqnarray}
and
\begin{IEEEeqnarray}{rCl}
\func_{\bl,\argpn} (x) \geq 1 + \func_{\bl,\argpn}'(\tangentp) x, \quad \forall \, x\in[0,\infty). \IEEEeqnarraynumspace
\label{eq:lb-q-n-gamma}
\end{IEEEeqnarray}
Here, $\func_{\bl,\argpn}'(\cdot)$ stands for the first derivative of the function $\func_{\bl,\argpn}(\cdot)$.

\item The infimum in~\eqref{eq:opt-problem-orig} is a minimum and the probability distribution $P_{\normx}^\ast$ that minimizes $\Ex{P_{\normx}}{\func_{\bl,\argpn}(\normx)}$ has the following structure:
\begin{itemize}
\item if $\snr <  \tangentp$, then $P_{\normx}^*$ has two mass points, one located at $0$ and the other located at $\tangentp$. Furthermore,
$P_{\normx}^{*}(0) = 1-{\snr}/{\tangentp}$ and $P_{\normx}^{*}( \tangentp) ={\snr}/{\tangentp}$.

\item If $\snr \geq \tangentp$, then  $ P_{\normx}^*$ has only one mass point located at $\snr$.
\end{itemize}

\end{enumerate}
\end{lemma}

\begin{IEEEproof}
To prove the first part of~Lemma~\ref{lemma:solusion-awgn-prob}, we observe that the function $\func_{\bl,\argpn}(\cdot)$ defined in~\eqref{eq:def-f-func} has the following properties (see Fig.~\ref{fig:fx-illustration}):
\begin{enumerate}
\item $\func_{\bl,\argpn}(0)\define \lim_{x\to 0}\func_{\bl,\argpn}(x) =1$ and  $\func_{\bl,\argpn}(e^\argpn -1 ) = 1/2$\label{item:condition1};
\item $\func_{\bl,\argpn}(\cdot)$ is differentiable and monotonically decreasing for every $\gamma>0$;
\item $\func_{\bl,\argpn}(\cdot)$ is strictly convex on $[e^\argpn-1, \infty)$\label{item:condition4};
%\item for every $\bl  > \frac{\pi}{2}\frac{e^\argpn+1}{e^\argpn-1}$, we have $|f_{\bl,\argpn}'(e^\argpn-1)| > 1/(2(e^\argpn-1))$
\item \label{item:condition5}
for every $\bl \geq 2\pi(e^{2\argpn}-1)\argpn^{-2} $ and for every~$x\in [0, e^\argpn-1]$, the function $\func_{\bl,\argpn}(x)$ lies above the line connecting the points $(0,1)$ and $(e^\argpn-1, 1/2)$, i.e,
\begin{equation}
\func_{\bl,\argpn}(x) \geq 1- \frac{1}{2}\frac{x}{e^\argpn-1}.
\label{eq:condition4}
\end{equation}
Furthermore,~\eqref{eq:condition4} holds with equality if $x=0$ or $x=e^\argpn-1$.
\end{enumerate}
Properties~\ref{item:condition1}--\ref{item:condition4} can be established through standard techniques.
 To prove Property~\ref{item:condition5}, we start by  noting that
\begin{IEEEeqnarray}{rCl}
%\IEEEeqnarraymulticol{3}{l}{
-\frac{C(x)-\gamma}{\sqrt{V(x)}} &=&
 - \frac{\ln(1+x) - \argpn}{\sqrt{x(x+2)(1+x)^{-2}}} \\
&=& \ln\mathopen{}\left(1+\frac{e^\argpn}{1+x}-1\right) \frac{1+x}{\sqrt{x^2+2x}} \IEEEeqnarraynumspace \label{eq:lower-bound-error-0}\\
&\geq& \frac{\argpn}{e^\argpn-1}\left(\frac{e^\argpn}{1+x} -1\right) \frac{1+x}{\sqrt{x^2+2x}} \IEEEeqnarraynumspace\label{eq:lower-bound-error-1}\\
&=&  \frac{\gamma (1- x/(e^\gamma-1))}{\sqrt{x^2 + 2x}}\\
%&=& \frac{\argpn(1-x/(e^\argpn-1))}{\sqrt{x^2+2x}}\\
&\geq& \frac{\argpn(1- \sqrt{x/(e^\argpn-1)})} { \sqrt{ (e^\argpn +1)x}} \label{eq:lower-bound-error2}\\
&= &\frac{\argpn}{\sqrt{ e^{2\argpn}-1}} \left(\sqrt{ \frac{e^\argpn-1}{x}} -1\right).
\label{eq:lower-bound-error}
\end{IEEEeqnarray}
Here,~\eqref{eq:lower-bound-error-1} follows because $\ln(1+a) \geq \argpn a/(e^\argpn-1)$ for every $a\in[0,e^\argpn-1]$ and by setting $a=e^\gamma/(1+x) -1$; in~\eqref{eq:lower-bound-error2} we used that $\sqrt{x^2+2x} \leq \sqrt{(e^\argpn + 1)x}$ and that $x/(e^\argpn-1) \leq \sqrt{x/(e^\argpn-1)}$ for every $x\in[0,e^\argpn-1]$.
Using~\eqref{eq:lower-bound-error}, we obtain that for every $\bl \geq 2\pi (e^{2 \argpn}-1) \argpn^{-2}$
\begin{IEEEeqnarray}{rCl}
\IEEEeqnarraymulticol{3}{l}{
q_{\bl,\argpn}(x) + \frac{1}{2}\frac{x}{e^\argpn-1} -1}\notag\\
\quad &=&  \frac{1}{2}\frac{x}{e^\argpn-1} - Q\mathopen{}\left(- \sqrt{\bl} \frac{\ln(1+x) - \argpn}{\sqrt{x(x+2)(1+x)^{-2}}}\right) \label{eq:lower-bound-diff-t-q-s1}\\
&\geq&   \frac{1}{2}\frac{x}{e^\argpn-1}  -Q\mathopen{}\left(\frac{\sqrt{\bl} \argpn}{\sqrt{e^{2\argpn}-1}} \left(\sqrt{ \frac{e^\argpn-1}{x}} -1\right)\right) \IEEEeqnarraynumspace \\
&\geq&  \frac{1}{2}\frac{x}{e^\argpn-1}  - Q\mathopen{}\left(\sqrt{2\pi} \left( \sqrt{ \frac{e^\argpn-1}{x}} -1\right)\right).
\label{eq:lower-bound-diff-t-q}
\end{IEEEeqnarray}
Here, in~\eqref{eq:lower-bound-diff-t-q-s1} we used that $Q(x)+Q(-x)=1$ for every $x\in \realset$, and in~\eqref{eq:lower-bound-diff-t-q} we used that $n\geq 2\pi (e^{2 \argpn}-1) \argpn^{-2}$ and that $Q(\cdot)$ is monotonically decreasing.
The RHS of~\eqref{eq:lower-bound-diff-t-q}  is nonnegative on the interval $ [0,e^\argpn-1]$ since it is equal to zero if $x\in\{ 0,e^\argpn-1\}$ and it first increases and then decreases on $(0,e^\argpn-1)$. Finally, it can be verified that~\eqref{eq:condition4} holds with equality at $x=0$ and at $x=e^{\gamma}-1$.

Properties~\ref{item:condition1}--\ref{item:condition5} guarantee that there exists a unique $\tangentp \in [e^\argpn-1,\infty)$ and a line $\setL_0$ passing through the point $(0,1)$ such that $\setL_0$ is tangent to $\func_{\bl,\argpn}(\cdot)$ at $(\tangentp, \func_{\bl,\argpn}(\tangentp) )$ and that $\setL_0$ lies below $\func_{\bl,\argpn}(x)$ for all $x\geq 0$ (see~Fig.~\ref{fig:fx-illustration}).
By construction, $\tangentp$ is the unique number in $[e^\argpn-1,\infty)$ that satisfies~\eqref{eq:def-x0-mass-point} and~\eqref{eq:lb-q-n-gamma}.
This concludes the first part of Lemma~\ref{lemma:solusion-awgn-prob}.

\begin{figure}[t]
\centering
\includegraphics[scale=0.7]{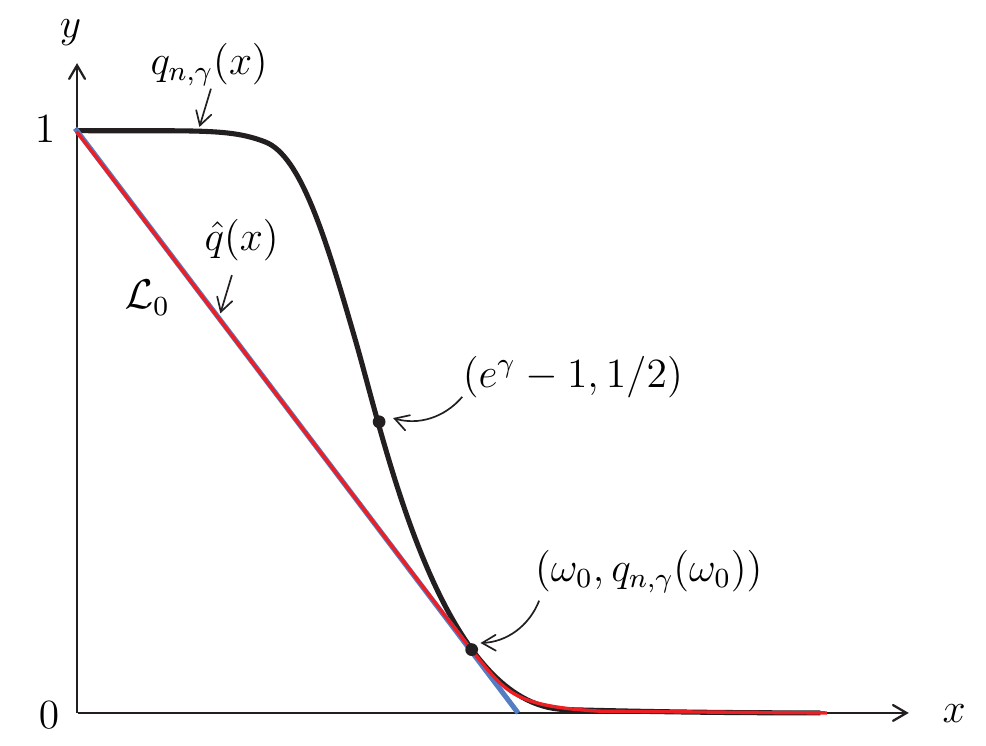}
\caption{\label{fig:fx-illustration} A geometric illustration of $\func_{\bl,\argpn}(\cdot)$ (black curve), of the tangent line $\setL_0$ (blue line), and of the convex envelope $\hat{\func}(\cdot)$ (red curve).} %\todo{WEI: update labels; $e^\gamma-1$, $\hat{\func}(x)$}}
\end{figure}

We proceed now to prove the second part of Lemma~\ref{lemma:solusion-awgn-prob}.
Let
\begin{IEEEeqnarray}{rCl}
\hat{\func}(x) \define \inf_{P_{\normx}: \Ex{}{\normx}\leq x} \Ex{P_{\normx}}{\func_{\bl,\gamma}(\normx)}
\label{eq:def-convex-envelope}
\end{IEEEeqnarray}
where the infimum is over all probability distributions~$P_{\normx}$ on~$\NonnegReal$ satisfying $\Ex{P_{\normx}}{\normx}\leq x$.
It follows that $\hat{\func}(\cdot)$ is convex, continuous, and nonincreasing.
In fact, $\hat{\func}(\cdot)$ is the \emph{convex envelope} (i.e., the largest convex lower bound)~\cite[p.~151]{bazaraa06} of $\func_{\bl,\gamma}(\cdot)$ over $\NonnegReal$.
Indeed, let $\hat{\setE}$ and $\setE$ denote the \emph{epigraph}\footnote{\label{footnote:epigraph}The epigraph of a function $f:\realset^n \mapsto \realset$ is the set of points lying on or above its graph~\cite[p.~104]{bazaraa06}.} of $\hat{\func}(\cdot)$ and of $\func_{\bl,\gamma}(\cdot)$ over $\NonnegReal$, respectively.
To show that  $\hat{\func}(\cdot)$ is the convex envelope of $\func_{\bl,\gamma}(\cdot)$, it suffices to show that $\hat{\setE}$ is the closure of the convex hull of~$\setE$ (see~\cite[Ex.~3.33]{bazaraa06}), i.e.,
\begin{IEEEeqnarray}{rCl}
\hat{\setE} = \mathrm{Cl}(\mathrm{Conv}(\setE))
\label{eq:set-eq}
\end{IEEEeqnarray}
where $\mathrm{Cl}(\setS)$ and $\mathrm{Conv}(\setS)$ stand for the closure and convex hull of a given set $\setS$, respectively.
Since $\func_{\bl,\gamma}(x) \geq \hat{\func}(x)$ for all $x \in \NonnegReal$, it follows that $\setE \subset \hat{\setE}$. Moreover, since $\hat{\func}(\cdot)$ is convex and continuous, its epigraph $\hat{\setE}$ is convex and closed.
This implies that $\mathrm{Cl}(\mathrm{Conv}(\setE)) \subset \hat{\setE}$.

We next show that $\hat{\setE} \subset \mathrm{Cl}(\mathrm{Conv}(\setE))$.
Consider an arbitrary $(x_0,y_0)\in \hat{\setE}$.
If $y_0 > \hat{\func}(x_0)$, then by~\eqref{eq:def-convex-envelope} there exists a probability distribution $P_{\Pi}$ satisfying $\Ex{P_{\Pi}}{\Pi} \leq x_0$ and $\Ex{P_\Pi}{\func_{\bl,\gamma}(\Pi)} <y_0$.
By the definition of convex hull, $(\Ex{}{\Pi}, \Ex{}{\func_{\bl,\gamma}(\Pi)}) \in  \mathrm{Conv}(\setE)$.
Since  $\func_{\bl,\gamma}(\cdot)$ is monotonically decreasing, we conclude that all points $(x,y)$ such that $x\geq\Ex{P_{\Pi}}{\Pi}$ and $y\geq \Ex{}{\func_{\bl,\gamma}(\Pi)}$ must lie in $\mathrm{Conv}(\setE)$. Hence, $(x_0,y_0) \in  \mathrm{Conv}(\setE)$.
If $y_0 = \hat{\func}(x_0)$, then we can find a sequence $\{(x_0,y_n)\}$ such that $y_n > y_0$ for all $n$, and  $\lim_{n\to\infty} y_n =y_0$.
Since $\{(x_0,y_n)\} \subset  \mathrm{Conv}(\setE)$, it follows that $(x_0,y_0) \in \mathrm{Cl}(\mathrm{Conv}(\setE))$.
This proves that $\hat{\setE} \subset \mathrm{Cl}(\mathrm{Conv}(\setE))$ and, hence,~\eqref{eq:set-eq}.

We next characterize $\hat{\func}(\cdot)$.
Properties~\ref{item:condition1}--\ref{item:condition5} imply that $\hat{\func}(x)$ coincides with the straight line connecting the points $(0,1)$ and $(\tangentp, \func_{\bl,\gamma}(\tangentp))$ for $x\in[0,\tangentp]$, and coincides with $\func_{\bl,\gamma}(x)$ for $x\in(\tangentp, \infty)$ (see Fig.~\ref{fig:fx-illustration}).
To summarize, we have that
\begin{IEEEeqnarray}{rCl}
\hat{\func}(x) = \left\{
                          \begin{array}{ll}
                            1-\frac{x}{\tangentp} + \frac{x}{\tangentp} q_{\bl,\gamma}(\tangentp), & \hbox{$x \in[0,\tangentp]$} \\
                            \func_{\bl,\gamma}(x), & \hbox{$x\in(\tangentp,\infty).$}
                          \end{array}
                        \right.
\end{IEEEeqnarray}
The proof is concluded by noting that the probability distribution~$P_{\normx}^*$ defined in Lemma~\ref{lemma:solusion-awgn-prob} satisfies
\begin{IEEEeqnarray}{rCl}
\Ex{P_{\normx}^*}{\func_{\bl,\gamma}(\normx)} = \hat{\func}(\snr)
\end{IEEEeqnarray}
i.e., it achieves the infimum in~\eqref{eq:opt-problem-orig}.
\end{IEEEproof}

We now use Lemma~\ref{lemma:solusion-awgn-prob} to further lower-bound the RHS of~\eqref{eq:prob-Sn-2nd-ineq}, and, hence, further upper-bound the RHS of~\eqref{eq:ub-R-PX-awgn}.
Let $\tangentp$ be as in Lemma~\ref{lemma:solusion-awgn-prob}.
Assume that $\gamma$ in~\eqref{eq:ub-R-PX-awgn} is chosen from the interval $\big(\Cawgn(\snr/(1-\error)) -\delta, \Cawgn(\snr/(1-\error)) + \delta\big)$ for some $0< \delta< \Cawgn(\snr/(1-\error))$ (recall that the upper bound~\eqref{eq:ub-R-PX-awgn} holds for every $\gamma>0$).
%\footnote{Recall that~\eqref{eq:ub-M-PX-awgn} holds for every $\gamma >0$.}
%
For such a $\gamma$, we have
\begin{IEEEeqnarray}{rCl}
\tangentp &\geq&  e^\gamma-1 \\
 &>& \exp\mathopen{}\left(\Cawgn\mathopen{}\left(\frac{\snr}{1-\error}\right)-\delta\right) -1\\
&=&  e^{-\delta}\left(1+ \frac{\snr}{1-\error}\right)  -1 .
\label{eq:lb-w0}
\end{IEEEeqnarray}
Note that the RHS of~\eqref{eq:lb-w0} can be made greater than $\snr$ by choosing $\delta$ sufficiently small.
Let
\begin{IEEEeqnarray}{rCl}
\bl_0 &\define& \frac{2\pi  \big(e^{2\Cawgn(\snr/(1-\error))+2\delta} -1\big)}{\big(\Cawgn(\snr/(1-\error))-\delta\big)^{2}} \\
&\geq&  2\pi  \big(e^{2\gamma} -1\big) \gamma^{-2} . \IEEEeqnarraynumspace
\label{eq:def-n0}
\end{IEEEeqnarray}
Using~\eqref{eq:lb-w0},~\eqref{eq:def-n0}, and Lemma~\ref{lemma:solusion-awgn-prob},  we conclude that for all $\gamma \in\big(\Cawgn(\snr/(1-\error)) -\delta, \Cawgn(\snr/(1-\error)) + \delta\big)$ with $\delta$ chosen so that $\snr<\tangentp$, and all $\bl\geq \bl_0$,
\begin{IEEEeqnarray}{rCl}
\inf\limits_{P_{\normx} \in \Omega }\Ex{}{\func_{\bl,\argpn}(\normx)} = 1- \frac{\snr}{\tangentp} + \frac{\snr}{\tangentp} \func_{\bl,\gamma}(\tangentp) \, .
\label{eq:lb-ex-fn}
\end{IEEEeqnarray}
%
%
%
%Furthermore, the third term on the RHS of~\eqref{eq:ub-R-PX-awgn} can be evaluated as
%\begin{IEEEeqnarray}{rCl}
%
%\label{eq:awgn-varrho-n}
%\end{IEEEeqnarray}
Substituting~\eqref{eq:lb-ex-fn} into~\eqref{eq:prob-Sn-2nd-ineq}, then~\eqref{eq:prob-Sn-2nd-ineq} into~\eqref{eq:ub-R-PX-awgn}, and using that $\varrho_\bl = \bigO\mathopen{}\big(\bl^{-1}\ln \bl\big)$, we obtain
\begin{IEEEeqnarray}{rCl}
\Rawgn^*(\bl,\error) &\leq& \gamma -\frac{1}{\bl}\ln\mathopen{}\left( 1- \frac{\snr}{\tangentp} + \frac{\snr}{\tangentp}   \func_{\bl,\gamma}(\tangentp)
\right. \IEEEeqnarraynumspace \notag\\  &&
\left. \quad\quad\quad \,\,
- \frac{6 \cdot 3^{3/2}}{\sqrt{\bl}} -\error \right)  +  \bigO\mathopen{}\left(\frac{\ln\bl}{\bl}\right). \IEEEeqnarraynumspace
\label{eq:ub-R-PX-awgn-2}
\end{IEEEeqnarray}
%
%The RHS of~\eqref{eq:ub-R-PX-awgn-2} is independent of $P_W$ and the chosen code.

We choose now $\gamma$ as the solution of
\begin{IEEEeqnarray}{rCl}
 1- \frac{\snr}{\tangentp} + \frac{\snr}{\tangentp} \func_{\bl,\gamma}(\tangentp) - \frac{6\cdot 3^{3/2}}{\sqrt{\bl}} -\error = \frac{1}{\sqrt{\bl}} \,.
\label{eq:def-tilde-gamma-n}
\end{IEEEeqnarray}
In words, we choose $\gamma$ so that the argument of the $\ln$ on the RHS of~\eqref{eq:ub-R-PX-awgn-2} is $1/\sqrt{\bl}$.
Evaluating~\eqref{eq:def-x0-mass-point} and~\eqref{eq:def-tilde-gamma-n} for large $n$, we conclude that $\omega_0$ and $\gamma$ must satisfy (see Appendix~\ref{app:proof-expan-gamma-awgn})
\begin{IEEEeqnarray}{rCl}
\sqrt{\bl} \frac{\Cawgn(\tangentp) -\gamma}{\sqrt{\Vawgn(\tangentp)}} = \sqrt{\ln n} + \littleo(1).
\label{eq:ratio-awgn-order}
\end{IEEEeqnarray}
%To partially motivate~\eqref{eq:ratio-awgn-order}, we notice that for a fixed $\gamma>0$, the function $q_{n,\gamma}(\cdot)$ converges to a step function centered at $e^{\gamma}-1$ as $n\to\infty$.
% %
% This implies that $q_{n,\gamma}(\omega_0) \to 0$ and that the LHS of~\eqref{eq:def-x0-mass-point} converges to $-1/(e^{\gamma}-1)$ as $n\to\infty$.
% %
% Using that $Q'(x) = -\frac{1}{\sqrt{2\pi}}e^{-\frac{x^2}{2}}$, the RHS of~\eqref{eq:def-x0-mass-point} is equal to
%\begin{equation}
%-\frac{1}{\sqrt{2\pi}}\exp\mathopen{}\left(-\frac{1}{2}\Big(\sqrt{n}\frac{C(\omega_0) - \gamma}{\sqrt{V(\omega_0)}}\Big)^2\right) \cdot \sqrt{n} \cdot  \frac{d}{dx}\Bigg|_{x=\omega_0}\left(\frac{C(x) -\gamma}{\sqrt{V(x)}}\right).
%\label{eq:intuition-omega}
%\end{equation}
%Assuming that the derivative term in~\eqref{eq:intuition-omega} is $\bigO(1)$, we conclude that $\omega_0$ must satisfy
%\begin{equation}
%\exp\mathopen{}\left(-\frac{1}{2}\mathopen{}\left(\sqrt{n}\frac{C(\omega_0) - \gamma}{\sqrt{V(\omega_0)}}\right)^2\right) = \bigO\mathopen{}\left(\frac{1}{\sqrt{n}}\right)
%\end{equation}
%which is equivalent to~\eqref{eq:ratio-awgn-order}.
%%
%To prove~\eqref{eq:ratio-awgn-order} rigorously, we need to account for the fact that~$\gamma$ changes with $n$, and to show that the derivative in~\eqref{eq:intuition-omega} is indeed $\bigO(1)$.
%%
%This is done in Appendix~\ref{app:proof-expan-gamma-awgn}.
%
Substituting~\eqref{eq:ratio-awgn-order} in~\eqref{eq:def-tilde-gamma-n} (recall the definition of $\func_{\bl,\argpn}(\cdot)$ in~\eqref{eq:def-f-func}), and using that $Q(-\sqrt{\ln n}+\littleo(1))=1-\littleo(1/\sqrt{n})$,   we have
\begin{IEEEeqnarray}{rCl}
\tangentp %&=& \frac{\snr (1-\littleo(1/\sqrt{n}))}{ 1-\error} + \bigO\mathopen{}\left(\frac{1}{\sqrt{\bl}}\right)\\
&=& \frac{\snr}{1-\error} + \bigO\mathopen{}\left(\frac{1}{\sqrt{n}}\right).\label{eq:expan-w_0-app-awgn-2}
\end{IEEEeqnarray}
Finally, solving~\eqref{eq:ratio-awgn-order} for $\gamma$, and using~\eqref{eq:expan-w_0-app-awgn-2}, we conclude that
\begin{IEEEeqnarray}{rCl}
\gamma &=& C(\tangentp) - \sqrt{V(\tangentp)}\sqrt{\frac{\ln\bl}{\bl}} + \littleo\mathopen{}\left(\frac{1}{\sqrt{\bl}}\right)\label{eq:expan-gamma-appen-1} \\
&=& \Cawgn\mathopen{}\left(\frac{\snr}{1-\error}\right)  - \sqrt{\Vawgn\mathopen{}\left(\frac{\snr}{1-\error}\right) } \sqrt{\frac{\ln\bl}{\bl}} + \bigO\mathopen{}\left(\frac{1}{\sqrt{\bl}}\right). \IEEEeqnarraynumspace
\label{eq:ub-tilde-gamma-n}
\end{IEEEeqnarray}
Observe now that $\gamma$ belongs indeed to the interval~$\big(\Cawgn(\snr/(1-\error)) -\delta, \Cawgn(\snr/(1-\error)) + \delta\big)$ for sufficiently large $\bl$.
The proof of the converse part of Theorem~\ref{thm:awgn-second-order} is concluded by substituting~\eqref{eq:def-tilde-gamma-n} and~\eqref{eq:ub-tilde-gamma-n} into~\eqref{eq:ub-R-PX-awgn-2}.

\subsection{Proof of the Achievability Part}
\label{sec:proof-awgn-sec-order-ach}
The proof is a refinement of the proof of~\cite[Th.~77]{polyanskiy10}.
Let $(\bl,\NumCode_\bl,\error_\bl)_{\st}$, where
\begin{equation}
\error_\bl = \frac{2}{\sqrt{\bl\ln\bl}}
\label{eq:def-error-n-awgn}
\end{equation}
be a code for the AWGN channel~\eqref{eq:channel-io-awgn} with codewords $\{\vecc_l\}$, $l=1,\ldots,\NumCode_\bl$,  satisfying the short-term power constraint
\begin{IEEEeqnarray}{rCl}
\frac{1}{\bl}\|\vecc_l\|^2 \leq \snr_\bl \define \snr \frac{1-\error_\bl}{1-\error}, \quad l=1,\ldots,\NumCode_\bl.
\label{eq:def-rho-n}
\end{IEEEeqnarray}
Set
\begin{equation}
\NumCode = \NumCode_\bl \frac{1- \error_\bl}{1-\error}
\label{eq:awgn-M-M1}
\end{equation}
and assume that $\bl$ is large enough so that $\NumCode > \NumCode_\bl$.
We construct a code with $\NumCode$ codewords for the case of long-term power constraint by adding $(\NumCode-\NumCode_\bl)$ all-zero codewords to the codewords of the~$(\bl,\NumCode_\bl,\error_\bl)_{\st}$ code.
However, we leave  the decoder unchanged in spite of the addition of extra codewords.
The resulting code satisfies the long-term power constraint. Indeed,
\begin{IEEEeqnarray}{rCl}
0 \cdot \frac{\NumCode-\NumCode_\bl}{\NumCode} + \snr_\bl\cdot \frac{\NumCode_\bl }{\NumCode} =  \snr.
\end{IEEEeqnarray}
At the same time, the average probability of error of the new code is upper-bounded by
%\footnote{Note that, the receiver does not need to detect whether the transmitted codeword is zero or not. Instead, it can always assume that a non-zero codeword is transmitted.}
\begin{IEEEeqnarray}{rCl}
1 \cdot \frac{\NumCode -\NumCode_\bl}{\NumCode} + \error_\bl\cdot \frac{\NumCode_\bl}{\NumCode}  = \error.
\end{IEEEeqnarray}
Therefore, by definition,
\begin{IEEEeqnarray}{rCl}
 \Rawgn^* (\bl,\error)  &\geq & \frac{\ln \NumCode}{ \bl}\\
&=& \frac{\ln \NumCode_\bl}{\bl} + \frac{1}{\bl}\ln\mathopen{}\left(\frac{1-\error_\bl}{1- \error}\right) \label{eq:awgn-nonasy-ach-bound}\\
&=&  \frac{\ln \NumCode_\bl}{\bl} +  \bigO\mathopen{}\left(\frac{1}{\bl}\right).\label{eq:awgn-nonasy-ach-bound-2}
\end{IEEEeqnarray}
Here,~\eqref{eq:awgn-nonasy-ach-bound} follows from~\eqref{eq:awgn-M-M1}, and in~\eqref{eq:awgn-nonasy-ach-bound-2} we used~\eqref{eq:def-error-n-awgn}.
As noted in Section~\ref{sec:intro}, the strategy just described is equivalent to concatenating the $(\bl,\NumCode_\bl,\error_\bl)_{\st}$ code with a power controller that zeroes the power of the transmitted codeword with probability
\begin{IEEEeqnarray}{rCL}
  \frac{\error-\error_n}{1-\error}=\error-\bigO\mathopen{}\left(\frac{1}{\sqrt{\bl\ln \bl}}\right)
\end{IEEEeqnarray}
and keep the power unchanged otherwise.

To conclude the proof, we show that there exists an $(\bl,\NumCode_\bl,\error_\bl)_{\st}$  code with $\error_\bl$ as in~\eqref{eq:def-error-n-awgn} and with codewords satisfying~\eqref{eq:def-rho-n},  for which
\begin{equation}
\frac{\ln \NumCode_\bl}{\bl} \geq \Cawgn\mathopen{}\left(\frac{\snr}{1-\error}\right) - \sqrt{\Vawgn\mathopen{}\left(\frac{\snr}{1-\error}\right)} \! \sqrt{\frac{\ln \bl}{\bl}} + \bigO \mathopen{}\left(\frac{1}{\sqrt{\bl\ln\bl}}\right).
\label{eq:awgn-bound-M1}
\end{equation}
Before establishing this inequality, we remark that a weaker version of~\eqref{eq:awgn-bound-M1}, with~$\bigO(1/\sqrt{\bl\ln\bl})$ replaced by~$\littleo(\sqrt{\bl^{-1}\ln\bl})$, follows directly from~\cite[Th.~96]{polyanskiy10}. %\footnote{This weaker version cannot be used to establish~\eqref{eq:thm-awgn-R}.}
The proof of~\cite[Th.~96]{polyanskiy10} is built upon a moderate-deviation analysis~\cite[Th.~3.7.1]{dembo98}.
To prove the tighter inequality~\eqref{eq:awgn-bound-M1} we use instead a Cramer-Esseen-type central limit theorem~\cite[Th.~VI.1]{petrov75}.

We proceed now with the proof of~\eqref{eq:awgn-bound-M1}.
By applying the $\kappa\beta$ bound~\cite[Th.~25]{polyanskiy10-05}, with $\tau = \error_\bl/2$, $\setF_\bl\define  \{\vecx \in \complexset^\bl : \|\vecx\|^2 =\bl\snr_\bl \}$,
and
$Q_{\randvecy} = \jpg(\mathbf{0}, (1+\snr_\bl) \matI_{\bl})$, we conclude that there exists an $(\bl,\NumCode_\bl,\error_\bl)_{\st}$ code with codewords in $\setF_{\bl}$ for which
\begin{IEEEeqnarray}{rCl}
\ln \NumCode_\bl &\geq& - \sup_{\vecx \in \setF_\bl} \mathopen{}\Big\{\ln \beta_{1-\error_\bl/2}(P_{\randvecy \given \randvecx = \vecx}, Q_{\randvecy})\Big\} \notag\\
 && + \, \ln \kappa_{\error_\bl/2}(\setF_\bl , Q_{\randvecy}).
\label{eq:awgn-kappa-beta}
\end{IEEEeqnarray}
Here, $\kappa_{\error_\bl/2}(\setF_\bl , Q_{\randvecy})$ is defined as follows~\cite[Eq.~(107)]{polyanskiy10-05}:
\begin{IEEEeqnarray}{rCl}
\kappa_{\error_\bl/2}(\setF_\bl , Q_{\randvecy}) \define \inf \int P_{Z\given \randvecy}(1\given \vecy) Q_{\randvecy} (d\vecy).
\label{eq:def-kappa-tau}
\end{IEEEeqnarray}
The infimum in~\eqref{eq:def-kappa-tau} is over all conditional probability distributions $P_{Z\given \randvecy}: \complexset^\bl\to \{0,1\}$ satisfying
\begin{IEEEeqnarray}{rCl}
 \int P_{Z\given \randvecy}(1\given \vecy) P_{\randvecy \given \randvecx =\vecx} (d\vecy) \geq \frac{\error_\bl}{2}, \quad \forall \vecx \in \setF_{\bl}.
\end{IEEEeqnarray}
Let $\vecx_0 \define [\sqrt{\snr_\bl},\cdots,\sqrt{\snr_\bl}] \in \setF_{\bl}$.
Using that $\kappa_{\error_\bl/2}(\setF_n , Q_{\randvecy}) \geq \big(\error_\bl/2 - e^{-c_2 \bl}\big)/c_1$ for some constants $c_1>0$ and $c_2>0$ (see~\cite[Lem.~61]{polyanskiy10-05}) and that $\beta_{1-\error_\bl/2}(P_{\randvecy \given \randvecx = \vecx}, Q_{\randvecy})$ takes the same value for all $\vecx \in \setF_\bl$ (see~\cite[Sec.~III.J]{polyanskiy10-05}), we get
\begin{IEEEeqnarray}{rCl}
\ln \NumCode_\bl &\geq& - \ln \beta_{1-\error_\bl/2}(P_{\randvecy \given \randvecx = \vecx_0}, Q_{\randvecy}) \notag\\
&&+ \, \ln\mathopen{}\left(\frac{1}{c_1}\left(\frac{1}{\sqrt{\bl\ln \bl}} - e^{-c_2\bl}\right)\right)\\
&=& - \ln \beta_{1-\error_\bl/2}(P_{\randvecy \given \randvecx = \vecx_0}, Q_{\randvecy})  + \bigO(\ln\bl). \IEEEeqnarraynumspace
\label{eq:lb-Mn-awgn}
\end{IEEEeqnarray}
We now further lower-bound the first term on the RHS of~\eqref{eq:lb-Mn-awgn} as follows~\cite[Eq.~(103)]{polyanskiy10-05}:
\begin{IEEEeqnarray}{rCl}
 - \ln \beta_{1-\error_\bl/2}(P_{\randvecy \given \randvecx = \vecx_0}, Q_{\randvecy}) &\geq&\bl \gamma_\bl
\label{eq:lb-M1}
\end{IEEEeqnarray}
where $\gamma_\bl$ satisfies
\begin{IEEEeqnarray}{rCl}
P_{\randvecy \given \randvecx = \vecx_0} \mathopen{}\left[ \ln \frac{dP_{\randvecy \given \randvecx = \vecx_0} }{dQ_{\randvecy}} \leq  \bl \gamma_\bl \right] \leq \frac{\error_\bl}{2} = \frac{1}{\sqrt{\bl\ln\bl}}. \IEEEeqnarraynumspace
\label{eq:awgn-ach-def-gamma-n}
\end{IEEEeqnarray}

To conclude the proof, we  show that, for sufficiently large~$\bl$, the choice
\begin{IEEEeqnarray}{rCl}
\gamma_\bl = C(\snr_\bl) - \sqrt{V(\snr_\bl)}\sqrt{\frac{\ln\bl }{\bl}}
\label{eq:awgn-ach-choose-gamma-n}
\end{IEEEeqnarray}
satisfies~\eqref{eq:awgn-ach-def-gamma-n}.
The desired result~\eqref{eq:awgn-bound-M1} then follows by substituting~\eqref{eq:awgn-ach-choose-gamma-n} into~\eqref{eq:lb-M1}, and~\eqref{eq:lb-M1} into~\eqref{eq:lb-Mn-awgn}, and by using that
\begin{IEEEeqnarray}{rCl}
&&C(\snr_\bl ) = \Cawgn\mathopen{}\left(\frac{\snr}{1-\error}\right) + \bigO\mathopen{}\left(\frac{1}{\sqrt{\bl\ln\bl}}\right)\\
&&V(\snr_\bl) = \Vawgn\mathopen{}\left(\frac{\snr}{1-\error}\right) + \bigO\mathopen{}\left(\frac{1}{\sqrt{\bl\ln\bl}}\right)\label{eq:expan-sqrt-loglogn}
%\\ &&\sqrt{\log (\bl \log\bl) } = \sqrt{\log\bl } + \littleo(1).
\end{IEEEeqnarray}
which follow from~\eqref{eq:def-error-n-awgn},~\eqref{eq:def-rho-n}, and from Taylor's theorem~\cite[Th.~5.15]{rudin76}.

To establish that~\eqref{eq:awgn-ach-def-gamma-n} holds when $\gamma_\bl$ is chosen as in~\eqref{eq:awgn-ach-choose-gamma-n}, we shall use a Cramer-Esseen-type central limit theorem on the LHS of~\eqref{eq:awgn-ach-def-gamma-n}.
We start by noting that, under $P_{\randvecy \given \randvecx =\vecx_0}$, the random variable $\ln\frac{dP_{\randvecy \given \randvecx = \vecx_0}}{dQ_{\randvecy}}$ has the same distribution as (see~\eqref{eq:info_density_awgn})
\begin{IEEEeqnarray}{rCl}
\bl C(\snr_\bl) + \sqrt{V(\snr_\bl)}\sum\limits_{i=1}^{\bl} T_i
\label{eq:info-den-ach-awgn}
\end{IEEEeqnarray}
where
\begin{IEEEeqnarray}{rCl}
T_i \define \frac{1}{\sqrt{V(\snr_\bl)}} \left(1-\frac{|\sqrt{\snr_\bl}Z_i-1|^2}{1+\snr_\bl}\right), \,\,i=1,\ldots,\bl \IEEEeqnarraynumspace
\end{IEEEeqnarray}
are i.i.d. random variables with zero mean and unit variance, and $\{Z_i\}$, $i=1,\ldots,\bl$, are i.i.d. $\jpg(0,1)$-distributed.
It follows that
\begin{IEEEeqnarray}{rCl}
\IEEEeqnarraymulticol{3}{l}{
\prob \mathopen{}\left[\ln\frac{dP_{\randvecy \given \randvecx = \vecx_0}}{dQ_{\randvecy}}\leq \bl\gamma_\bl\right]}\notag\\
\quad  &=& \prob\mathopen{}\left[\frac{1}{\sqrt{\bl}} \sum\limits_{i=1}^{\bl} T_i \leq \sqrt{\bl} \frac{\gamma_\bl -C(\snr_\bl)}{\sqrt{V(\snr_\bl)}}\right]\label{eq:prob-sn-ach-941}\\
&=&\prob\mathopen{}\left[\frac{1}{\sqrt{\bl}} \sum\limits_{i=1}^{\bl} T_i \leq - \sqrt{\ln\bl} \right]\label{eq:prob-sn-ach}
\end{IEEEeqnarray}
where the last step follows by choosing $\gamma_\bl$ as specified in~\eqref{eq:awgn-ach-choose-gamma-n}.
To upper-bound the RHS of~\eqref{eq:prob-sn-ach}, we shall need the following version of Cramer-Esseen-type central-limit theorem.
\begin{thm}[\!\!{\cite[Th.~VI.1]{petrov75}\cite[Th.~15]{yang14-07a}}]
\label{thm:refine-be-lt}
Let $X_1,\ldots,X_\bl$ be a sequence of i.i.d. real random variables having zero mean and unit variance.
%Let $\cdfF_X(x)$ be the cdf of one such random variables.
Furthermore, let
\begin{IEEEeqnarray}{rCl}
\varphi(t)&\define&\Ex{}{e^{itX_1}}\,\,\text{and}\,\,
F_\bl(\xi) \define \prob\mathopen{}\left[\frac{1}{\sqrt{\bl}} \sum\limits_{j=1}^{\bl}X_j \leq \xi\right].\IEEEeqnarraynumspace
\end{IEEEeqnarray}
If $\Ex{}{|X_1|^4} < \infty$ and if $\sup_{|t|\geq \zeta} |\varphi(t)| \leq k_0$ for some $k_0 <1$, where $\zeta\define1/({12\Ex{}{|X_1|^3}})$, then for every $\xi$ and $\bl$
\begin{IEEEeqnarray}{rCl}
\IEEEeqnarraymulticol{3}{l}{
\left|F_\bl(\xi) - Q(-\xi) - k_1(1-\xi^2)e^{-\xi^2/2}\frac{1}{\sqrt{\bl}}\right|
}\notag\\ \quad
&\leq& k_2\left\{\frac{\Ex{}{|X_1|^4} }{\bl}+ \bl^6\left(k_0+\frac{1}{2\bl}\right)^{\bl}\right\}.\IEEEeqnarraynumspace
\label{eq:thm-osipov-refine}
\end{IEEEeqnarray}
Here, $k_1 \define \Ex{}{X_1^3}/(6\sqrt{2\pi})$, and $k_2$ is a positive constant independent of $\{X_i\}$ and $\xi$.
\end{thm}

To apply Theorem~\ref{thm:refine-be-lt}, we need first to verify that the conditions under which this theorem holds are satisfied, i.e., that
\begin{IEEEeqnarray}{rCl}
\Ex{}{T_1^4} <\infty
\label{eq:cond1-cramer-esseen}
\end{IEEEeqnarray}
and that
\begin{IEEEeqnarray}{rCl}
\sup\limits_{|t|\geq 1/(12\Ex{}{|T_1|^3}) } \left| \Ex{}{e^{i t T_1}} \right|\leq k_0
\label{eq:cond2-cramer-esseen}
\end{IEEEeqnarray}
for some $k_0<1$. Both~\eqref{eq:cond1-cramer-esseen} and~\eqref{eq:cond2-cramer-esseen} follow as special cases of the more general results provided in~\cite[App.~IV.A]{yang14-07a}.
Applying Theorem~\ref{thm:refine-be-lt} to the RHS of~\eqref{eq:prob-sn-ach}, we obtain
\begin{IEEEeqnarray}{rCl}
\IEEEeqnarraymulticol{3}{l}{
 \prob\mathopen{}\left[\frac{1}{\sqrt{\bl}} \sum\limits_{i=1}^{\bl} T_i \leq - \sqrt{\ln\bl} \right]}\notag\\
\quad  &\leq& Q\mathopen{}\left( \sqrt{\ln\bl} \right)   + \underbrace{\frac{\Ex{}{T_1^3}}{6\sqrt{2\pi} \sqrt{\bl}} \Big(1- \ln\bl\Big) e^{-\frac{\ln\bl }{2}}}_{=\bigO(\ln(\bl)/\bl)} \notag\\
&&+\, \underbrace{k_2 \bigg(\frac{\Ex{}{T_1^4}}{\bl (1+\sqrt{\ln \bl})^4} + \bl^6\Big( k_0 + \frac{1}{2\bl}\Big)^\bl\bigg)}_{=\littleo(1/\bl)}{}\label{eq:bound-prob-sn-awgn-1}\IEEEeqnarraynumspace\\
&=& Q\mathopen{}\left( \sqrt{\ln\bl } \right) + \bigO\mathopen{}\left(\frac{\ln\bl}{\bl}\right) \label{eq:bound-prob-sn-awgn-0}\\
%&\leq& \frac{1}{\sqrt{2\pi}}\frac{e^{-(\log\bl)/2}}{  \sqrt{\log\bl}}  + \bigO\mathopen{}\left(\frac{\log\bl}{\bl}\right) \label{eq:bound-prob-sn-awgn-1}\\
&\leq & \frac{1}{\sqrt{2\pi}\sqrt{\bl \ln\bl }}  +\bigO\mathopen{}\left(\frac{\ln\bl}{\bl}\right)\label{eq:bound-prob-sn-awgn}
\end{IEEEeqnarray}
where $k_2>0$ in~\eqref{eq:bound-prob-sn-awgn-1} is a constant that does not depend on~$T_1$ and~$\bl$.
Here, in~\eqref{eq:bound-prob-sn-awgn-0} we used~\eqref{eq:cond1-cramer-esseen}, that, by Lyapunov's inequality,~$|\Ex{}{T_1^3}|\leq \Ex{}{|T_1|^3}\leq (\Ex{}{T_1^4})^{3/4} <\infty$, and that
\begin{IEEEeqnarray}{rCl}
n^6\left(k_0 + \frac{1}{2\bl}\right)^\bl = \littleo\mathopen{}\left(\frac{1}{\bl}\right).
\end{IEEEeqnarray}
Furthermore, \eqref{eq:bound-prob-sn-awgn} follows because
\begin{IEEEeqnarray}{rCl}
Q(x)\leq \frac{1}{\sqrt{2\pi}x}e^{-x^2/2},  \quad \forall x >0.
\end{IEEEeqnarray}
The bound~\eqref{eq:bound-prob-sn-awgn} implies that for the choice of~$\gamma_\bl$ in~\eqref{eq:awgn-ach-choose-gamma-n}, the inequality~\eqref{eq:awgn-ach-def-gamma-n} holds  for sufficiently large $\bl$. This concludes the proof of the achievability part of Theorem~\ref{thm:awgn-second-order}.

\subsection{Convergence to Capacity}
\label{sec:conv-capacity-awgn}
For AWGN channels subject to a short-term power constraint, it follows from~\cite[Sec.~IV.B]{polyanskiy10-05} that the finite-blocklength rate penalty compared to channel capacity is approximately proportional to $1/\sqrt{\bl}$.
In contrast, Theorem~\ref{thm:awgn-second-order} in Section~\ref{sec:awgn-case} shows that for AWGN channels subject to a long-term power constraint, this rate penalty is approximately proportional to $\sqrt{\bl^{-1}\ln \bl}$.
%
%
%
%this does not necessarily mean that $\Rawgn^*(\bl,\error)$ converges to capacity slower in the case of long-term power constraint than in the case of short-term power constraint.
%In fact, the term $Q^{-1}(\error)$ multiplying $1/\sqrt{\bl}$ in~\eqref{eq:approx_R_intro_awgn_st} is comparable to $\sqrt{\log\bl}$ for practical values of $\error$ and $\bl$.
%%
%For example, for $\epsilon=10^{-3}$, $\bl=1000$, and $\snr=0$ dB, we have
%\begin{equation}
%\sqrt{\frac{\Vawgn(\snr)}{\bl}}Q^{-1}(\error) = 0.085 > \sqrt{V\mathopen{}\left(\frac{\snr}{1-\error}\right)}\sqrt{\frac{\log\bl}{\bl}}= 0.072.
%\end{equation}
%%
To understand the implications of this asymptotic difference in convergence speed, we next complement our asymptotic characterization of $\Rawgn^*(\bl,\error)$ with numerical results and an easy-to-evaluate approximation that is more accurate than~\eqref{eq:intro-R-awgn-lt}.

\subsubsection{Normal Approximation}
We  start by developing a normal approximation for $\Rawgn^*(\bl,\error)$ along the lines of~\cite[Eq.~(296)]{polyanskiy10-05}.
We will then show through numerical results that this approximation is useful to characterize the speed at which $\Rawgn^*(\bl,\error)$ converges to  $C(\snr/(1-\epsilon))$ as $\bl\to\infty$.
We define the normal approximation $\Rawgnnalt(\bl,\error)$ of $\Rawgn^*(\bl,\error)$ to be the solution of\footnote{The term~$(2\bl)^{-1}\ln \bl$ in~\eqref{eq:awgn-na-lt} is motivated by the normal approximation in~\cite[Eq.~(296)]{polyanskiy10-05} for the short-term power constraint case.}
 %
%Note that the results reported in~\cite[Eq.~(296)]{polyanskiy10-05} and~\cite{tan13-11} pertain to the real AWGN channel.
%Extension to the complex case follows by identifying a complex AWGN channel with blocklength $n$ with a real AWGN channel with the same SNR and blockength $2n$.}
%
\begin{equation}
\inf_{P_{\normx} \in \Omega } \!\Ex{}{Q\mathopen{}\left(\!\sqrt{\bl}\frac{\Cawgn(\normx) - \Rawgnnalt(\bl,\error)+(2n)^{-1}\ln\bl}{ \sqrt{\Vawgn( \normx)}}\right)} =\error.
\label{eq:awgn-na-lt}
\end{equation}
Note that this optimization problem is a special case of~\eqref{eq:opt-problem-orig} (set $\gamma = \Rawgnnalt(\bl,\error)- (2\bl)^{-1}\ln\bl$).
It then follows from Lemma~\ref{lemma:solusion-awgn-prob} that the  probability distribution that minimizes the LHS of~\eqref{eq:awgn-na-lt} has two forms depending on $\bl$, $\error$, and $\snr$.
For small values of~$\bl$ or~$\error$,  the optimal probability distribution has only one mass point located at~$\snr$.
In this case, the resulting approximation~$\Rawgnnalt(\bl,\error)$ coincides with the normal approximation for the case of short-term power constraint, which we denote by $\Rawgnnast(\bl,\error)$, and is given by~\cite[Eq.~(296)]{polyanskiy10-05}
\begin{IEEEeqnarray}{rCl}
\Rawgnnast (\bl,\error) \define \Cawgn(\snr)-\sqrt{\frac{\Vawgn(\snr)}{\bl}}Q^{-1}(\error) + \frac{\ln\bl}{2\bl}. \IEEEeqnarraynumspace
\end{IEEEeqnarray}
This suggests that a long-term power constraint is not beneficial in this scenario.
Conversely, the long-term power constraint may be beneficial when the $P_{\Pi}$ solving~\eqref{eq:awgn-na-lt} has two mass points, in which case we have
%$\bl$, $\error$, and $\snr$ satisfy
\begin{IEEEeqnarray}{rCl}
\Rawgnnalt(\bl,\error) > \Rawgnnast (\bl,\error).
\label{eq:lt-benificial}
\end{IEEEeqnarray}

%\begin{table}[t]
%%
%   \caption{Minimum blocklength required for the long-term power constraint to be beneficial on an AWGN channel.\label{tab:awgn-bl}}
%\begin{center}
%\begin{tabular}{|l|c|}
%\hline
%SNR, error probability& Blocklength\\
%\hline
%$\snr = 0$ dB, $\error = 10^{-1}$ & $\bl > 86$\\
%\hline
%$\snr= 0$ dB, $\error = 10^{-3}$ & $\bl > 2.65\times 10^{5}$\\
%\hline
%$\snr = 10$ dB, $\error = 10^{-4}$ & $\bl>7.66 \times 10^{6}$ \\
%\hline
%$\snr = 20$ dB, $\error =10^{-6}$ & $\bl>4.17 \times 10^{10}$\\ \hline
%\end{tabular}
%\end{center}
%\end{table}

\begin{table}[t]
   \caption{Minimum blocklength required for the long-term power constraint to be beneficial on an AWGN channel.\label{tab:awgn-bl}}
\begin{center}
  \renewcommand{\arraystretch}{1.3}
\begin{tabular}{|l|c|c|}
\hline
& $\error=0.1$ & $\error = 10^{-3}$\\
\hline
$\snr = -10$ dB & $n\gtrsim 10^3$ & $\bl \gtrsim 2\times 10^6$\\
\hline
$\snr = 0$ dB & $n\gtrsim 10^2$ & $\bl \gtrsim 3\times 10^5$\\
\hline
$\snr= 10$ dB& $n \gtrsim 30$ & $\bl \gtrsim 10^5$\\
\hline
$\snr = 20$ dB & $n\gtrsim 30$ & $\bl \gtrsim 9\times 10^4$\\ \hline
\end{tabular}
\end{center}
\end{table}

Next, we establish a sufficient condition for~\eqref{eq:lt-benificial} to hold.
Set $\gamma_0 = \Rawgnnast (\bl,\error) - (2\bl)^{-1}\ln\bl$. By Lemma~\ref{lemma:solusion-awgn-prob},~\eqref{eq:lt-benificial} holds if
\begin{IEEEeqnarray}{rCl}
\bl \geq 2\pi(e^{2\gamma_0} -1)\gamma_0^{-2}
\label{eq:cond-lt-eq-st2}
\end{IEEEeqnarray}
and if
\begin{IEEEeqnarray}{rCl}
\snr < \tangentp
\label{eq:cond-lt-benifitial-2}
\end{IEEEeqnarray}
where $\tangentp$ is the solution of~\eqref{eq:def-x0-mass-point} with $\gamma$ replaced by $\gamma_0$.
Since $\func_{\bl,\gamma_0}(\cdot)$ defined in~\eqref{eq:def-f-func} is convex and strictly decreasing on $[e^{\gamma_0}-1,\infty)$, and since
\begin{IEEEeqnarray}{rCl}
e^{\gamma_0} -1 \leq  e^{C(\snr)} -1  =\snr
\end{IEEEeqnarray}
the inequality~\eqref{eq:cond-lt-benifitial-2} holds if and only if
\begin{IEEEeqnarray}{rCl}
\frac{\func_{\bl,\gamma_0}(\snr)-1}{\snr} > \func_{\bl,\gamma_0}'(\snr).
\label{eq:cond-lt-benificial-3}
\end{IEEEeqnarray}
A direct computation shows that~\eqref{eq:cond-lt-benificial-3} is equivalent to
\begin{equation}
\bl > \! \bigg( \! \frac{1+\snr}{\snr}\sqrt{2\pi V(\snr)}  (1-\error)e^{\frac{(Q^{-1}(\error))^2}{2}} + \frac{Q^{-1}(\error)}{(1+\snr)^2\sqrt{V(\snr)}} \! \bigg)^{2} \!.\,
\label{eq:cond-lt-eq-st3}
\end{equation}
In Table~\ref{tab:awgn-bl}, we list the minimum blocklength required for the long-term power constraint to be beneficial for different values of $\snr$ and $\error$, according to the normal approximation.
%The numbers are computed using~\eqref{eq:cond-lt-eq-st2} and~\eqref{eq:cond-lt-eq-st3}, which are based on comparing the normal approximations.

%In fact, for the parameters considered in Fig.~\ref{fig:bounds-awgn-0001}, the minimization
%
%

\subsubsection{Numerical Results}

\begin{figure}[t]
	\centering
\includegraphics[scale=0.8]{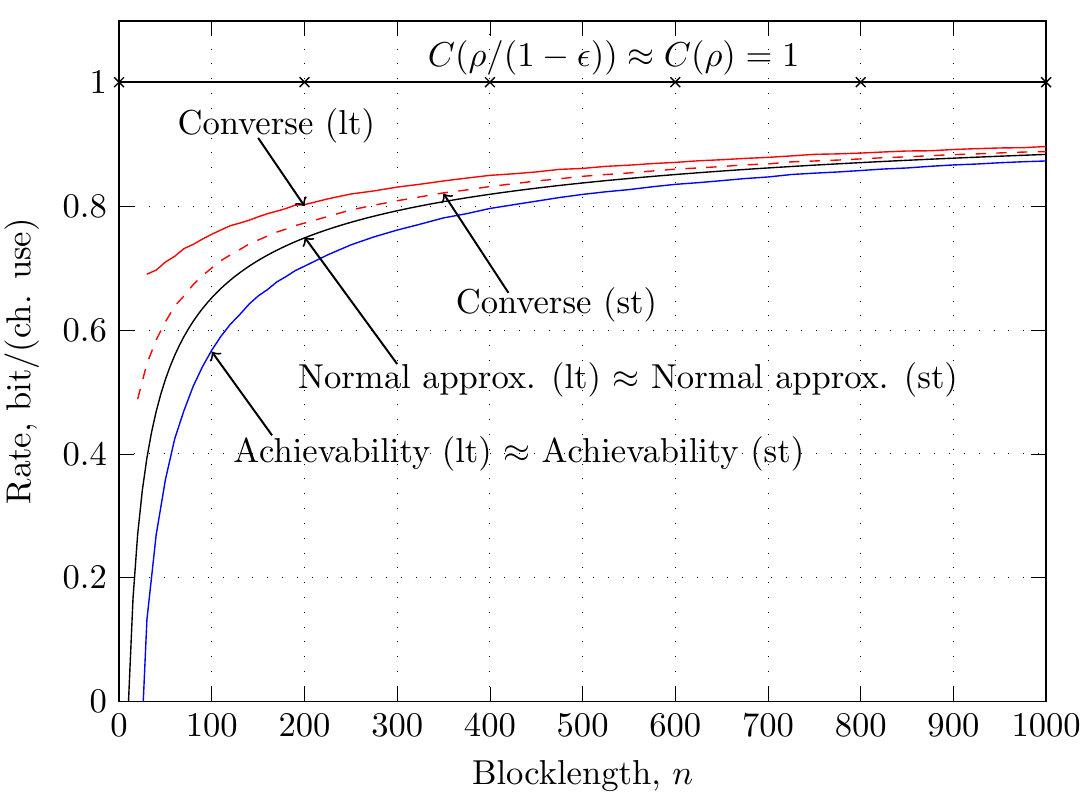}
\caption{Nonasymptotic bounds on $\Rawgn^*(\bl,\error)$ and normal approximation for the case $\snr = 0$ dB, and $\error=10^{-3}$.
Two nonasymptotic bounds for the case of short-term power constraint and the corresponding normal approximation are also depicted. Here, $\lt$ stands for long-term and $\st$ stands for short-term.
%\todo{Wei: ``Approximation'' should be lower case.}
\label{fig:bounds-awgn-0001}}
\end{figure}
\begin{figure}[t]
	\centering
\includegraphics[scale=0.8]{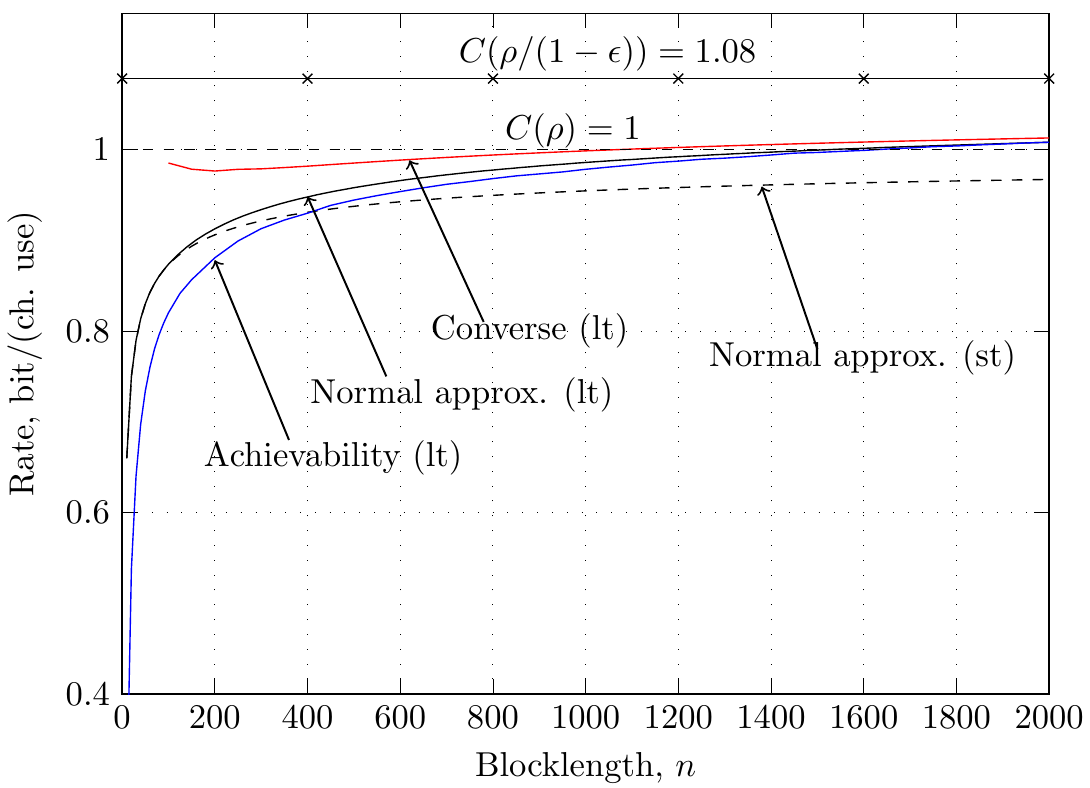}
\caption{Nonasymptotic bounds on $\Rawgn^*(\bl,\error)$ and normal approximation for the case $\snr = 0$ dB, and $\error=0.1$.
The normal approximation for the case of short-term power constraint is also depicted. Here, $\lt$ stands for long-term and $\st$ stands for short-term.
%\todo{Wei: ``Approximation'' should be lower case.}
\label{fig:bounds-awgn-01}}
\end{figure}

In Fig.~\ref{fig:bounds-awgn-0001}, we compare\footnote{The numerical routines used to obtain these results are available at https://github.com/yp-mit/spectre} the normal approximation~\eqref{eq:awgn-na-lt} against nonasymptotic converse and achievability bounds for the case $\rho = 0$ dB and $\error = 10^{-3}$.
The achievability bound is computed by (numerically) maximizing~\eqref{eq:awgn-nonasy-ach-bound} over $\error_\bl \in (0,\error)$ with $\ln\NumCode_\bl$ given in~\eqref{eq:awgn-kappa-beta}.
The converse bound is computed by using~\eqref{eq:ub-R-PX-awgn}.
Note that the infimum
\begin{equation}
\inf\limits_{P_{\normx}\in \Omega} \prob[S_\bl(\normx) \leq \bl\gamma]
\end{equation}
on the RHS of~\eqref{eq:ub-R-PX-awgn} can be  solved analytically using the same technique as in the proof of Lemma~\ref{lemma:solusion-awgn-prob}.
For comparison, we also plot the achievability bound ($\kappa\beta$ bound~\cite[Th.~25]{polyanskiy10-05}) and converse bound (meta-converse bound~\cite[Th.~41]{polyanskiy10-05}) as well as the normal approximation~\cite[Eq.~(296)]{polyanskiy10-05} for an AWGN channel with the same SNR and error probability, but subject to a short-term power constraint.
We observe that for the parameters considered in this figure, the achievability bounds for the long-term power constraint and the short-term power constraint  coincide numerically.
The same observation holds also for the normal approximation.
This is not surprising, since~\eqref{eq:cond-lt-eq-st3} implies that a blocklength $\bl> 2.65\times 10^{5}$ is required for $\Rawgnnalt(\bl,\epsilon)$ to be larger than $\Rawgnnast (\bl,\error)$.

In Fig.~\ref{fig:bounds-awgn-01}, we consider the case $\rho = 0$ dB and $\error = 10^{-1}$.
In this scenario, having a long-term power constraint yields a rate gain compared to the case of short-term power constraint (about $4\%$ when $\bl=1000$).
Observe that the blocklength required to achieve $90\%$ of the $\epsilon$-capacity for the long-term constraint case is approximately~$650$.
For the case of short-term power constraint, this number is approximately $320$.
Hence, for the parameters chosen in Fig.~\ref{fig:bounds-awgn-01}, the maximal channel coding rate converges more slowly to the $\epsilon$-capacity when a long-term power constraint is present.
To conclude, we note that the approximation for the maximal channel coding rate obtained by omitting the $\bigO(1/\sqrt{n})$ term in~\eqref{eq:intro-R-awgn-lt} is often less accurate than~\eqref{eq:awgn-na-lt}.

%This agrees with the result in Theorem~\ref{thm:awgn-second-order}, where we showed that the second-order term in the large-\bl expansion of $\Rawgn^*(\bl,\error)$ is proportional to $\sqrt{\log(\bl)/\bl}$, whereas it is proportional to $1/\sqrt{\bl}$ in the short-term power-constraint case.
%
%%
%%
%Finally, observe that the normal approximation~\eqref{eq:awgn-na-lt} is accurate for the parameters considered in both figures.
%

%\subsection{Numerical Results}
%
%
%
%However, this does not necessarily mean that $\Rawgn^*(\bl,\error)$ converges to capacity slower in the case of long-term power constraint than in the case of short power constraint.
%In fact, the term $Q^{-1}(\error)$ multiplying $1/\sqrt{\bl}$ in~\eqref{eq:approx_R_intro_awgn_st} is comparable to $\sqrt{\log\bl}$ for practical values of $\error$ and $\bl$.
%%
%For example, for $\epsilon=10^{-3}$, $\bl=1000$, and $\snr=0$ dB, we have
%\begin{IEEEeqnarray}{rCl}
%\sqrt{\frac{\Vawgn(\snr)}{\bl}}Q^{-1}(\error) = 0.085 > \! \sqrt{V\!\mathopen{}\left(\!\frac{\snr}{1-\error}\!\right)}\sqrt{\frac{\log\bl}{\bl}}= 0.072.\IEEEeqnarraynumspace
%\end{IEEEeqnarray}
%To better characterize the speed of convergence to capacity, our result needs to be complemented with plottable nonasymptotic bounds on $\Rawgn^*(\bl,\error)$ similar to the ones reported in~\cite[Sec.~III.J]{polyanskiy10-05}.

\section{The Quasi-Static Fading Channel}
\label{sec:qs-case}
We move now to the quasi-static fading channel~\eqref{eq:channel-io-qs}.
%
%
% %$\vecx \in \complexset^{\bl}$ contains the transmitted symbols; $\vecnoise\sim \jpg(\mathbf{0}, \matI_{\bl})$ is the additive white gaussian noise vector independent of $H$ and~$\vecx$.
%%
%We assume that the realizations of $H$ are known to both the transmitter and the receiver and denote the channel gain by $G=|H|^2$.
%
%
An $(\bl, \NumCode,\error)_{\avg}$ code for the quasi-static fading channnel~\eqref{eq:channel-io-qs} consists of:
\begin{enumerate}
\item an encoder $\encoder$: $ \{1,\ldots,\NumCode\}\times \complexset \to \complexset^\bl$ that maps the message $\msg \in \{1,\ldots,\NumCode\}$ and the channel coefficient $H$ to a codeword $\vecx = \encoder (\msg, H)$ satisfying the long-term power constraint
\begin{IEEEeqnarray}{rCl}
\Ex{}{ \| \encoder (\msg, H) \|^2} \leq \bl\snr.
 \label{eq:power-constr-qs}
\end{IEEEeqnarray}
Here, $\msg$ is equiprobable on $\{1,\ldots,\NumCode\}$ and the average in~\eqref{eq:power-constr-qs} is with respect to the joint probability distribution of $\msg$ and $H$.
\item A decoder $\decoder$: $\complexset^{\bl} \times \complexset \to\{1,\ldots,\NumCode\}$ satisfying the average error probability constraint
\begin{equation}
\prob[\decoder(\randvecy, H) \neq \msg ] \leq \error
\label{eq:avg-error-qs}
\end{equation}
where $\randvecy$ is the channel output induced by the transmitted codeword $\vecx = \encoder(\msg, H)$ according to~\eqref{eq:channel-io-qs}.
\end{enumerate}

%\begin{IEEEeqnarray}{rCl}
%\|\encoder(J,H)\|^2 \leq \bl \snr.
%\end{IEEEeqnarray}
%
%
The maximal channel coding rate is defined as
\begin{IEEEeqnarray}{rCl}
\Rquasi^*(\bl,\error) \define \sup\mathopen{}\left\{\frac{\ln\NumCode}{\bl}: \,\, \exists\, (\bl,\NumCode,\error)_{\avg} \,\,\,\text{code}\right\}.
\end{IEEEeqnarray}
As discussed in Section~\ref{sec:intro}, the $\error$-capacity of the quasi-static fading channel~\eqref{eq:channel-io-qs} is
\begin{IEEEeqnarray}{rCl}
%\ceqslt(\snr) =
\lim \limits_{\bl\to\infty} \Rquasi^*(\bl,\error) = C\mathopen{}\big(\snr/\avgg\big)
\label{eq:ce-quasi}
\end{IEEEeqnarray}
where $C(\cdot)$ is defined in~\eqref{eq:def-capacity-awgn} and $\avgg$ in~\eqref{eq:avgg-def}.
Note that, for the AWGN case, a long-term power constraint yields a higher $\epsilon$-capacity compared to the short-term case only under the average probability of error formalism (and not under a maximal probability of error---see Section~\ref{sec:awgn-case}).
For the quasi-static fading case, the situation is different and~\eqref{eq:ce-quasi} holds
also if the average error probability constraint~\eqref{eq:avg-error-qs} is replaced by the maximal error probability constraint
\begin{equation}
\max\limits_{1\leq j\leq \NumCode} \prob[\decoder(\randvecy, H ) \neq \msg \given \msg =j ] \leq \error
\label{eq:max-error-qs}
\end{equation}
provided that $H$ is a continuous random variable.
Indeed, one way to achieve~\eqref{eq:ce-quasi} under the maximal error probability formalism~\eqref{eq:max-error-qs} is to employ the channel coefficient $H$ as the common randomness shared by the transmitter and the receiver. Using this common randomness, we can convert the average probability of error into a maximal probability of error by applying a ($H$-dependent) relabeling of the codewords.

If we replace~\eqref{eq:power-constr-qs} with the short-term power constraint
\begin{IEEEeqnarray}{rCL}
  %\IEEEeqnarraymulticol{3}{l}{...}
  % a & = & b +c
  \| \encoder (j, h) \|^2 \leq \bl\snr, \quad \forall j\in\{1,\dots,M\}, \,\,\, \forall h\in\complexset \IEEEeqnarraynumspace
  \label{eq:st-power-constr-qs}
\end{IEEEeqnarray}
then~\eqref{eq:ce-quasi} ceases to be valid  and the $\epsilon$-capacity is given by the well-known expression~$\Cawgn\mathopen{}\left(\snr \Finvs(\error)\right)$ (see, e.g.,~\cite{yang14-07a}).

Theorem~\ref{thm:quasi-static-second-order} below characterizes the first two terms in the asymptotic expansion of $ \Rquasi^\ast (\bl,\error)$ for fixed $0<\error<1$ and large $\bl$.

%The main result of this section is the following theorem.
\begin{thm}
\label{thm:quasi-static-second-order}
Assume that the input of the quasi-static fading channel~\eqref{eq:channel-io-qs} is subject to the long-term  power constraint $\snr$.
Let $0<\error <1$ be the average probability of error and assume that
\begin{enumerate}
\item $\Ex{}{G} <\infty$, where $G \define |H|^2$ is the channel gain;\label{item:cond-finite}
\item CSI is available at both the transmitter and the receiver;\label{item:cond-csirt}
\item $\Finvs(\cdot)$ defined in~\eqref{eq:def-finvs} is strictly positive in a neighborhood of $\error$, namely, $\exists\, \delta\in(0, \error)$ such that $\Finvs(\error-\delta)>0$.
\label{item:cond-conti}
\end{enumerate}
Then
\begin{IEEEeqnarray}{rCl}
 \Rquasi^\ast (\bl,\error)  = \Cawgn\mathopen{}\left(\frac{\snr}{\avgg}\right) - \sqrt{\Vawgn \mathopen{}\left(\frac{\snr}{\avgg} \right)} \sqrt{\frac{\ln\bl}{\bl}}+ \bigO\mathopen{}\left(\frac{1}{\sqrt{\bl}}\right)\IEEEeqnarraynumspace
 \label{eq:thm-qs}
\end{IEEEeqnarray}
where~$\Cawgn(\cdot)$ and~$\Vawgn(\cdot)$ are defined in~\eqref{eq:def-capacity-awgn} and~\eqref{eq:def-dispersion-awgn}, respectively, and $\bar{g}_{\error}$ is given in~\eqref{eq:avgg-def}.
%The asymptotic expansion~\eqref{eq:thm-qs}  holds also if the average error probability constraint~\eqref{eq:avg-error-qs} is replaced by the maximal error probability constraint~\eqref{eq:max-error-qs}.
\end{thm}
\begin{rem}
The AWGN channel~\eqref{eq:channel-io-awgn}, which can be viewed as a quasi-static channel with $H=1$ with probability one, satisfies all conditions in Theorem~\ref{thm:quasi-static-second-order}.
Indeed, Conditions~\ref{item:cond-finite} and~\ref{item:cond-csirt} in Theorem~\ref{thm:quasi-static-second-order} are trivially satisfied.
Condition~\ref{item:cond-conti} is also satisfied, since for an AWGN channel $\Finvs(\error)=1$ for every $\error \in(0,1)$.
Therefore, Theorem~\ref{thm:quasi-static-second-order} implies Theorem~\ref{thm:awgn-second-order} (for the AWGN case, we have that $\avgg = 1-\error$).
\end{rem}
%
%\begin{rem}
%\textcolor{blue}{Note that, if $\Finvs(\cdot)$ is  continuous and strictly positive at $\error$, then it satisfies Condition~\ref{item:cond-conti}. }
%\end{rem}
\begin{IEEEproof}
See Sections~\ref{sec:quasi-static-proof-conv} and~\ref{sec:proof-qs-ach} below.
\end{IEEEproof}

Before proving Theorem~\ref{thm:quasi-static-second-order}, we motivate the validity of~\eqref{eq:thm-qs} through a heuristic argument, which also illustrates the similarities and the differences between the AWGN and the quasi-static fading case.
Fix an arbitrary code with rate $R$ that satisfies the long-term power constraint~\eqref{eq:power-constr-qs}, and let $P_{\normx\given G}$ be the conditional probability distribution induced by the code on the normalized codeword power $\normx =\|\randvecx\|^2/\bl$ given $G$.
We shall refer to $P_{\normx\given G}$ as (stochastic) \emph{power controller}.
Note that $P_{\normx\given G}$ must be chosen so that (see~\eqref{eq:power-constr-qs})
\begin{IEEEeqnarray}{rCl}
\Ex{P_{\normx , G}}{ \normx } \leq \snr.
\label{eq:power-constraint-heur-qs}
\end{IEEEeqnarray}
For the quasi-static fading channel~\eqref{eq:power-constr-qs}, the effective power seen by the decoder is $\normx G$.
Thus, the minimum error probability $\error(P_{\normx\given G}) $ achievable with the power controller  $P_{\normx\given G}$ is roughly (cf.~\eqref{eq:norm-app-heuristic})
\begin{IEEEeqnarray}{rCl}
\epsilon(P_{\normx\given G}) \approx \Ex{P_{\normx, G}}{Q\mathopen{}\left(\sqrt{\bl}\frac{\Cawgn(\normx G) - R }{\sqrt{\Vawgn(\normx G)}}\right)}. \IEEEeqnarraynumspace
\label{eq:norm-app-heuristic-qs}
\end{IEEEeqnarray}
As in the AWGN case, we need to minimize the RHS of~\eqref{eq:norm-app-heuristic-qs} over all power controllers $P_{\normx\given G}$ satisfying~\eqref{eq:power-constraint-heur-qs}.
Because of Lemma~\ref{lemma:solusion-awgn-prob}, it is tempting to conjecture that, for sufficiently large~$\bl$, the optimal power controller should be such that $\normx G$ has two mass points, located at~$0$ and~$\tangentp$, respectively,  with~$\tangentp$ satisfying~\eqref{eq:omega_approx}.
This two-mass-point distribution can be achieved by choosing $\Pi(g)$ to be equal to $\omega_0/g$ with probability one if $g>\gth$, and to be $0$ with probability one if $g<\gth$.
For the case that the distribution of $G$ has a  mass point at $\gth$, i.e., $\prob[G=\gth]>0$, we need to choose $\Pi(\gth)$ to be a discrete random variable supported on $\{0,\omega_0/\gth\}$.
Here, the threshold~$\gth$ is chosen so as to guarantee that~\eqref{eq:power-constraint-heur-qs} holds with equality.
The resulting power controller corresponds to truncated channel inversion.\footnote{\label{footnote:tci-practical}For given $R$ and $\snr$, the truncated channel inversion scheme depends on the fading statistics only through the threshold $\gth$. For unknown fading statistics, the threshold can be estimated through the fading samples (see~\cite{caire99-05} for a detailed discussion).} Indeed, the fading channel is inverted if the fading gain is above $\gth$. Otherwise, transmission is silenced.
Although this truncated channel inversion power controller turns out to be optimal up to second order, in general it does not minimize the RHS of~\eqref{eq:norm-app-heuristic-qs} for any finite $n$.
This implies that some technicalities, which do not arise in the AWGN case, need to be taken care of in the proof of Theorem~\ref{thm:quasi-static-second-order}.

Using the truncated channel inversion power controller in~\eqref{eq:norm-app-heuristic-qs}, and then making use of~\eqref{eq:omega_approx}, we obtain (assuming for simplicity that $\prob[G=\gth]=0$)
\begin{IEEEeqnarray}{rCL}
  %\IEEEeqnarraymulticol{3}{l}{...}
  % a & = & b +c
  \epsilon&\approx& Q\mathopen{}\left(\sqrt{\ln n}\right)\Pr\{G\geq \gth\} +\Pr\{G< \gth\}  \IEEEeqnarraynumspace  \\
  &\approx& \Pr\{G< \gth\}
\end{IEEEeqnarray}
where the last approximation holds when $n$ is large. Using~\eqref{eq:def-finvs},
we conclude that the minimum error probability must satisfy
\begin{IEEEeqnarray}{rCl}
\gth \approx \Finvs(\error).
\label{eq:intui-gth}
\end{IEEEeqnarray}
Furthermore, combining~\eqref{eq:intui-gth} with~\eqref{eq:power-constraint-heur-qs}, we conclude that
\begin{IEEEeqnarray}{rCl}
\tangentp \approx \snr/\avgg
\label{eq:intui-app-tangentp}
\end{IEEEeqnarray}
where $\avgg$ was defined in~\eqref{eq:avgg-def}.
Finally, the desired result follows from~\eqref{eq:omega_approx} and~\eqref{eq:intui-app-tangentp} as follows
\begin{IEEEeqnarray}{rCl}
\Rquasi^*(\bl,\error) &\approx& \Cawgn(\tangentp) - \sqrt{\frac{\Vawgn(\tangentp)}{\bl}} \sqrt{\ln \bl}\\
&\approx& \Cawgn\mathopen{}\left(\frac{\snr}{\avgg}\right) - \sqrt{\Vawgn\mathopen{}\left(\frac{\snr}{\avgg} \right)} \sqrt{\frac{\ln \bl}{\bl}}. \IEEEeqnarraynumspace
\end{IEEEeqnarray}
 We next provide a rigorous justification for these heuristic steps.

\subsection{Proof of the Converse Part}
\label{sec:quasi-static-proof-conv}
%For simplicity of presentation, we shall assume throughout that~$G$ is a continuous random variable, i.e., that the probability density function (pdf) of~$G$ exists.
%
%Under this assumption, $\avgg$ in~\eqref{eq:avgg-def} takes the following simpler form
%\begin{IEEEeqnarray}{rCl}
%\avgg = \Ex{}{\frac{1}{G} \indfun{ G > \Finvs(\error) }}.
%\label{eq:def-avgg-continuous}
%\end{IEEEeqnarray}
%All steps can be easily extended to the case where~$G$ is not continuous by proceeding as in the proof of~\cite[Prop.~4]{caire99-05}.

The proof follows closely that of the converse part of Theorem~\ref{thm:awgn-second-order}.
We shall avoid repeating the parts that are in common with the AWGN case, and focus instead on the novel parts.
For the channel~\eqref{eq:channel-io-qs} with CSI at both the transmitter and the receiver, the input is the pair~$(\randvecx, H)$ and the output is the pair~$(\randvecy, H)$.
Consider an arbitrary $(\bl,\NumCode,\error)_\avg$ code.
To upper-bound $ \Rquasi^\ast (\bl,\error) $, we use the meta-converse theorem~\cite[Th.~26]{polyanskiy10-05}.
As auxiliary channel $Q_{\randvecy \given \randvecx H}$, we take a channel that passes~$H$ unchanged and generates $\randvecy$ according to the following distribution
\begin{IEEEeqnarray}{rCl}
Q_{\randvecy \given  \randvecx=\vecx,H=h} = \jpg\mathopen{}\left(\mathbf{0}, \left(1+ \frac{\|\vecx\|^2|h|^2}{\bl}\right)\matI_{\bl}\right). \IEEEeqnarraynumspace
\end{IEEEeqnarray}
Then,~\cite[Th.~26]{polyanskiy10-05}
\begin{IEEEeqnarray}{rCl}
%\inf_{P_{\randvecx \given H}}
\beta_{1-\error}(P_{\randvecx \randvecy H}, P_{H} P_{\randvecx\given H} Q_{\randvecy\given \randvecx H}) \leq 1-\error'  \IEEEeqnarraynumspace
\label{eq:meta-converse-quasi}
\end{IEEEeqnarray}
where $\error'$ is the average probability of error incurred by using the selected $(\bl,\NumCode,\error)_\avg$ code over the auxiliary channel $Q_{\randvecy \given \randvecx H}$, and $P_{\randvecx \given H}$ denotes the conditional probability distribution  on $\randvecx$ induced by the  encoder.

As in Section~\ref{sec:proof-awgn-sec-order-conv}, we next lower-bound the LHS of~\eqref{eq:meta-converse-quasi} using~\cite[Eq.~(102)]{polyanskiy10-05} as follows: for every $\gamma>0$
\begin{IEEEeqnarray}{rCl}
\IEEEeqnarraymulticol{3}{l}{
\beta_{1-\error}(P_{\randvecx \randvecy H}, P_{H} P_{\randvecx\given H} Q_{\randvecy\given \randvecx H}) }\notag\\
 \quad
  &\geq& e^{-\bl\gamma} \big|\prob[S_\bl(\normx G) \leq \bl\gamma] -\error\big|^{+} \IEEEeqnarraynumspace
\label{eq:lower-bd-beta-PXYH}
\end{IEEEeqnarray}
where  $S_\bl(\cdot)$ was defined in~\eqref{eq:info_density_awgn}, $\normx \define \|\randvecx\|^2/\bl$, and $G \define |H|^2$.
%
%%
%Under $  P_{\randvecx \randvecy H}$, the random variable $ \ln \frac{ d  P_{\randvecx \randvecy H}}{ d(P_{H} P_{\randvecx\given H} Q_{\randvecy\given \randvecx H})} $ has the same distribution as $S_\bl(\normx G)$, where
%%
%Using, we obtain
%
%
%
The RHS of~\eqref{eq:meta-converse-quasi} can be lower-bounded as follows (see Appendix~\ref{sec:proof-conv-q-qs})
\begin{equation}
1- \error' \leq \frac{ 1}{\NumCode}\bigg(1+ \sqrt{\frac{\bl}{2\pi}} \Ex{}{ \big| \ln G - \ln\eta_0 \big|^{+}} \bigg)
\label{eq:converse-q-quasi}
\end{equation}
where $\eta_0$ is the solution of
\begin{IEEEeqnarray}{rCl}
\Ex{}{ \Big|\frac{1}{\eta_0} - \frac{1}{G}\Big|^{+}} = \NumCode\snr.
\label{eq:def-eta_0-quasi-conv-q}
\end{IEEEeqnarray}
Let
\begin{IEEEeqnarray}{rCl}
\varrho_{\bl}(\NumCode) \define \frac{1}{\bl} \ln\mathopen{}\left( 1 + \sqrt{\frac{\bl}{2\pi}}\Ex{}{ \big| \ln G - \ln\eta_0 \big|^{+}} \right) \IEEEeqnarraynumspace
\label{eq:def-varrho-m}
\end{IEEEeqnarray}
where the dependence on $M$ is through $\eta_0$.
Substituting~\eqref{eq:lower-bd-beta-PXYH} and~\eqref{eq:converse-q-quasi} into~\eqref{eq:meta-converse-quasi}, taking the logarithm of both sides of~\eqref{eq:meta-converse-quasi}, and using~\eqref{eq:def-varrho-m}, we obtain
\begin{IEEEeqnarray}{rCl}
 \ln \NumCode  \leq \bl \gamma -\ln\mathopen{}\big| \prob[S_\bl(\normx G) \leq \bl\gamma] -\error \big|^{+} + \bl\varrho_{\bl}(\NumCode) . \IEEEeqnarraynumspace
 \label{eq:conv-quasi-static-code}
\end{IEEEeqnarray}
Note that the RHS of~\eqref{eq:conv-quasi-static-code} depends on the chosen $(\bl,\NumCode,\error)_{\lt}$ code only through the conditional probability distribution $P_{\normx \given G}$ that the encoder induces on $\normx = \|\randvecx\|^2/\bl$.
Maximizing the RHS of~\eqref{eq:conv-quasi-static-code} over all $P_{\normx\given G}$ satisfying~\eqref{eq:power-constraint-heur-qs},
we conclude that every~$(\bl, \NumCode, \error)_{\lt}$ code for the quasi-static fading channel~\eqref{eq:channel-io-qs} must satisfy
\begin{equation}
\ln \NumCode  \leq  \bl\gamma - \ln\mathopen{}\Big| \inf\limits_{P_{\normx \given G}}  \prob[S_\bl(\normx G) \leq \bl\gamma] -\error \Big|^{+} + \bl\varrho_{\bl}(\NumCode).
\label{eq:ub-R-quasi}
\end{equation}

We next characterize the asymptotic behavior of the RHS of~\eqref{eq:ub-R-quasi} for large $\bl$.
We start by analyzing $\varrho_{\bl}(\NumCode)$.
Choose an arbitrary $g_0>0$ such that $\prob[G> g_0] >0$.
If $\eta_0 \ge g_0$, we have
\begin{IEEEeqnarray}{rCl}
\varrho_{\bl}(\NumCode) &\leq& \frac{1}{\bl} \ln\mathopen{}\bigg( 1 + \sqrt{\frac{\bl}{2\pi}} \Ex{}{\big| \ln G - \ln g_0 \big|^{+} }\bigg)  \IEEEeqnarraynumspace
\label{eq:ub-varrho-qs-case1-1}\\
&\leq & \frac{1}{\bl} \ln\mathopen{}\bigg(1+\sqrt{\frac{\bl}{2\pi}} \frac{\Ex{}{G}}{g_0}\bigg) \label{eq:ub-varrho-qs-case1-2}\\
&=& \bigO\mathopen{}\left(\frac{\ln\bl}{\bl}\right). \label{eq:ub-varrho-qs-case1}
\end{IEEEeqnarray}
Here, in~\eqref{eq:ub-varrho-qs-case1-2} we used that $\ln x < x$ for every $x\in \NonnegReal$, in~\eqref{eq:ub-varrho-qs-case1} we used that $\Ex{}{G}<\infty$.
If $\eta_0 < g_0$,  we have
\begin{IEEEeqnarray}{rCl}
\Ex{}{\Big| \frac{1}{\eta_0} - \frac{1}{G} \Big|^{+} } &\geq& \Ex{}{\Big( \frac{1}{\eta_0} - \frac{1}{G}\Big)  \cdot\indfun{G > g_0 }} \IEEEeqnarraynumspace \\
&\geq& \frac{\prob[G>g_0]}{\eta_0} - \frac{\prob[G>g_0]}{g_0}. \IEEEeqnarraynumspace
\label{eq:lb-eta-0-qs}
\end{IEEEeqnarray}
%The restriction $\eta_0 <g_0$ comes without loss of generality because $\eta_0 $ decreases to $0$ as $\NumCode \to \infty$ (see~\eqref{eq:def-eta_0-quasi-conv-q}).
%
Combining~\eqref{eq:def-eta_0-quasi-conv-q} with~\eqref{eq:lb-eta-0-qs}, we obtain
\begin{IEEEeqnarray}{rCl}
\eta_0 \geq \left(\frac{\NumCode\snr }{\prob[G> g_0]} + \frac{1}{g_0}\right)^{-1}.
\end{IEEEeqnarray}
Since $\ln M\leq \bl \Cawgn(\snr/\avgg) + \littleo(\bl)$ (see~\eqref{eq:ce-quasi}), we have
\begin{IEEEeqnarray}{rCl}
\ln \eta_0 \geq - \bl \Cawgn(\snr/\avgg) +\littleo(\bl).
\label{eq:expan-eta_0-quasi}
\end{IEEEeqnarray}
Substituting~\eqref{eq:expan-eta_0-quasi} into~\eqref{eq:def-varrho-m},
\begin{IEEEeqnarray}{rCl}
\varrho_{\bl}(\NumCode) &\leq  &  \frac{1}{\bl}\ln\mathopen{}\left(1 + \sqrt{\frac{\bl}{2\pi}} \Ex{}{\log G \cdot\indfun{G > \eta_0}}  \right.\notag\\
&& \quad\, \left. + \, \sqrt{\frac{\bl}{2\pi}} \Big(\bl \Cawgn(\snr/\avgg) + \littleo(\bl) \Big)\prob[G> \eta_0]\right) \IEEEeqnarraynumspace \\
&\leq&  \frac{1}{\bl}\ln\mathopen{}\Big(\sqrt{\frac{\bl}{2\pi}}  \Ex{}{G}  + \bigO(\bl) \Big) \IEEEeqnarraynumspace\label{eq:converse-Q-quasi-expan-1}\\
&= & \bigO\mathopen{}\left(\frac{\ln\bl}{\bl}\right).
\label{eq:converse-Q-quasi-expan}
\end{IEEEeqnarray}
Here,~in~\eqref{eq:converse-Q-quasi-expan-1} we used again that $\ln x <x$ for every $x\in \NonnegReal$;~\eqref{eq:converse-Q-quasi-expan} follows because $\Ex{}{G} <\infty$.
Combining~\eqref{eq:ub-varrho-qs-case1} and~\eqref{eq:converse-Q-quasi-expan}, we conclude that%\footnote{Note that if the probability distribution of $G$ is discrete, then~\eqref{eq:converse-Q-quasi-expan-total} follows directly from Fano's inequality as in the AWGN case.}
\begin{IEEEeqnarray}{rCl}
\varrho_{\bl}(\NumCode)&\leq  &  \bigO\mathopen{}\left(\frac{\ln\bl}{\bl}\right).
\label{eq:converse-Q-quasi-expan-total}
\end{IEEEeqnarray}
Substituting~\eqref{eq:converse-Q-quasi-expan-total} into~\eqref{eq:ub-R-quasi} and dividing each side of~\eqref{eq:ub-R-quasi} by \bl, we obtain
\begin{IEEEeqnarray}{rCl}
\Rquasi^*(\bl,\error) &\leq & \gamma - \frac{1}{\bl} \ln\mathopen{}\Big| \inf\limits_{P_{\normx \given G}}\prob[S_\bl(\normx G) \leq \bl\gamma] -\error \Big|^{+} \quad \notag\\
\quad && +\,\bigO\mathopen{}\left(\frac{\ln\bl}{\bl}\right).  \qquad\qquad\qquad\qquad
\label{eq:ub-R-quasi2}
\end{IEEEeqnarray}

Next, we evaluate the second term on the RHS of~\eqref{eq:ub-R-quasi2}.
Applying the Berry-Esseen theorem and following similar steps as the ones reported in~\eqref{eq:prob-Sn-first-ineq}--\eqref{eq:prob-Sn-2nd-ineq}, we obtain that
\begin{IEEEeqnarray}{rCl}
\prob[S_\bl(\normx G) \leq \bl \argpn] \geq \Ex{}{\func_{\bl,\argpn}(\normx G)} -\frac{6 \cdot 3^{3/2}}{\sqrt{\bl}}   \IEEEeqnarraynumspace
\label{eq:prob-Sn-2nd-ineq-quasi}
\end{IEEEeqnarray}
where the function~$\func_{\bl,\argpn}(\cdot)$ was defined in~\eqref{eq:def-f-func}.
The infimum of $\Ex{}{\func_{\bl,\argpn}(\normx G)}$ over $P_{\normx\given G}$ can be computed exactly via the convex envelope $\hat{\func}(\cdot)$ of $\func_{\bl,\argpn}(\cdot)$ (see Section~\ref{sec:non-asy-qs}).
In particular, if the distribution of $G$ is discrete and takes finitely many (say $m$) values, then the minimizer $P^*_{\Pi\given G}$ is such that $\Pi G $ takes at most $m+1$ different values, and
the RHS of~\eqref{eq:ub-R-quasi2} can be analyzed using a similar approach as in the AWGN case.
However, the analysis becomes more involved when $G$ is nondiscrete.
%
%unlike the AWGN case, the resulting value of $\Ex{}{\func_{\bl,\argpn}(\normx G)}$ is complicated to analyze asymptotically.
%
%
To circumvent this difficulty, we next derive a lower bound on $\Ex{}{\func_{\bl,\argpn}(\normx G)}$, which is easier to analyze and is sufficient to establish~\eqref{eq:thm-qs}. Furthermore, as we shall see shortly, the resulting lower bound is minimized by truncated channel inversion.
Let $\gamma$ belong to the interval $(\Cawgn(\snr/\avgg)-\delta, \Cawgn(\snr/\avgg ) +\delta)$ for some $0<\delta<\Cawgn(\snr/\avgg)$ (recall that~\eqref{eq:ub-R-quasi2} holds for every $\gamma>0$).
Furthermore, let
\begin{IEEEeqnarray}{rCl}
\bl_1 &\define& \frac{2\pi \big(e^{2(\Cawgn(\snr/\avgg) +\delta)} -1\big) }{(\Cawgn(\snr/\avgg)-\delta)^2}\\
&\geq& 2\pi(e^{2\gamma} -1)\gamma^{-2}.
\end{IEEEeqnarray}
Using Lemma~\ref{lemma:solusion-awgn-prob}, we obtain that for all $\bl>\bl_1$ there exists a unique $\tangentp \in [e^{\argpn}-1,\infty)$ satisfying~\eqref{eq:def-x0-mass-point}  and~\eqref{eq:lb-q-n-gamma}.
Let
\begin{equation}
k(\bl,\argpn) \define - q_{\bl,\gamma}'(\tangentp).
\label{eq:def-slope-f-w0}
\end{equation}
Using~\eqref{eq:lb-q-n-gamma} and that $q_{\bl,\gamma}(x)\geq 0$, $\forall x\geq 0 $, we conclude that
\begin{IEEEeqnarray}{rCl}
\func_{\bl,\argpn}(x) \geq \big|1 - k(\bl,\argpn) x\big|^{+}, \quad\quad \forall x\geq 0.
\label{eq:lb-f-n-x}
\end{IEEEeqnarray}
Note that the lower bound  $\big|1 - k(\bl,\argpn) x\big|^{+}$ differs from the convex envelope $\hat{\func}(x)$ of $q_{n,\gamma}(x)$ by at most $1/\sqrt{n}$. Indeed, as it can be seen from Fig.~\ref{fig:fx-illustration}, for every $x\geq 0$,
\begin{IEEEeqnarray}{rCl}
\big| \hat{\func}(x) - |1-k(n,\argpn)x|^{+} \big| &\leq& q_{n,\gamma}(\omega_0)
  \approx Q(\sqrt{\ln n})
  \approx 1/\sqrt{n}.\notag\\
\end{IEEEeqnarray}
This suggests that if $\func_{\bl,\argpn}(x)$ is replaced with the lower bound $\big|1 - k(\bl,\argpn) x\big|^{+}$, then the RHS of~\eqref{eq:prob-Sn-2nd-ineq-quasi} is changed only by $1/\sqrt{n}$, which is immaterial for the purpose of establishing~\eqref{eq:thm-qs}.

%
%Substituting~\eqref{eq:lb-f-n-x} into~\eqref{eq:prob-Sn-2nd-ineq-quasi}, we obtain
%%
%\begin{IEEEeqnarray}{rCl}
%\prob[S_\bl(\normx G) \leq \bl \argpn] \geq \Ex{}{[1-k(\bl,\argpn) \normx G]^{+}} -\frac{6 \cdot 3^{3/2}}{\sqrt{\bl}}\IEEEeqnarraynumspace
%\label{eq:prob-Sn-final-ineq-quasi}
%\end{IEEEeqnarray}
%for all $\bl >\bl_1$.
%Through these steps, we replaced the function $\func_{\bl,\argpn}(\cdot) $ on the RHS of~\eqref{eq:norm-app-heuristic-qs} with a piecewise linear function, which is easier to analyze.
%

%To eliminate the dependency of the RHS of~\eqref{eq:prob-Sn-final-ineq-quasi} on $P_{\normx \given G}$,
We proceed to consider the following optimization problem
\begin{IEEEeqnarray}{rCl}
\inf\limits_{P_{\normx\given G} } \Ex{P_{\normx,G}}{\big|1- k(\bl,\argpn) \normx G\big|^{+}}
\label{eq:opt-prob-quasi-static}
\end{IEEEeqnarray}
where the infimum is over all conditional probability distributions $P_{\normx \given G}$ satisfying~\eqref{eq:power-constraint-heur-qs}.
The solution of~\eqref{eq:opt-prob-quasi-static} is given in the following lemma.
\begin{lemma}\label{lemma:inf}
%Let $\snr, k$  be positive constants and let $G$ be a nonnegative random variable.
Let
\begin{IEEEeqnarray}{rCl}
\gth\define \inf\left\{ t > 0:\,\Ex{}{\frac{1}{k(n,\gamma)  G}\indfun{G \geq  t} } \leq \snr \right\} \IEEEeqnarraynumspace
\label{eq:def-gth-quasi}
\end{IEEEeqnarray}
and\footnote{If $\prob[G=\gth]=0$ then $p^*(\gth)$ can be defined arbitrarily. }
\begin{IEEEeqnarray}{rCl}
p^*(g) \define \left\{
                 \begin{array}{ll}
                   1, & \hbox{if $g>\gth$} \\
                   \dfrac{\gth \big(\snr k - \Ex{}{G^{-1} \indfun{G > \gth} }\big)}{\prob[G=\gth]}, & \hbox{if $g=\gth$} \\
                   0, & \hbox{if $g<\gth$.}
                 \end{array}
               \right.\notag\\
%\IEEEeqnarraynumspace
\label{eq:def-gth-quasi}
\end{IEEEeqnarray}
Then, the conditional probability distribution $P^\ast_{\normx\given G}$ that minimizes~\eqref{eq:opt-prob-quasi-static} satisfies
\begin{IEEEeqnarray}{rCl}
    P^\ast_{\normx\given G}\mathopen{}\Big(\frac{1}{k(n,\gamma)g} \Big|  g\Big) =  p^*(g)\,\,\text{and}\,\,
    P^\ast_{\normx\given G}\mathopen{}\big(0 \given g\big) = 1-p^*(g).\,\,\,\,\notag\\
\label{eq:opt-powalloc-quasi}
\end{IEEEeqnarray}
\end{lemma}
\begin{IEEEproof}
See Appendix~\ref{app:proof-lemma-inf-qs}.
\end{IEEEproof}
Note that the minimizer~\eqref{eq:opt-powalloc-quasi} is precisely truncated channel inversion.
By Lemma~\ref{lemma:inf}, we have
\begin{IEEEeqnarray}{rCl}
\IEEEeqnarraymulticol{3}{l}{
\inf\limits_{P_{\normx\given G} } \Ex{P_{\normx, G}}{\big|1- k(\bl,\argpn) \normx G\big|^{+}} }\notag\\
\quad &=& \prob[G< \gth] + (1- p^*(\gth))\prob[G=\gth].
\label{eq:opt-prob-quasi-static-answer}
\end{IEEEeqnarray}
Substituting~\eqref{eq:opt-prob-quasi-static-answer},~\eqref{eq:lb-f-n-x}, and~\eqref{eq:prob-Sn-2nd-ineq-quasi} into~\eqref{eq:ub-R-quasi2}, we obtain
\begin{IEEEeqnarray}{rCl}
\IEEEeqnarraymulticol{3}{l}{
\Rquasi^*(\bl,\error)}\notag \\
\quad &\leq&  \gamma - \frac{1}{\bl} \ln\mathopen{}\bigg( \prob[G < \gth] +  (1- p^*(\gth))\prob[G=\gth]  \IEEEeqnarraynumspace \notag\\
&& \qquad\qquad\,\,\, \, - \,\frac{6\cdot 3^{3/2}}{\sqrt{\bl}}-\error \bigg)+ \bigO\mathopen{}\left(\frac{\ln\bl}{\bl}\right).
\label{eq:ub-R-quasi-simp}
\end{IEEEeqnarray}

We next choose $\gamma$ to be the solution of
\begin{IEEEeqnarray}{rCl}
\prob[G < \gth] +  (1- p^*(\gth))\prob[G=\gth] - \frac{6\cdot 3^{3/2}}{\sqrt{\bl}}-\error = \frac{1}{\sqrt{\bl}} \notag\\
\label{eq:def-gamma-quasi}
\end{IEEEeqnarray}
where the $\gth$ on the LHS of~\eqref{eq:def-gamma-quasi} depends on $\gamma$ through $k(n,\gamma)$.
Assume for a moment that the following relation holds
\begin{equation}
\rho k(n,\gamma) = \avgg + \bigO(1/\sqrt{n}), \quad n\to\infty.
\label{eq:evaluate-g-th-final}
\end{equation}
Combining~\eqref{eq:evaluate-g-th-final} with~\eqref{eq:def-slope-f-w0} and~\eqref{eq:def-x0-mass-point}, we obtain
\begin{IEEEeqnarray}{rCl}
\avgg - \frac{\snr}{\tangentp} (1-\func_{\bl,\gamma}(\tangentp)) =  \bigO\mathopen{}\left(\frac{1}{\sqrt{\bl}}\right).
\label{eq:condtion-gamma-quasi-static}
\end{IEEEeqnarray}
Solving~\eqref{eq:def-x0-mass-point} and~\eqref{eq:condtion-gamma-quasi-static} for $\tangentp$ and $\gamma$ by proceeding as in the converse proof for the AWGN case (see Appendix~\ref{app:proof-expan-gamma-awgn}), we conclude that
\begin{IEEEeqnarray}{rCl}
%\gamma_\bl(P_{\normx\given G}) \leq
\gamma = \Cawgn\mathopen{}\left(\frac{\snr}{\avgg}\right) - \sqrt{ \Vawgn\mathopen{}\left(\frac{\snr}{\avgg}\right) } \sqrt{\frac{\ln\bl}{\bl}} + \bigO\mathopen{}\left(\frac{1}{\sqrt{\bl}}\right). \IEEEeqnarraynumspace
\label{eq:asy-expan-gamma-quasi}
\end{IEEEeqnarray}
Observe now that $\gamma$ in~\eqref{eq:asy-expan-gamma-quasi}  belongs indeed to the interval $(\Cawgn(\snr/\avgg)-\delta, \Cawgn(\snr/\avgg) +\delta)$ for sufficiently large $\bl$.
The converse part of Theorem~\ref{thm:quasi-static-second-order} follows by substituting~\eqref{eq:asy-expan-gamma-quasi} and~\eqref{eq:def-gamma-quasi} into~\eqref{eq:ub-R-quasi-simp}.

To conclude the proof, it remains to prove~\eqref{eq:evaluate-g-th-final}. By~\eqref{eq:def-gamma-quasi} and~\eqref{eq:def-finvs}, we have that
\begin{equation}
\Finvs(\error) \leq \gth \leq \Finvs(\error + c_1/\sqrt{n})
\label{eq:gth-def-value}
\end{equation}
where $c_1 \define 1+ 6\cdot 3^{3/2}$.
If the LHS of~\eqref{eq:gth-def-value} holds with equality, i.e., if $\Finvs(\error) = \gth $, then we have
\begin{IEEEeqnarray}{rCl}
\snr \cdot k(\bl,\argpn) &=&
\Ex{}{\frac{1}{G}\indfun{G>\gth}} + \frac{p^*(\gth) \prob[G=\gth]}{\gth}\label{eq:evaluate-g-th-00} \IEEEeqnarraynumspace \\
&=&\Ex{}{\frac{1}{G}\indfun{G>\Finvs(\error)}}  + \frac{\prob[G < \Finvs(\error)] }{\Finvs(\error)} \notag\\
&&+ \,\frac{\prob[G=\Finvs(\error)] -\error -c_1/\sqrt{n} }{\Finvs(\error)} \label{eq:evaluate-g-th-01} \IEEEeqnarraynumspace\\
&=&\avgg - \frac{c_1}{\Finvs(\error)\sqrt{n}} \, .\label{eq:evaluate-g-th-0}
\end{IEEEeqnarray}
Here,~\eqref{eq:evaluate-g-th-00} follows from~\eqref{eq:def-gth-quasi};~\eqref{eq:evaluate-g-th-01} follows from~\eqref{eq:def-gamma-quasi}; and~\eqref{eq:evaluate-g-th-0} follows from~\eqref{eq:avgg-def}.

If the RHS of~\eqref{eq:gth-def-value} holds with strict inequality, i.e., $\Finvs(\error) < \gth$, then we have
\begin{IEEEeqnarray}{rCl}
\IEEEeqnarraymulticol{3}{l}{
\snr \cdot k(\bl,\argpn)  }\notag\\
\quad &=&  \Ex{}{\frac{1}{G}\indfun{G>\gth}}  + \frac{p^*(\gth)\prob[G=\gth]}{\gth} \,\,\,\IEEEeqnarraynumspace \\
&=& \underbrace{ \Ex{}{\frac{1}{G}\indfun{G>\Finvs(\error)}} + \frac{\prob[G\leq \Finvs(\error)] -\error}{\Finvs(\error)}}_{=\avgg} \IEEEeqnarraynumspace \notag\\
&&+\, \underbrace{\frac{p^*(\gth)\prob[G=\gth]}{\gth} -\frac{\prob[G\leq \Finvs(\error)]-\error}{\Finvs(\error)}} _{\define \delta_{1,n}} \notag\\
&& - \, \underbrace{  \Ex{}{ \frac{1}{G}\indfun{G\in(\Finvs(\error),\gth]} }  }_{\define \delta_{2,n}}\,.
\label{eq:evaluate-g-th-4} \IEEEeqnarraynumspace
\end{IEEEeqnarray}
The terms $\delta_{1,n}$  and $\delta_{2,n}$ defined on the RHS of~\eqref{eq:evaluate-g-th-4} can be evaluated as follows
\begin{IEEEeqnarray}{rCl}
0 &\geq &\delta_{1,n} -\delta_{2,n}\\
 &\geq& \frac{p^*(\gth)\prob[G=\gth]}{\gth} -  \Ex{}{\frac{1}{G}\indfun{G\in(\Finvs(\error),\gth)}} \notag\\
 &&-\, \frac{\prob[G=\gth]}{\gth} -\frac{\prob[G\leq \Finvs(\error)]-\error}{\Finvs(\error)}
 \\
&=&  - \mathopen{}\left(\frac{\error-\prob[G<\gth] + c_1/\sqrt{n}}{\gth} \right) - \frac{\prob[G<\gth] -\error}{\Finvs(\error)}\label{eq:evaluate-g-th-1} \IEEEeqnarraynumspace \\
&\geq & -\frac{c_1}{\Finvs(\error)\sqrt{n}} \,.\label{eq:evaluate-g-th-3}
\end{IEEEeqnarray}
Here,~\eqref{eq:evaluate-g-th-1} follows from~\eqref{eq:def-gamma-quasi}, and~\eqref{eq:evaluate-g-th-3} follows because, by~\eqref{eq:opt-prob-quasi-static-answer} and~\eqref{eq:gth-def-value}, $\error \leq  \prob[G<\gth] \leq  \error + c_1/\sqrt{n}$.
Since~$\Finvs(\error)\geq \Finvs(\error-\delta)>0$ by assumption,~\eqref{eq:evaluate-g-th-0},~\eqref{eq:evaluate-g-th-4},~\eqref{eq:evaluate-g-th-3} imply~\eqref{eq:evaluate-g-th-final}.

\subsection{Proof of the Achievability Part}
\label{sec:proof-qs-ach}
We build upon the proof of the achievability part of Theorem~\ref{thm:awgn-second-order} in Section~\ref{sec:proof-awgn-sec-order-ach}.
In the quasi-static case, the effective power seen by the decoder is $\normx G$, where $\normx = \|f(\msg,H)\|^2/\bl$ denotes the normalized power of the codeword $f(\msg,H)$.
The encoder uses the randomness in $G$ to shape the effective power distribution---i.e., the probability distribution of $\normx G$---to a two-mass-point probability distribution with mass points located at $0$ and $\snr/\avgg + \bigO(1/\sqrt{\bl\ln\bl})$, respectively.

Let
\begin{IEEEeqnarray}{c}
\error_\bl \define \frac{2}{\sqrt{\bl\ln \bl}} \, \text{ and }\, \error'_\bl \define \frac{\error - \error_\bl}{1- \error_\bl}.
\label{eq:set-error-n-qs}
\end{IEEEeqnarray}
For sufficiently large $\bl$, we have $\error_\bl <\error$ and, hence, $\error_\bl' >0$.
Let
\begin{IEEEeqnarray}{c}
\avggn \define \Ex{}{\frac{1}{G} \indfun{ \vphantom{\Big( \big)} \!G >  \Finvs(\error'_\bl)\!}\!} + \frac{\prob[G\leq \Finvs(\error'_n)] -\error_n'}{\Finvs(\error'_n)} \IEEEeqnarraynumspace
\label{eq:avggn-def}
\end{IEEEeqnarray}
and let
\begin{equation}
\snr_{\bl} \define \snr/ \avggn.
\label{eq:snr-n-def}
\end{equation}
We define a randomized truncated channel inversion power-allocation function $\Pi^{*}(g)$ for each $g \in \NonnegReal$ such that the conditional distribution of $\Pi^*$ given $G$ coincides with the one given in~\eqref{eq:opt-powalloc-quasi} with $k(n,\gamma)$ and $\gth$ replaced by $1/\snr_n$ and $\Finvs(\error_n')$, respectively.
%\begin{IEEEeqnarray}{c}
%\normxre(g) \define \frac{1}{g} \indfun{\vphantom{\big(}g > \Finvs(\error'_\bl)}.
%\label{eq:pc-qs-def}
%\end{IEEEeqnarray}
%
%This power allocation function corresponds to truncated channel inversion. Indeed, the fading channel is inverted if the gain is above $\Finvs(\error'_\bl)$. Otherwise, transmission is silenced.
%
Let $\NumCode_\bl$ denote the maximal number of length-$\bl$ codewords that can be decoded with \emph{maximal} probability of error not exceeding~$\error_\bl$ over the AWGN channel~\eqref{eq:channel-io-awgn} subject to the short-term power constraint~$\snr_\bl$.
Let the corresponding code be $(\bl, \NumCode_\bl, \error_\bl)_\st$  and its codewords be $\{\vecc_1,\ldots,\vecc_{\NumCode_\bl}\}$.

Consider now a code whose encoder $\encoder$ has the following structure
\begin{IEEEeqnarray}{rCl}
\encoder(j,h) = \sqrt{\frac{\Pi^{*}(|h|^2)}{\snr_n}} \vecc_j , \quad j\in\{1,\ldots, \NumCode_\bl\}, \,\, h\in \complexset. \IEEEeqnarraynumspace
\label{eq:truncate-ci}
\end{IEEEeqnarray}
%
%
%Note that the encoder in~\eqref{eq:truncate-ci} may be random when $|h|^2 = \Finvs(\error_n')$ and $\prob[|H|^2 = \Finvs(\error_n')]>0$.
%
This encoder can be made deterministic by assigning power $\snr_n/\Finvs(\error_n')$ to the first $M_n p^*(\Finvs(\error_n'))$ codewords, where $p^*(\cdot)$ is given in~\eqref{eq:def-gth-quasi}, and allocating zero power to the remaining codewords.
The resulting code satisfies the long-term power constraint. Indeed,
\begin{IEEEeqnarray}{rCl}
\IEEEeqnarraymulticol{3}{l}{
\frac{1}{\NumCode_\bl} \Ex{H}{\sum\limits_{j=1}^{\NumCode_\bl} \|\encoder(j,H) \|^2}  }\notag\\
\quad &=&\frac{1}{\NumCode_\bl} \Ex{P_{G}P_{\Pi\given G}^*}{\normx}  \sum\limits_{j=1}^{\NumCode_\bl} \|\vecc_j \|^2\label{eq:bound-power-qs-ach-1}  \IEEEeqnarraynumspace \\
&\leq& \snr.\label{eq:bound-power-qs-ach-3}
\end{IEEEeqnarray}
Here,~\eqref{eq:bound-power-qs-ach-1} follows from~\eqref{eq:truncate-ci},~\eqref{eq:bound-power-qs-ach-3} follows from~\eqref{eq:def-gth-quasi},~\eqref{eq:avggn-def}, and~\eqref{eq:snr-n-def}.
The maximal probability of error of the code is upper-bounded by
\begin{IEEEeqnarray}{rcl}
1 \cdot \error'_\bl + \error_\bl (1-\error'_\bl) = \error.
\label{eq:error-prob-qs}
\end{IEEEeqnarray}
Indeed, channel inversion is performed with probability
\begin{IEEEeqnarray}{rCl}
\IEEEeqnarraymulticol{3}{l}{
\prob[G>\Finvs(\error_n')] + \frac{\prob[G \leq \Finvs(\error_n')]-\error_n'}{\prob[G=\Finvs(\error_n')]}\cdot \prob[G=\Finvs(\error_n')] } \notag\\
\quad\quad &=& 1-\error'_n.
\end{IEEEeqnarray}
Channel inversion transforms the quasi-static fading channel into an AWGN channel.
Hence, the conditional error probability given that channel inversion is performed is upper-bounded by~$\error_\bl$.
When channel inversion is not performed, transmission is silenced and an error occurs with probability 1.
This shows that the code that we have just constructed is an $(\bl, \NumCode_\bl, \error)_\lt$ code, which implies that
\begin{equation}
\Rquasi^*(\bl,\error) \geq \frac{\ln \NumCode_\bl(n,\error_n)}{\bl}.
\label{eq:lb-Rquasi-1}
\end{equation}
From~Section~\ref{sec:proof-awgn-sec-order-ach}, we know that
\begin{IEEEeqnarray}{rCl}
\frac{\ln M_\bl}{\bl} \geq \Cawgn\mathopen{}\left(\snr_\bl\right) -\sqrt{\Vawgn(\snr_\bl)} \sqrt{\frac{\ln \bl}{\bl}}  + \bigO\mathopen{}\left(\frac{1}{\sqrt{\bl\ln\bl}}\right). \IEEEeqnarraynumspace
\label{eq:ub-logM-qs}
\end{IEEEeqnarray}

We next show that
\begin{equation}
\snr_\bl = \frac{\snr}{\avgg} + \bigO\mathopen{}\left(\frac{1}{\sqrt{\bl\ln\bl}}\right).
\label{eq:asy-behavior-snr-bl}
\end{equation}
The achievability part of Theorem~\ref{thm:quasi-static-second-order} follows then by substituting~\eqref{eq:asy-behavior-snr-bl} into~\eqref{eq:ub-logM-qs} and by a Taylor series expansion of $\Cawgn(\cdot)$ and $\Vawgn(\cdot)$
around $\snr/\avgg$.
To prove~\eqref{eq:asy-behavior-snr-bl}, we proceed as in~\eqref{eq:gth-def-value}--\eqref{eq:evaluate-g-th-3}, and obtain that
\begin{IEEEeqnarray}{rCl}
|\avggn - \avgg| \leq \frac{\error-\error_n'}{\Finvs(\error_n')}.\label{eq:diff-avgg-avggn}
\end{IEEEeqnarray}
Since $\error  - \error_n'= \bigO(1/\sqrt{n\ln n})$, and since $\Finvs(\cdot)$ is nondecreasing and positive at $\error-\delta$ for some $\delta>0$, we conclude that the RHS of~\eqref{eq:diff-avgg-avggn} is $\bigO(1/\sqrt{n \ln n})$.
This together with~\eqref{eq:snr-n-def} establishes~\eqref{eq:asy-behavior-snr-bl}.

\subsection{Convergence to Capacity}
\label{sec:converge-to-capacity-qs}
Motivated by the asymptotic expansion~\eqref{eq:thm-qs}, we define the normal approximation~$\Rqsltna(\bl,\error)$ of $\Rquasi^*(\bl,\error)$ as follows
\begin{IEEEeqnarray}{rCl}
\Rqsltna(\bl,\error) = \Cawgn\mathopen{}\left(\frac{\snr}{\avgg}\right) - \sqrt{\Vawgn \mathopen{}\left(\frac{\snr}{\avgg} \right)} \sqrt{\frac{\ln\bl}{\bl}}.  \IEEEeqnarraynumspace
\label{eq:normal-app-qs}
\end{IEEEeqnarray}
As for the AWGN case, we now compare the approximation~\eqref{eq:normal-app-qs} against nonasymptotic bounds.

\subsubsection{Nonasymptotic Bounds}
\label{sec:non-asy-qs}
An achievability bound can be obtained by numerically maximizing~\eqref{eq:lb-Rquasi-1} over all $\error_\bl \in (0,\error)$ with $\ln\NumCode_\bl(n, \error_\bl)$ given in~\eqref{eq:awgn-kappa-beta}. %
To obtain a nonasymptotic converse bound, we compute numerically the largest $M$ that satisfies~\eqref{eq:ub-R-quasi}, i.e.,
\begin{IEEEeqnarray}{rCl}
\Rquasi^*(\bl,\error) \leq \max\mathopen{}\left\{\frac{1}{\bl}\ln M: \,M\text{ satisfies~\eqref{eq:ub-R-quasi}}\right\}. \IEEEeqnarraynumspace
\end{IEEEeqnarray}
To this end, we need to solve the optimization problem $\inf_{P_{\normx\given G}} \Ex{P_{\normx,G}}{\prob[S_\bl(\normx G) \leq \bl\gamma]}$ on the RHS of~\eqref{eq:ub-R-quasi}. Next, we briefly explain how this is done.
%
%
%%
%The reason for doing so is that the infimum on the RHS of~\eqref{eq:ub-R-quasi} is difficult to evaluate numerically and replacing $\prob[S_\bl(\cdot) \leq \bl\gamma]$ by a piecewise linear function makes the resulting infimum analytically tractable.
Let
\begin{equation}
f_{n,\gamma} (x) \define \prob[S_\bl(x) \leq \bl\gamma], \quad x\in\NonnegReal. 
\end{equation}
For $\gamma$ values sufficiently close to $\Cawgn(\snr/\avgg)$ and for sufficiently large $\bl$, the function $f_{n,\gamma} (x) $ has a similar shape as $q_{n,\gamma}(x)$ (see Fig.~\ref{fig:fx-illustration}).
More precisely, $f_{\bl,\gamma}(0)=1$, $f_{\bl,\gamma}(\cdot)$ is monotonically decreasing, and there exists an $x_0>0$ such that $f_{n,\gamma} (x) $ is concave on $(0,x_0)$ and is convex on $(x_0,\infty)$.
Let $\hat{f}(x)$ be the convex envelope of $f_{n,\gamma} (x)$ over $\NonnegReal$.
It follows that $\hat{f}(x)$ coincides with the straight line connecting $(0,1)$ and $(x_1, f_{\bl,\gamma}(x_1))$ for $x\in[0,x_1]$, and equals $f_{\bl,\gamma}(x)$ for $x\in(x_1,\infty)$ for some $x_1>0$.
By Lemma~\ref{lemma:solusion-awgn-prob}, if $G$ is a continuous random variable, then
\begin{IEEEeqnarray}{rCl}
\inf_{P_{\normx\given G}} \Ex{P_{\normx,G}}{ f_{n,\gamma} (\normx G)} = \inf_{\normxre: \Ex{}{\normxre(G)}\leq \snr} \Ex{G}{\hat{f}(\normxre(G) G)} \IEEEeqnarraynumspace
\label{eq:inf-P-Sn-exact}
\end{IEEEeqnarray}
where the infimum on the RHS of~\eqref{eq:inf-P-Sn-exact} is over all functions $\normxre: \NonnegReal \to \NonnegReal$ satisfying $\Ex{}{\normxre(G)}\leq \snr$.
Since $\hat{f}$ is convex by construction, the minimization problem on the RHS of~\eqref{eq:inf-P-Sn-exact} can be solved using standard convex optimization tools~\cite[Sec.~5.5.3]{boyd04}.
In particular, if $G$ is a continuous random variable, then the solution $\normxre^*(\cdot)$ of~\eqref{eq:inf-P-Sn-exact} satisfies
%\begin{itemize}
%\item if $g < \tilde{g}_{\mathrm{th}}$ then $\omega^*(g) =0$ ;
%\item if $g  \geq \tilde{g}_{\mathrm{th}}$ then  $\omega^*(g)$ satisfies
%\begin{IEEEeqnarray}{rCl}
%f'_{\mathrm{c}} (\omega^*(g) g ) g + \mu =0
%\end{IEEEeqnarray}
%\end{itemize}
\begin{equation}
\normxre^*(g) = ({\hat{f}}')^{-1} \mathopen{}\left(\mu/g\right) \indfun{g \geq \tilde{g}_{\mathrm{th}}}
\end{equation}
where  $({\hat{f}}' )^{-1}$ denotes the inverse of the derivative of the function ${\hat{f}}(\cdot)$, and $\tilde{g}_{\mathrm{th}}>0$ and $\mu<0$ are the solution of
\begin{IEEEeqnarray}{rCl}
\normxre^*(\tilde{g}_{\mathrm{th}})\tilde{g}_{\mathrm{th}} = x_1
\end{IEEEeqnarray}
and
\begin{IEEEeqnarray}{rCl}
\int_{\tilde{g}_{\mathrm{th}}}^{\infty}  ({\hat{f}}')^{-1} \mathopen{}\left(\mu/g\right)  f_{G}(g) dg =\snr.
\end{IEEEeqnarray}

\begin{figure}[t]
	\centering
\includegraphics[scale=0.78]{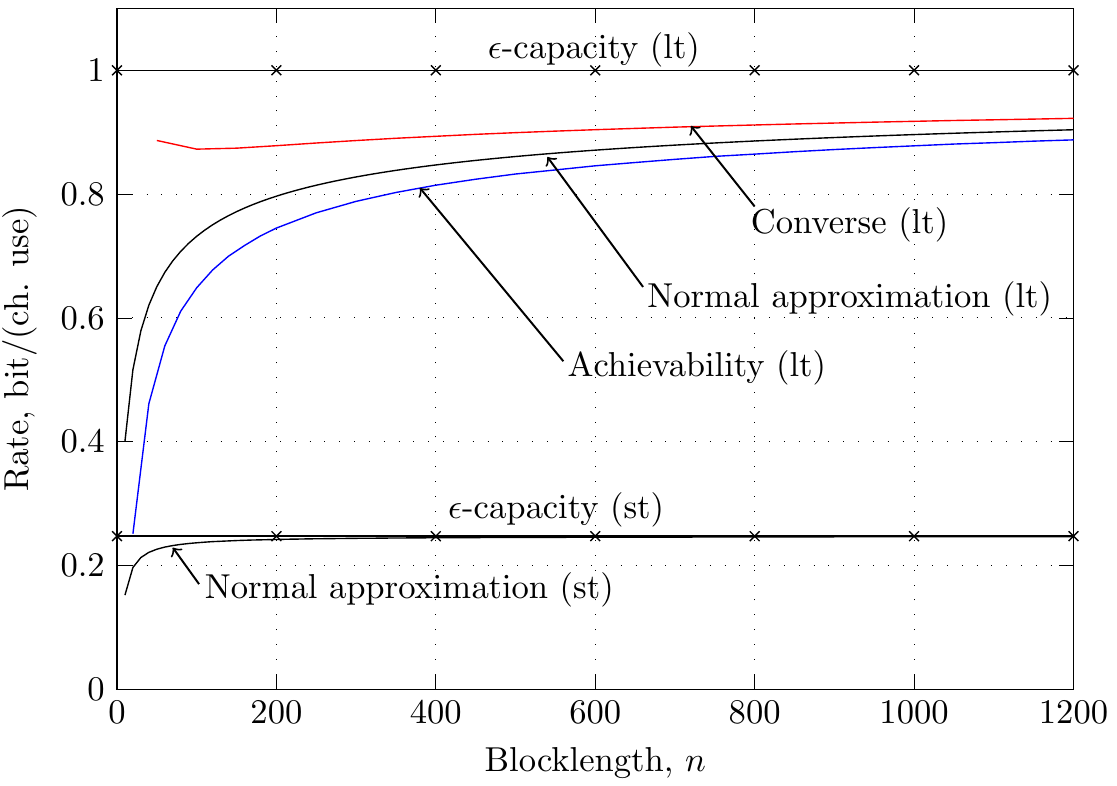}
\caption{Nonasymptotic bounds on~$\Rquasi^*(\bl,\error)$ and normal approximation for a quasi-static Rayleigh-fading channel with $\snr=2.5$ dB, and $\error=0.1$.
The normal approximation for the case of short-term power constraint is also depicted. Here, $\lt$ stands for long-term and $\st$ stands for short-term.
%\todo{Wei: ``Bound'' and ``Approximation'' should be lower case.}
\label{fig:bounds-qs-01}}
\end{figure}

\subsubsection{Numerical Results}
\label{sec:numerical-set}
In Fig.~\ref{fig:bounds-qs-01}, we compare the normal approximation~\eqref{eq:normal-app-qs} against the converse and achievability bounds for a quasi-static Rayleigh fading channel with $\snr =2.5$~dB and $\error=0.1$.
For comparison, we also show the normal approximation for the same channel with inputs subject to a short-term power constraint~(see~\cite[Eq.~(59)]{yang14-07a}).
As we can see from Fig.~\ref{fig:bounds-qs-01}, the gap between the normal approximation~\eqref{eq:normal-app-qs} and the achievability and converse bounds is less than $0.04$ bit$/$(ch. use) for blocklengths larger than 500.
We also observe that having a long-term power constraint in this scenario yields a significant rate gain\footnote{Note that the assumption of perfect CSIT is crucial to exploit the benefit of a long-term power constraint.} compared to the case of short-term power constraint already at short blocklengths.

\section{Conclusion}
\label{sec:conclusion}
In this paper, we studied the maximal channel coding rate for a given blocklength and error probability, when the codewords are subject to a long-term power constraint.
We showed that the second-order term in the large-\bl expansion of the maximal channel coding rate is proportional to $\sqrt{\bl^{-1}\ln\bl}$ for both AWGN channels and quasi-static fading channels with perfect CSI at the transmitter and the receiver.
This is in contrast to the case of short-term power constraint, where the second-order term is $\bigO(1/\sqrt{\bl})$  for AWGN channels and $\bigO(\bl^{-1}\ln\bl)$ for quasi-static fading channels.
We developed simple approximations for the maximal channel coding rate of both channels. We also discussed the accuracy of these approximations by comparing them to non-asymptotic achievability and converse bounds.

For AWGN channels, our results imply that a long-term power constraint is beneficial only when the blocklength or the error probability is large.
For example, for an AWGN channel with SNR of $0$ dB and block error probability equal to $10^{-3}$, a blocklength of $265\, 000$ is needed in order to benefit from a long-term power constraint.

For quasi-static fading channels, we showed that truncated channel inversion is both first- and second-order optimal.
This result is particularly appealing for practical wireless communication systems, since it is a common practice in such systems to maintain a certain target rate through power control.
Finally, numerical evidence shows that the rate gain resulting from CSIT and long-term power constraint occurs already at short blocklengths.

There are several possible generalizations of the results in this paper.
\begin{itemize}
\item One generalization is to consider the maximal achievable rate of codes under both short-term and long-term power constraints. In~\cite[Prop.~5]{caire99-05}, it is shown that to achieve the $\error$-capacity, one of the power constraints is always redundant. Using the approach developed in this paper, it is not difficult to show that this is also true if one wants to achieve the second-order term in the expansion of $R^*(n,\error)$.

\item Another direction is to consider a total energy constraint on $K$ successive packets for some finite $K$.
This constraint lies in between the short-term and the long-term ones (the short-term and the long-term power constraints correspond to $K=1$ and $K=\infty$, respectively).
Assuming that the channel gain is known \emph{causally} at the transmitter, the power-control policy that maximizes the outage capacity is obtained through dynamic programming, and no closed-form solutions are available in general~\cite{negi02-09a,caire04-10}.\footnote{If the $K$ channel gains are known noncausally at the transmitter, then the scheme that maximizes the outage probability is a variation of water-filling~\cite{caire99-05}.}
Determining the optimal power-control strategy under such a power constraint in the finite-blocklength regime is an open problem.

\item In this paper, we assume that perfect CSI is available at the transmitter.
A more realistic assumption is that the transmitter is provided with a noisy (or quantized) version of the fading coefficient.
The impact of nonperfect CSIT on the outage probability of quasi-static fading channels are studied in~\cite{kim10-11a,kim07-04}. In both papers, it is shown that nonperfect CSIT still yields substantial gains over the no-CSIT case.
Whether this remains true in the finite-blocklength regime requires further investigations.

\end{itemize}

\appendices

\section{Proof of~\eqref{eq:converse-q-awgn}}
\label{app:proof-awgn-converse-q}
%In order to prove~\eqref{eq:converse-q-awgn}, we need to establish a converse bound on the auxiliary channel~$Q_{\randvecy\given \randvecx}$.
%
According to~\eqref{eq:def-Q-awgn}, the output of the channel $Q_{\randvecy \given \randvecx}$ depends on the input~$\randvecx$ only through $\normx= \|\randvecx\|^2/\bl$.
Let $\normy  \define \|\randvecy\|^2/\bl$. Then,~$\normy$ is a sufficient statistic for the detection of $\randvecx $ from~$\randvecy$. Therefore, to establish~\eqref{eq:converse-q-awgn}, it suffices to lower-bound the average probability of error $\error'$ over the channel $Q_{\normy \given \normx}$ defined by
\begin{equation}
\normy = \frac{1+ \normx}{\bl} \sum\limits_{i=1}^{\bl} |Z_i|^2
\label{eq:eq-channel-q-u}
\end{equation}
where $\{Z_i\}$, $i=1,\ldots,\bl$, are i.i.d. $\jpg(0,1)$-distributed.
By taking the logarithm of both sides of~\eqref{eq:eq-channel-q-u}, the multiplicative noise in~\eqref{eq:eq-channel-q-u} can be converted into an additive noise.
This results in the following input-output relation
\begin{IEEEeqnarray}{rCl}
\lognormy \define \ln \normy = \ln(1+\normx) + \ln \sum\limits_{i=1}^{\bl} |Z_i|^2 -\ln\bl. \IEEEeqnarraynumspace
\label{eq:q-channel-log}
\end{IEEEeqnarray}
Given $\normx = \normxre$, the random variable $\lognormy$ is Log-Gamma distributed, i.e., its pdf is~\cite[Eq.~(2)]{lawless1980-08a}
\begin{IEEEeqnarray}{rCl}
q_{\lognormy\given \normx  }(\normyre \given \normxre)  &=&
\frac{\bl^\bl e^{ \bl \normyre - \bl \cdot e^\normyre/(1+\normxre)}}{ (1+\normxre)^\bl (n-1)!}.
 %\frac{\bl \normyre^{\bl-1} }{(1+\normxre)^\bl\Gamma(\bl)}\exp\mathopen{}\left(-\frac{\bl \normyre}{1+\normxre}\right)
\IEEEeqnarraynumspace
\end{IEEEeqnarray}
For later use, we note that  $q_{\lognormy\given \normx }(\normyre \given \normxre)$ can be upper-bounded as
\begin{IEEEeqnarray}{rCl}
q_{\lognormy\given \normx}(\normyre \given \normxre) &\leq & \frac{\bl^\bl e^{-\bl}}{  (n-1)!} \leq  \sqrt{\frac{\bl}{2\pi}}, \quad\forall \normyre \geq 0 \IEEEeqnarraynumspace
\label{eq:ub-pdf-q-2}
\end{IEEEeqnarray}
where the first inequality follows because $q_{\lognormy\given \normx }(\normyre \given \normxre)$ is a unimodal function with maximum at $\normyre = \ln (1+\normxre)$, and the second inequality follows from Stirling's formula~\cite[Eq.~(1)]{robbins55-01}.
%\begin{equation}
%q_{\normy\given \normx = \normxre}(\normyre) \leq \frac{\sqrt{\bl}}{1+ \normxre}, \quad\quad\quad \forall \normyre \geq 0.
%\label{eq:ub-pdf-q-awgn}
%\end{equation}
Note that the upper bound in~\eqref{eq:ub-pdf-q-2} is uniform in both~$u$ and~$w$.

Consider now the code for the channel $Q_{\lognormy \given \normx}$ induced by the $(\bl,\NumCode,\error)_{\lt}$ code chosen at the beginning of Section~\ref{sec:proof-awgn-sec-order-conv}.
By definition, the codewords $\{c_1, \ldots,c_{\NumCode}\} \subset \NonnegReal$ of the induced code (which are scalars) satisfy
\begin{IEEEeqnarray}{c}
\frac{1}{ \NumCode}\sum\limits_{j=1}^{\NumCode} c_j \leq \snr.
\label{eq:condition-sum-awgn}
\end{IEEEeqnarray}
Without loss of generality, we assume that the codewords are labeled so that
\begin{equation}
0\leq c_1 \leq \cdots\leq c_\NumCode.
\label{eq:condition-increase-awgn}
\end{equation}
Let $\{\setD_j\}$, $j=1,\ldots,\NumCode$, be the disjoint decoding sets, determined by the maximum likelihood (ML) criterion, corresponding to each of the $\NumCode$ codewords $\{c_j\}$.
For simplicity, we assume that all codewords are distinct. If two or more codewords coincide, we assign the decoding set to only one of the coinciding codewords  and choose the decoding set of the other codewords to be the empty set.

\begin{figure}[t]
\centering
\includegraphics[scale=0.75]{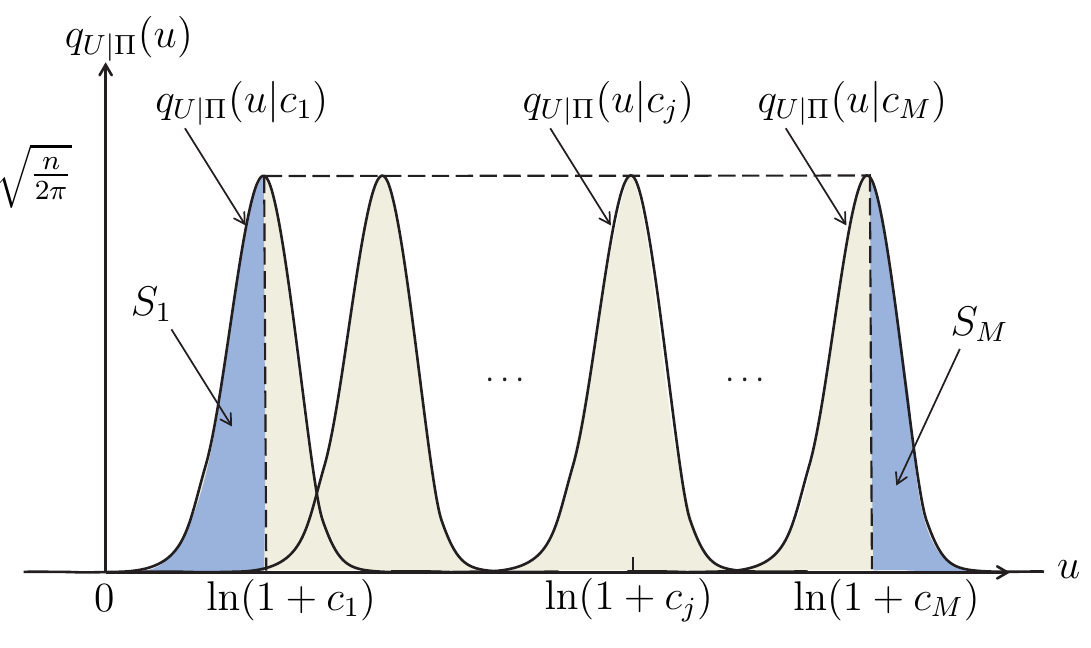}
\caption{\label{fig:pdfs-illustration} A geometric illustration of the probability of successful decoding under the ML criterion.
The average probability of success is equal to the area of the shaded regions (both grey and blue) divided by the number of codewords $\NumCode$. Note that the area of the shaded regions is upper-bounded by the sum of the area of the  dashed rectangle and the area of the blue-shaded regions~$S_1$ and~$S_\NumCode$.
%\todo{Wei: update the labels: get rid of $\log e$ and substitute $\log$ with $\ln$ on the x-axis. Also substitute $W$ with $\Pi$.}
}
\end{figure}

Next, we show that the interval $(-\infty,\ln(1+c_1))$ is included in the decoding set $\setD_1$, and that the interval
$(\ln(1+c_{\NumCode}), \infty)$ is included in~$\setD_{\NumCode}$.
Indeed, consider an arbitrary codeword $c_j$, $j\neq 1$.
The conditional pdf $q_{\lognormy\given \normx }(u \given c_j)$ can be obtained from $q_{\lognormy \given \normx}(u \given c_1)$ by a translation (see~\eqref{eq:q-channel-log} and Fig.~\ref{fig:pdfs-illustration})
\begin{equation}
q_{\lognormy\given \normx}(\normyre \given c_j) = q_{\lognormy \given \normx }\big(\normyre +\ln(1+c_{1})- \ln(1+c_j) \given c_1\big).
\label{eq:pdf-shift}
\end{equation}
Since $q_{\lognormy \given \normx}(u \given c_1)$ is strictly increasing on $(-\infty,\ln(1+c_1))$, we have
\begin{IEEEeqnarray}{rCl}
 q_{\lognormy \given \normx} \big(\normyre +\ln(1+c_{1})- \ln(1+c_j) \given c_1\big) &<& q_{\lognormy\given \normx }(\normyre \given c_1) \IEEEeqnarraynumspace
\end{IEEEeqnarray}
for all $u< \ln(1+c_1)$,  which implies that $q_{\lognormy\given \normx }(\normyre \given c_j) < q_{\lognormy\given \normx}(u\given c_1)$ on~$(-\infty,\ln(1+c_1))$.
Therefore,
\begin{equation}
(-\infty,\ln(1+c_1))\subset \setD_1 .
\label{eq:decoding-set-D1}
\end{equation}
The relation
\begin{equation}
(\ln(1+c_{\NumCode}), \infty) \subset \setD_{\NumCode}
\label{eq:decoding-set-DM}
\end{equation}
 can be proved in a similar way.

%\begin{IEEEeqnarray}{rCl}
%1-\error'
%\end{IEEEeqnarray}
%which follows because ML decoding minimizes the average probability of error for a given code.

The average probability of successful decoding $1-\error'$ is then upper-bounded as (see Fig.~\ref{fig:pdfs-illustration} for a geometric illustration)
\begin{IEEEeqnarray}{rCl}
1-\error' &\leq& \frac{1}{\NumCode} \sum\limits_{j=1}^{\NumCode} \int\nolimits_{\setD_j} q_{\lognormy\given \normx }(\normyre \given c_j) d\normyre\label{eq:ub-1-errorp-awgn-1}\\
&=&\frac{1}{\NumCode} \bigg(\int\nolimits_{-\infty}^{\ln(1+c_1)} q_{\lognormy\given \normx }(\normyre \given c_1) d\normyre \notag\\
&&\quad\quad +  \sum\limits_{j=1}^{\NumCode} \int\nolimits_{\setD_j \bigcap [\ln(1+c_1), \ln(1+c_{\NumCode})]} q_{\lognormy\given \normx }(\normyre \given c_j) d\normyre\notag
\\
&&    \quad\quad+  \int\nolimits_{\ln(1+c_{\NumCode})}^{\infty} q_{\lognormy\given \normx }(\normyre \given c_\NumCode) d\normyre \bigg) \label{eq:ub-1-errorp-awgn-2}\\
&\leq& \frac{1}{\NumCode}\bigg( 1 + \int\nolimits_{\ln(1+c_1)}^{\ln(1+c_{\NumCode})} \sqrt{\frac{\bl}{2\pi}} du \bigg)\label{eq:ub-1-errorp-awgn-3}\\
&\leq&\frac{1}{\NumCode}\bigg( 1 + \int\nolimits_{0}^{\ln(1+\NumCode\snr)} \sqrt{\frac{\bl}{2\pi}} du \bigg)\label{eq:ub-1-errorp-awgn-35}\\
& = &\frac{1}{\NumCode}\left(1+  \sqrt{\frac{\bl}{2\pi}} \ln(1+\NumCode \snr)\right).\label{eq:ub-1-errorp-awgn-4}
\end{IEEEeqnarray}
Here, \eqref{eq:ub-1-errorp-awgn-1} follows because ML decoding minimizes the average probability of error for a given code;~\eqref{eq:ub-1-errorp-awgn-2} follows from~\eqref{eq:decoding-set-D1} and~\eqref{eq:decoding-set-DM};~\eqref{eq:ub-1-errorp-awgn-3} follows from~\eqref{eq:ub-pdf-q-2} and because
\begin{equation}
\int\nolimits_{-\infty}^{\ln(1+c_1)} q_{\lognormy\given \normx }(\normyre \given c_1) d\normyre  + \int\nolimits_{\ln(1+c_{\NumCode})}^{\infty} q_{\lognormy\given \normx }(\normyre \given c_\NumCode) d\normyre =1;
\end{equation}
and~\eqref{eq:ub-1-errorp-awgn-35} follows because $0\leq c_1< c_{\NumCode} \leq \NumCode\snr$.
This concludes the proof of~\eqref{eq:converse-q-awgn}.

\section{Proof of~\eqref{eq:ratio-awgn-order}}
\label{app:proof-expan-gamma-awgn}
To prove~\eqref{eq:ratio-awgn-order}, we evaluate~\eqref{eq:def-x0-mass-point} and~\eqref{eq:def-tilde-gamma-n} for large $\bl$.
Let
\begin{IEEEeqnarray}{rCl}
y_0 \define \sqrt{\bl} \frac{\Cawgn(\tangentp) -\gamma}{\sqrt{\Vawgn(\tangentp)}}.
\label{eq:app-def-y-0-awgn}
\end{IEEEeqnarray}
Since $\tangentp \geq e^\gamma-1$ (see Lemma~\ref{lemma:solusion-awgn-prob}), we have $y_0\geq 0$, which implies that
\begin{equation}
\frac{1}{2} \leq Q(-y_0) = 1- \func_{\bl,\gamma}(\tangentp) \leq  1.
\label{eq:bound-q--y0}
\end{equation}
Solving~\eqref{eq:def-tilde-gamma-n} for $\tangentp$, we obtain
\begin{IEEEeqnarray}{rCl}
\tangentp &=& \frac{\snr \, Q(-y_0) }{1-\error -(6\cdot 3^{3/2} + 1 )/\sqrt{\bl} } \\
&=& \frac{\snr \, Q(-y_0)}{ 1-\error} + \bigO\mathopen{}\left(\frac{1}{\sqrt{\bl}}\right).
\label{eq:expan-w_0-app-awgn}
\end{IEEEeqnarray}
Next, we solve~\eqref{eq:def-x0-mass-point} for $y_0$. The first derivative of $\func_{\bl,\gamma}(\tangentp)$ is given by
\begin{IEEEeqnarray}{rCl}
\func'_{\bl,\gamma}(\tangentp) = -\frac{\sqrt{\bl}}{\sqrt{2\pi}} e^{-y_0^2/2} \varphi(\tangentp, \gamma)
\label{eq:deriv-func-gamma-n}
\end{IEEEeqnarray}
where
\begin{IEEEeqnarray}{rCl}
\varphi(\tangentp, \gamma) \define \frac{\Vawgn(\tangentp)(1+\tangentp)^2 - (\Cawgn(\tangentp)-\gamma)}{\sqrt{\Vawgn^3(\tangentp)} (1+\tangentp)^3}.
\label{eq:def-varphi-w-0-gamma}
\end{IEEEeqnarray}
Substituting~\eqref{eq:deriv-func-gamma-n} into~\eqref{eq:def-x0-mass-point}, we obtain
\begin{IEEEeqnarray}{rCl}
\sqrt{\bl}  e^{-y_0^2/2} = \frac{\sqrt{2\pi}\, Q(-y_0)}{ \tangentp \,\varphi(\tangentp, \gamma)}.
\label{eq:exp-miny_0-2}
\end{IEEEeqnarray}
Assume for a moment that
\begin{IEEEeqnarray}{rCl}
 k_1 +\littleo(1) \leq \varphi(\tangentp, \gamma)  \leq k_2 + \littleo(1)
\label{eq:expan-varphi-w0}
\end{IEEEeqnarray}
for some finite constants $0< k_1<k_2<\infty$.
Then, using~\eqref{eq:bound-q--y0},~\eqref{eq:expan-w_0-app-awgn}, and~\eqref{eq:expan-varphi-w0} in~\eqref{eq:exp-miny_0-2}, and then taking the logarithm of both sides of~\eqref{eq:exp-miny_0-2},  we obtain the sought-after
\begin{IEEEeqnarray}{rCl}
y_0 =\sqrt{ \ln\bl + \bigO(1)} = \sqrt{\ln \bl} + \littleo(1).
\label{eq:expan-y-0}
\end{IEEEeqnarray}
%Combining~\eqref{eq:expan-y-0} with~\eqref{eq:expan-w_0-app-awgn} and using that $Q(-\sqrt{\ln \bl}) = 1- \littleo(1/\sqrt{\bl})$, we have
%\begin{IEEEeqnarray}{rCl}
%\tangentp &=& \frac{\snr}{1-\error} + \bigO\mathopen{}\left(\frac{1}{\sqrt{\bl}}\right).
%\label
%\end{IEEEeqnarray}
%

To conclude the proof of~\eqref{eq:ratio-awgn-order}, it remains to demonstrate~\eqref{eq:expan-varphi-w0}.
We establish the upper bound in~\eqref{eq:expan-varphi-w0} through the following steps:
%
% Since $\tangentp \geq e^{\gamma}-1$ by assumption, $\varphi(\tangentp,\gamma)$ is upper-bounded by
\begin{IEEEeqnarray}{rCl}
\IEEEeqnarraymulticol{3}{l}{
\varphi(\tangentp,\gamma) }\notag\\
  &\leq& \frac{1}{\sqrt{\Vawgn(\tangentp)}(1+\tangentp)} \label{eq:ub-varphi-1-0}\\
&\leq& \frac{1}{1+ \tangentp} \! \left(1 - \frac{1}{\big(1+\snr/(2-2\error) + \bigO(1/\sqrt{\bl})\big)^2}\right)^{-1/2}\label{eq:ub-varphi-1} \IEEEeqnarraynumspace\\
%&=& \frac{1}{1+ \tangentp} \left(1-\frac{1}{\big(1+\snr/(1-\error)\big)^2}\right)^{-1/2} + \bigO\mathopen{}\left(\frac{1}{\sqrt{\bl}}\right)
&\leq & \left(1-\frac{1}{\big(1+\snr/(2-2\error)\big)^2}\right)^{-1/2} + \bigO\mathopen{}\left(\frac{1}{\sqrt{\bl}}\right)\label{eq:ub-varphi-1-2}.
\end{IEEEeqnarray}
Here,~\eqref{eq:ub-varphi-1-0} follows because $\tangentp \geq e^{\gamma}-1$;~\eqref{eq:ub-varphi-1} follows from~\eqref{eq:def-dispersion-awgn},~\eqref{eq:expan-w_0-app-awgn}, and the lower bound in~\eqref{eq:bound-q--y0}.

Next, we establish the lower bound in~\eqref{eq:expan-varphi-w0}.
Substituting both the lower bound in~\eqref{eq:bound-q--y0} and~\eqref{eq:ub-varphi-1} into~\eqref{eq:exp-miny_0-2}, we obtain
\begin{IEEEeqnarray}{rCl}
\IEEEeqnarraymulticol{3}{l}{
\sqrt{\bl} e^{-y_0^2/2} }\notag\\
&\geq& \frac{\sqrt{2\pi}}{2} \frac{1+\tangentp}{\tangentp} \left(1-\frac{1}{\big(1+\snr/(2-2\error)\big)^2}\right)^{1/2} \! + \bigO\mathopen{}\left(\frac{1}{\sqrt{\bl}}\right)  \notag\\
\\
 &\geq & \underbrace{\frac{\sqrt{2\pi}}{2} \left(1-\frac{1}{\big(1+\snr/(2-2\error)\big)^2}\right)^{1/2}}_{\define k_3}{} + \bigO\mathopen{}\left(\frac{1}{\sqrt{\bl}}\right).
 \label{eq:lb-e-y02}
\end{IEEEeqnarray}
Since $k_3 >0$, it follows from~\eqref{eq:lb-e-y02} that
\begin{IEEEeqnarray}{rCl}
y_0 &\leq& \sqrt{\ln \bl -2 \ln k_3 + \bigO(1/\sqrt{\bl})} \\
&=& \sqrt{\ln \bl} + \littleo(1). \IEEEeqnarraynumspace
\label{eq:ub-y0-app-awgn}
\end{IEEEeqnarray}
Using~\eqref{eq:ub-y0-app-awgn} in~\eqref{eq:app-def-y-0-awgn}, we obtain
\begin{IEEEeqnarray}{rCl}
C(\tangentp) -\gamma \leq \frac{\sqrt{\ln \bl} \sqrt{\Vawgn(\tangentp)}}{\sqrt{\bl} } + \littleo\mathopen{}\left(\frac{1}{\sqrt{\bl}}\right).
\label{eq:bound-C-w0-gamma}
\end{IEEEeqnarray}
Finally, utilizing~\eqref{eq:bound-C-w0-gamma}, we establish the desired lower bound on $\varphi(\tangentp,\gamma)$ as follows:
\begin{IEEEeqnarray}{rCl}
%\IEEEeqnarraymulticol{3}{l}{
\varphi(\tangentp,\gamma) &\geq& \frac{1}{\sqrt{\Vawgn(\tangentp)}(1+\tangentp)} - \frac{1}{\Vawgn(\tangentp) (1+\tangentp)^3} \sqrt{\frac{\ln\bl}{\bl}} \,\,\,
 \notag\\
&&+ \, \littleo\mathopen{}\left({1}/{\sqrt{\bl}}\right) \\
&\geq & \frac{1}{1+ \snr/(1-\error) + \bigO(1/\sqrt{\bl})}  + \littleo(1)\label{eq:lb-varphi-1}\\
&=& \frac{1}{1+\snr/(1-\error)} + \littleo(1).
\end{IEEEeqnarray}
Here, in~\eqref{eq:lb-varphi-1} we used~\eqref{eq:expan-w_0-app-awgn}, the upper bound in~\eqref{eq:bound-q--y0}, and that $\Vawgn(\tangentp)\leq 1$ for all $\tangentp \geq 0$.

\section{Proof of~\eqref{eq:converse-q-quasi}}
\label{sec:proof-conv-q-qs}
%We next relate $\error'$ to $\Rquasi^*(\bl,\error)$. The following lemma serves this purpose.
%\begin{lemma}
%\label{lemma:converse-q-quasi}
%For every code with $\NumCode$ codewords, blocklength $\bl$, and with long-term power constraint, the average probability of error $\error'$ over the channel $\outdist_{\randvecy H \given\randvecx H}$ satisfies
%
%\end{lemma}
%\begin{IEEEproof}
As in the proof of~\eqref{eq:converse-q-awgn}, it suffices to analyze the average probability of error~$\error'$ over the channel $Q_{\normy \given \normx G}$  with input-output relation (recall that $G=|H|^2$, $\normx = \|\randvecx\|^2/\bl$, and ~$\normy = \|\randvecy\|^2/\bl$)
\begin{equation}
\normy = \frac{1+ \normx G}{\bl} \sum\limits_{i=1}^{\bl} |Z_i|^2.
\end{equation}
Here, $\{Z_i\}$, $i=1,\ldots,\bl$, are i.i.d. $\jpg(0,1)$-distributed.

%Consider now the code for the channel $Q_{\normy G \given \normx G}$ induced by the $(\bl,\NumCode,\error)_{\lt}$ code introduced at the beginning of Section~\ref{sec:quasi-static-proof-conv}.
%
Let $P_{\normx \given G}$ be the conditional distribution of $\normx$ given $G$ induced by the $(\bl,\NumCode,\error)_{\lt}$ code introduced at the beginning of Section~\ref{sec:quasi-static-proof-conv}.
By assumption, $P_{\normx \given G}$ satisfies~\eqref{eq:power-constraint-heur-qs}.
Furthermore, let $\bar{\error}(g)$ be the conditional average probability of error over the channel $Q_{\normy \given \normx G}$ given $G=g$,
and let $\bar{\pi}(g) \define \Ex{}{\Pi \given G=g}$.
It follows from~\eqref{eq:converse-q-awgn} that
\begin{IEEEeqnarray}{rCl}
1 - \bar{\error}(g)\leq \frac{ 1}{\NumCode}\Big(1+\sqrt{\frac{\bl}{2\pi}}\ln(1+\NumCode \bar{\pi}(g) g)\Big).\IEEEeqnarraynumspace
\end{IEEEeqnarray}
Hence,
\begin{IEEEeqnarray}{rCl}
1-\error' &=& 1 - \Ex{}{\bar{\error}(G) } \\
&\leq& \frac{ 1}{\NumCode}\Big(1+ \sqrt{\frac{\bl}{2\pi}}\Ex{}{ \ln\mathopen{}\big(1+\NumCode \bar{\pi}(G) G\big)}  \Big). \IEEEeqnarraynumspace
\end{IEEEeqnarray}
The proof is concluded by noting that~\cite[Eq.~(7)]{goldsmith97-11a}
\begin{IEEEeqnarray}{rCl}
\sup\limits_{ \bar{\pi}: \Ex{}{\bar{\pi}(G)} \leq \snr} \!\!  \Ex{}{ \ln\mathopen{}\big(1+\NumCode \bar{\pi}(G) G\big)} = \Ex{}{ \big| \ln G - \ln\eta_0 \big|^{+} }  \IEEEeqnarraynumspace %\notag\\
\end{IEEEeqnarray}
where $\eta_0$ is defined in~\eqref{eq:def-eta_0-quasi-conv-q}.
%\end{IEEEproof}

\section{Proof of Lemma~\ref{lemma:inf}}
\label{app:proof-lemma-inf-qs}

To keep the mathematical expressions in this appendix compact, we shall indicate $k(\bl,\argpn)$ simply as $k$ throughout this appendix.
Let $P_{G}$ denote the probability distribution of the channel gain $G$.
We start by observing that the conditional probability distribution $P^*_{\normx \given G}$ specified in Lemma~\ref{lemma:inf} satisfies the constraint~\eqref{eq:power-constraint-heur-qs} with equality, i.e.,
\begin{IEEEeqnarray}{rCl}
\Ex{P_{G}P_{\Pi\given G}^*}{\Pi} =\snr.
\label{eq:opt-PW-pc}
\end{IEEEeqnarray}
Furthermore, it results in
\begin{IEEEeqnarray}{rCl}
\IEEEeqnarraymulticol{3}{l}{
 \Ex{P_G P^*_{\normx\given G}}{[1- k \normx G]^{+}} }\notag\\
\quad  &=& \prob[G < \gth] + (1-p^*(\gth))\prob[G=\gth] \define \error^*.\IEEEeqnarraynumspace
\end{IEEEeqnarray}
Consider now an arbitrary $P_{\normx \given G}$.
Let
\begin{IEEEeqnarray}{rCl}
\hat{\error}(g) &\define& \Ex{P_{\normx\given G=g}}{[1 - k \normx g]^{+} } \\
&=& \int\nolimits_{[0,1/(kg))}\!\!\! \left( 1 -  k \normxre g \right) dP_{\normx \given G} (\normxre\given g).\IEEEeqnarraynumspace
\label{eq:def-hat-error-g-quasi}
\end{IEEEeqnarray}
To prove Lemma~\ref{lemma:inf}, it suffices to show that if $\Ex{}{\hat{\error}(G)} $ is smaller than $\error^*$, then $P_{\normx\given G}$ must violate~\eqref{eq:power-constraint-heur-qs}.
Indeed, assume that $\Ex{}{\hat{\error}(g)} <\error^*  $. Then
\begin{IEEEeqnarray}{rCl}
\IEEEeqnarraymulticol{3}{l}{
\int_{0}^{\infty}\int_{0}^{\infty} \normxre dP_{\normx\given G}(\normxre\given g) \pdfG(g)  -\snr}\notag\\
&\geq&  \int_{0}^{\infty} \left(\int\nolimits_{[0,1/(k g))} \normxre  dP_{\normx\given G}(\normxre\given g) \right. \notag\\
&& \qquad \,\,\, \quad +\left.\int\nolimits^{\infty}_{1/(k g)} \frac{1}{k g}  dP_{\normx\given G}(\normxre \given g)\right) \pdfG(g) -\snr \IEEEeqnarraynumspace\\
& = & \int\nolimits_{0}^{\infty} \frac{1- \hat{\error}(g)}{k g} \pdfG(g) -\snr\label{eq:lb-power-W-snr-2}\\
%& =& \int\nolimits_{0}^{\infty} \frac{1- \hat{\error}(g)}{k g} \pdfG(g)  - \int\!\!\!\int \normxre dP_{\normx\given G}^*(\normxre\given g) \pdfG(g) \label{eq:lb-power-W-snr-3} \\
&=& \int\nolimits_{0}^{\infty} \frac{1- \hat{\error}(g)}{k g} \pdfG(g)  - \int\nolimits_{(\gth,\infty) }\!\frac{1}{kg} \pdfG(g)  \notag\\
&& -\, \frac{p^*(\gth) P_{G}[G=\gth]}{k\gth} \label{eq:lb-power-W-snr-3.25}\\
& =& \int\nolimits_{[0,\gth)}\frac{ 1- \hat{\error}(g)}{k g} \pdfG(g)  -  \int\nolimits_{(\gth,\infty )}\frac{\hat{\error}(g)}{k g} \pdfG(g) \notag\\
&&  + \, \frac{\prob[G=\gth]}{k\gth} \Big(1-p^*(\gth) - \hat{\error}(\gth)\Big)\IEEEeqnarraynumspace\label{eq:lb-power-W-snr-3.5} \\
&\geq &  \int\nolimits_{[0,\gth)}\frac{1-\hat{\error}(g)}{ k \gth}  \pdfG(g)  - \int_{(\gth,\infty)}\frac{\hat{\error}(g)}{k \gth}  \pdfG(g) \notag\\
&&+ \, \frac{\prob[G=\gth]}{k\gth} \Big(1-p^*(\gth) - \hat{\error}(\gth)\Big)
\label{eq:lb-power-W-snr-4}\IEEEeqnarraynumspace\\
&=&  \frac{1}{k \gth} \Big(\!\underbrace{\prob[G < \gth] + (1-p^*(\gth))\prob[G=\gth]}_{=\error^*}  -   \Ex{}{\hat{\error}(G)} \Big) \notag\\
\\
&>& 0.
\end{IEEEeqnarray}
Here,~\eqref{eq:lb-power-W-snr-2} follows from~\eqref{eq:def-hat-error-g-quasi}, and ~\eqref{eq:lb-power-W-snr-3.25} follows from~\eqref{eq:opt-PW-pc} and~\eqref{eq:opt-powalloc-quasi}.
This concludes the proof.

%
%\bibliographystyle{IEEEtran}
%\bibliography{IEEEabrv,publishers,confs-jrnls,../../WeiBib}


\begin{IEEEbiographynophoto}
{Wei Yang}(S'09)
received the B.E. degree in communication engineering and M.E. degree in communication and information systems from the Beijing University of Posts and Telecommunications, Beijing, China, in 2008 and 2011, respectively. He is currently pursuing a Ph.D. degree in electrical engineering at Chalmers University of Technology, Gothenburg, Sweden. In 2012 (July--August) and 2014 (July--September) he was a visiting student at the Laboratory for Information and Decision Systems, Massachusetts Institute of Technology, Cambridge, MA.

Mr. Yang is the recipient of a Student Paper Award at the 2012 IEEE International Symposium on Information Theory (ISIT), Cambridge, MA, and the 2013 IEEE Sweden VT-COM-IT joint chapter best student conference paper award.
His research interests are in the areas of information and communication theory.
\end{IEEEbiographynophoto}

\begin{IEEEbiographynophoto}{Giuseppe  Caire} (S'92--M'94--SM'03--F'05)
was born in Torino, Italy, in 1965. He received the B.Sc. in Electrical Engineering  from Politecnico di Torino (Italy), in 1990,
the M.Sc. in Electrical Engineering from Princeton University in 1992   and the Ph.D. from Politecnico di Torino in 1994.
He has been a post-doctoral research fellow with the European Space Agency (ESTEC, Noordwijk, The Netherlands) in 1994-1995,
Assistant Professor in Telecommunications at the Politecnico di Torino, Associate Professor at the University of Parma, Italy,
Professor with the Department of Mobile Communications at the Eurecom Institute,  Sophia-Antipolis, France, and he is currently a professor of Electrical Engineering with the Viterbi School of Engineering, University of Southern California, Los Angeles and an Alexander von Humboldt Professor
with the Electrical Engineering and Computer Science Department of the Technical University of Berlin, Germany.

He served as Associate Editor for the IEEE Transactions on Communications in 1998-2001 and as Associate Editor for the IEEE Transactions on Information Theory in 2001-2003.  He received the Jack Neubauer Best System Paper Award from the IEEE Vehicular Technology Society in 2003,  the
IEEE Communications Society \& Information Theory Society Joint Paper Award in 2004 and in 2011, the Okawa Research Award in 2006,
the Alexander von Humboldt Professorship in 2014, and the Vodafone Innovation Prize in 2015.
Giuseppe Caire is a Fellow of IEEE since 2005.  He has served in the Board of Governors of the IEEE Information Theory Society from 2004 to 2007,
and as officer from 2008 to 2013. He was President of the IEEE Information Theory Society in 2011.
His main research interests are in the field of communications theory, information theory, channel and source coding
with particular focus on wireless communications.
\end{IEEEbiographynophoto}

\begin{IEEEbiographynophoto}{Giuseppe Durisi}(S'02--M'06--SM'12)
 received the Laurea degree summa cum laude and the Doctor degree both from Politecnico di Torino, Italy, in 2001 and 2006, respectively. From 2002 to 2006, he was with Istituto Superiore Mario Boella, Torino, Italy. From 2006 to 2010 he was a postdoctoral researcher at ETH Zurich, Zurich, Switzerland. Since 2010, he has been with Chalmers University of Technology, Gothenburg, Sweden, where is now associate professor. He is also guest researcher at Ericsson, Sweden.

Dr. Durisi is a senior member of the IEEE. He is the recipient of the 2013 IEEE ComSoc Best Young Researcher Award for the Europe, Middle East, and Africa Region, and is co-author of a paper that won a "student paper award" at the 2012 International Symposium on Information Theory, and of a paper that won the 2013 IEEE Sweden VT-COM-IT joint chapter best student conference paper award.  From 2011 to 2014 he served as publications editor for the IEEE Transactions on Information Theory. His research interests are in the areas of communication and information theory, and compressed sensing.

\end{IEEEbiographynophoto}

\begin{IEEEbiographynophoto}{Yury Polyanskiy}(S'08--M'10--SM'14)
is an Associate Professor of Electrical Engineering and Computer Science and a member of LIDS at MIT.
Yury received the M.S. degree in applied mathematics and physics from the Moscow Institute of Physics and Technology, Moscow, Russia in 2005 and the Ph.D. degree in electrical engineering from Princeton University, Princeton, NJ in 2010. In 2000-2005 he lead the development of the embedded software in the Department of Surface Oilfield Equipment, Borets Company LLC (Moscow). Currently, his research focuses on basic questions in information theory, error-correcting codes, wireless communication and fault-tolerant and defect-tolerant circuits.
Dr. Polyanskiy won the 2013 NSF CAREER award and 2011 IEEE Information Theory Society Paper Award.
\end{IEEEbiographynophoto}

\end{document}